\documentclass[twocolumn,showpacs,preprintnumbers,amsmath,amssymb,nofootinbib]{revtex4-1}
\usepackage{graphicx}
\usepackage{amsmath}
\usepackage{color}
\usepackage{colordvi}
\usepackage{graphics}
\usepackage{amssymb}
\usepackage{natbib}
\usepackage{dcolumn}
\usepackage{bm}

\def\apj{Astrophys.~J.}

\def\apjs{{ApJ\ Suppl.}}

\def\aa{{A\&A}}

\def\mnras{{MNRAS}}

\def\plb{{Phys.\ Lett.\ B}}

\def\prd{{Phys. Rev. D}}
\def\prl{{Phys.~Rev.~Lett.}}

\newcommand{\eV}{\mathrm{eV}}
\newcommand{\keV}{\mathrm{keV}}
\newcommand{\MeV}{\mathrm{MeV}}
\newcommand{\GeV}{\mathrm{GeV}}
\newcommand{\cm}{\mathrm{cm}}
\newcommand{\fabs}{f_{\mathrm{abs.}}}
\newcommand{\ignore}[1]{}
\newcommand{\phgion}{\phi_{\gamma,\mathrm{ion.}}}
\newcommand{\phgcom}{\phi_{\gamma,\mathrm{Com.}}}
\newcommand{\phgpp}{\phi_{\gamma,\mathrm{pair}}}
\newcommand{\pheion}{\phi_{e,\mathrm{ion.}}}
\newcommand{\phecom}{\phi_{e,\mathrm{Com.}}}

\newcommand{\Hy}{\mathrm{H}}
\newcommand{\He}{\mathrm{He}}
\newcommand{\nb}{n_b}
\newcommand{\nH}{n_\Hy}
\newcommand{\nHe}{n_\He}
\newcommand{\eps}{\epsilon}
\newcommand{\CMB}{\mathrm{CMB}}
\newcommand{\sv}{\langle\sigma\upsilon\rangle}
\newcommand{\epm}{e^{\pm}}

\begin{document}

\title[Signatures of clumpy dark matter in the global 21\,cm background signal]{Signatures of clumpy dark matter in the global 21\,cm background signal}

\author{Daniel T. Cumberbatch}
\email{D.Cumberbatch@sheffield.ac.uk}
\affiliation{Astroparticle Theory \& Cosmology Group, Department of Physics \& Astronomy, The University of Sheffield, Hicks Building, Hounsfield Road, Sheffield, S3 7RH, U.K}
\author{Massimiliano Lattanzi}%
\affiliation{International Centre for Relativistic Astrophysics and Dipartimento di Fisica, Universit\`a di Roma "La Sapienza", Piazzale A. Moro 5, 00183, Roma, Italy.}
\email{lattanzi@icra.it}
\author{Joseph Silk}
\affiliation{Oxford Astrophysics, Denys Wilkinson Building, Keble Road, Oxford, OX1 3RH, U.K.}%
\email{silk@astro.ox.ac.uk}

\label{firstpage}

\begin{abstract}
\noindent We examine the extent to which the self-annihilation of supersymmetric neutralino dark matter, as well as light dark matter, influences the rate of heating, ionisation and Lyman-$\alpha$ pumping of interstellar hydrogen and helium and the extent to which this is manifested in the 21\,cm global background signal. We fully consider the enhancements to the annihilation rate from DM halos and substructures within them. We find that the influence of such structures can result in significant changes in the differential brightness temperature, $\delta T_b$. The changes at redsfhits $z<25$ are likely to be undetectable due to the presence of the astrophysical signal; however, in the most favourable cases, deviations in $\delta T_b$, relative to its value in the absence of self-annihilating DM, of up to $\simeq20\,{\rm mK}$ at $z=30$ can occur. 
Thus we conclude that, in order to exclude these models, experiments measuring the global 21\,cm signal, such as EDGES and CORE, will need to reduce the systematics at 50\,MHz to below 20\,mK.
\end{abstract}


\maketitle


\section{Introduction}
\noindent The standard cosmological model, motivated by measurements of temperature anisotropies in the Cosmic Microwave Background (CMB) \citep{Spergel:2003cb,Spergel:2006hy,Komatsu:2008hk,Dunkley:2008ie}, 
the large scale distribution of galaxies \citep{Cole:2005sx,Tegmark:2006az}, and by evidence of the accelerated expansion of the Universe from supernova observations \citep{Astier:2005qq,WoodVasey:2007jb}, requires that the Universe possesses a flat spatial geometry with a corresponding critical density, approximately 27 percent of which consists of physical matter. However these observations also indicate that only 4 percent of this matter is baryonic in nature, implying that the remaining 23 percent consists of an elusive, non-baryonic component called dark matter (DM) owing to the severe constraints that current astronomical data sets on its radiative capabilities.

Despite this compelling evidence for the existence of DM, its precise nature is still a topic of debate. Particle physicists have independently supported DM by postulating the existence of a variety of exotic particles with wide-ranging properties that may potentially solve problems in particle physics whilst resulting in a thermal relic particle density that is consistent with current observational constraints. 

The most intensely studied DM candidate is the lightest neutralino \citep*{bertone_review}, a weakly-interacting massive particle (WIMP) motivated by supersymmetric extensions of the Standard Model of
particle physics. In many of these extensions the neutralino is the lightest supersymmetric particle (LSP). In theories where the LSP is stable, for example theories where R-Parity is a conserved quantum number \citep*{Weinberg:1981wj,
Hall:1983id,Allanach:1999ic}, the neutralino is thus a highly-motivated DM candidate. Furthermore, an attractive feature of  neutralinos is that a large region of the relevant supersymmetric parameter space can be investigated using CERN's Large Hadron Collider (LHC)\footnote{www.cern.ch/LHC}. 

Whilst neutralino DM is ``cold'', owing to its negligible free-streaming length (i.e. the length scale below which fluctuations in DM density are suppressed), warm DM (WDM) is typically lighter and possesses a much longer free-streaming length. WDM is a viable alternative to cold dark matter (CDM) models which may potentially resolve several shortfalls of the standard CDM model, such as for example the over-prediction of low mass satellites and the existence of cuspy halos \citep*{hogan2000,Dalcanton:2000hn,avilareese,colin2008}. Among WDM candidates, there are sterile neutrinos \citep*{Dodelson:1993je,Asaka:2005an,Asaka:2005pn}, majorons \citep*{Akhmedov:1992hh,Berezinsky:1993fm,Lattanzi:2007ux}
and light DM (LDM) particles \citep{Boehm:2003hm}. What makes LDM interesting for this study is the fact that it can self-annihilate, as opposed to other forms of WDM, and therefore its annihilation rate can be enhanced by overdensities.

In this paper,  we re-examine the influence of neutralino and LDM annihilations on the thermal history of the Universe at times between the epochs of recombination and reionisation, commonly referred to as the ``Dark Ages'', when gas existed in a nearly uniform, dark, neutral state. The investigation of the Dark Ages is one of the frontiers of modern cosmology, and will be carried on by a new generation of radio interferometers such as LOFAR\footnote{
http://www.lofar.org
}, 
MWA\footnote{
http://www.haystack.mit.edu/ast/arrays/mwa
},
21\,CMA\footnote{
http://web.phys.cmu.edu/past/
},
and SKA\footnote{
http://www.skatelescope.org
}, as well as single antenna experiments such as EDGES\footnote{
http://www.haystack.mit.edu/ast/arrays/Edges/index.html
} 
and CORE.

These experiments will look for the redshifted 21\,cm signal associated with the hyperfine triplet-singlet transition of neutral hydrogen. If DM annihilates or decays, the resulting products subsequently collide and heat the surrounding gas, increasing its kinetic temperature and ionisation fraction. This is in turn manifested as distinct features in the 21\,cm background signal that can be used to constrain the properties of DM \cite{furlanetto,Shchekinov:2006eb,valdes}. 

With few exceptions (e.g.~\cite{Chuzhoy:2007fg,Yuan:2009xq}), 
past studies proclaim that the heating effects associated with the annihilation of SUSY WIMP DM are too small to be detected by current radio interferometers. However, these studies overlook the enhancements to the DM annihilation rate in galactic halos and in their substructures \citep{taylor2003}, which could be large enough to make the DM signature detectable by the next generation radio telescopes. In this paper, we calculate the effect of neutralino and LDM annihilations on the 21\,cm signal when accounting for the effect of DM clustering.\\

\noindent The rest of this paper is organized as follows. In \S\,\ref{sec:Basic_Physics} we elaborate on the basic properties of neutralinos and LDM. We also discuss the basic physics describing the way in which energy from annihilations is injected into the intergalactic medium (IGM). In \S\,\ref{sec:EG_DM_ann_rate} and \S\,\ref{sec:DM_ann_rate} we calculate the enhancement in the DM annihilation rate caused by the presence of halos and their substructures. In \S\,\ref{sec:Absorbed_Fraction} we estimate how much of the energy produced in a single DM annihilation is
actually injected into the IGM. In \S\,\ref{sec:21cm_Background} we discuss the modifications to the differential equations describing the evolution of the ionised fraction and kinetic temperature of the IGM and subsequently use these equations to calculate the modified 21\,cm background. In \S\,\ref{sec:Results} we calculate the predicted 21\,cm background for our benchmark neutralino and LDM models and 
discuss the potential for a detection. Finally, in \S\,\ref{sec:Discussion} we summarise our results and draw our conclusions.


\section{Dark matter candidates}
\label{sec:Basic_Physics}

\noindent The lightest supersymmetric (SUSY) neutralino is a superposition of higgsinos, winos and binos. 
Consequently, neutralinos are electrically neutral and colourless, only interacting weakly and gravitationally, and hence very difficult to detect directly. 
In SUSY models that conserve R-parity, the LSP is stable \citep{Weinberg:1981wj,Hall:1983id,Allanach:1999ic}. Consequently, in a scenario where present-day CDM exists as a result of thermal freeze-out, the dominant species of CDM could quite possibly include the LSP.
The relic density of the LSP will then heavily depend on its mass and annihilation cross section. 
Throughout this paper we assume that the LSP is the lightest SUSY neutralino. 
The neutralino is a popular candidate for CDM because the theoretically-motivated values of these parameters yield a corresponding value of the relic density that is in good agreement with observations (for a more detailed review of the various properties and motivations for neutralino DM see, e.g.,~\citet{bertone_review}).

Neutralinos possess a wide-range of annihilation spectra owing to the vast extent of currently unexcluded SUSY parameter space. Owing to the Majorana nature of the neutralino, its annihilation to fermionic channels is suppressed by a factor proportional to the square of the mass of the final state. This means that, if the neutralino is lighter than the $W^{\pm}$ and $Z$ bosons, annihilations will be dominated by the process $\chi\chi\rightarrow b\bar{b}$ with a minor contribution by $\chi\chi\rightarrow\tau^+\tau^-$. Assuming annihilations are 
dominated by the former process, the resulting spectrum will depend entirely on the LSP mass. For heavier 
LSPs, the annihilation products become more complex, often determined by several dominant annihilation modes, including 
$\chi\chi\rightarrow W^+W^-$, $\chi\chi\rightarrow ZZ$ or $\chi\chi\rightarrow t\bar{t}$ as well as $\chi\chi\rightarrow 
b\bar{b}$ and $\chi\chi\rightarrow\tau^+\tau^-$.

The other DM candidate we consider is LDM, consisting of MeV mass particles, which annihilate to electron-positron pairs
\footnote{
MeV LDM particles can also potentially annihilate directly into neutrinos and photons. However most theories suppress this emission in order to be consistent with observational constraints. Here we only consider scenarios where LDM annihilates entirely to electron-positron pairs, so that our results can be considered as an upper limit to the more general case.
}
and consequently were considered to be a possible source of the positrons contributing to the 511\,keV positronium decay signature from the bulge of the Galaxy observed by SPI/INTEGRAL \citep{knodlseder}. 
While the current view favours the interpretation of the 511 keV feature as due to $e^+e^-$ injection  by a population of astrophysical sources, there is nevertheless continued interest in  reviving a dark matter interpretation because of the possible connection with other anomalous spatially extended signals seen from the innermost  Galaxy, specifically the WMAP and the FERMI hazes \cite{Dobler:2009xz}.  More exotic dark matter models  are required in this case, most specificaly some form of multicomponent dark matter (see e.g. Refs. \cite{Boehm:2003ha,Feldman:2010wy}).

Relevant analyses of the 511 keV  emission impose the constraint on the LDM mass $m_{\rm DM}<20\,{\rm MeV}$ in order not to overproduce detectable gamma-rays from inner bremsstrahlung processes \citep*{beacom2005} (although see \citet{boehm2006}). A stronger, albeit less conservative constraint, $m_{\rm DM}<3\,{\rm MeV}$ can be obtained if one considers the generation of gamma-rays from the in-flight annihilation between positrons produced from LDM annihilation and electrons residing in the interstellar medium of our Galaxy \citep{beacom2006}.

Both in the case of neutralinos and LDM, the average rate of energy absorption per hydrogen atom in the IGM at a redshift $z$ is given by
\begin{equation}
\label{eq:ann_rate}
{\dot\epsilon}(z)=\frac{1}{2}f_{\rm abs.}(z)\frac{n_{\rm DM, 0}^2}{n_{\rm H, 0}} \langle \sigma_{\rm ann.}\upsilon\rangle m_{\rm DM} (1+z)^3\:C(z)
\end{equation}
\noindent where $m_{\rm DM}$ is the mass of the DM particle, $\langle \sigma_{\rm ann.}\upsilon\rangle$ is the thermally-averaged DM annihilation cross section, $n_{\rm DM, 0}$ and $n_{\rm H, 0}$ are the current {\em average} number densities of DM and hydrogen respectively, and $f_{\rm abs.}$ is the fraction of energy which is absorbed by the IGM. The ``clumping factor'' $C(z)$ is the redshift-dependent enhancement of the annihilation rate owing to the presence of DM structures, relative to a completely homogeneous Universe\footnote{The factor of 1/2 in Eq.(\ref{eq:ann_rate}) assumes Dirac DM particles; for Majorana particles this should be further multiplied by a factor of 2.}.


\section{Extragalactic dark matter Annihilation Rate}
\label{sec:EG_DM_ann_rate}

\noindent In the standard cosmological model, all structure in the Universe originated from small amplitude quantum fluctuations during an epoch of inflationary expansion shortly after the Big Bang. The 
linear growth of the resulting density fluctuations is then completely determined by their initial power spectrum, which for $\Lambda$CDM is usually assumed to be a power law with spectral index $n$.
Current limits on $n$ from observations of temperature fluctuations in the CMB conducted by the Wilkinson Microwave Anisotropy Probe, $n_{\rm WMAP}=0.963\pm0.012$ (at 68\% confidence level) \citep{Dunkley:2008ie,Komatsu:2010fb}, support the existence of a power spectrum consistent with inflation.

During the expansion of the Universe, the aforementioned small initial density fluctuations will eventually grow and produce the structures that we observe today.
In the currently accepted cosmological model, smaller structures form first and then merge to form larger ones in a process of ``bottom-up'' hierarchical structure formation.
The mass distribution at any given redshift can potentially be determined through the use of numerical simulations. 

As a first approximation, the smaller progenitors forming larger isolated structures are completely disrupted after merging and the resulting ``smooth'' DM density distribution can be described by a continuous function, conventionally of the form
\begin{equation}
\rho(r)=\frac{\rho_s}{ (r/r_s)^{\gamma} \left[ 1+(r/r_s)^{\alpha} \right] ^{(\beta-\gamma)/\alpha} },
\label{rho}
\end{equation}
\noindent where $r$ is the distance from the centre of the halo, $r_s$ is a scale radius, $\rho_s$ is a normalisation factor, and $\alpha,\,\beta$ and $\gamma$ are free parameters.

However, N-body simulations of CDM halos reveal that a wealth of substructure halos (henceforth referred to as subhalos) exist within such halos. 
Moreover, utilising results from the Via Lactea II simulations, \citet{diemand2008} claimed that a further generation of sub-subhalos exist with a near self-similar mass distribution relative to their parent subhalo. This suggests the possibility that if one were to conduct simulations with sufficiently high resolution, one would find a long nested near self-similar series of halos within halos within halos etc., all the way down to the smallest halos\footnote{Although there are results from the more recent Aquarius simulations \cite{Aq1, Aq2}, conducted by the Virgo consortium, that are in contention with these results (see \S\,\ref{sec:Discussion}).}.
This has significant implications for the indirect detection of annihilating DM since the rate of DM annihilations is proportional to the square of the local density, and hence the presence of over-densities can significantly increase the annihilation rate relative to that obtained with a smooth DM distribution. 

The above scenario applies to structures formed in a CDM-dominated Universe. In a WDM-dominated Universe, the significant damping of small-scale density fluctuations, due to the larger free-streaming length, should be taken into account. Following \citet{bardeen}, this can be accounted for by using the modified power spectrum 
$P(k)=T^2_{\rm WDM}(k)P_{\Lambda{\rm CDM}}(k)$,
where the WDM transfer function is approximated by
\begin{equation}
T_{\rm WDM}(k)=\exp\left[-\frac{kR_f}{2}-\frac{(kR_f)^2}{2}\right],
\end{equation}
\noindent where $R_f$ is the free-streaming length. 

For WDM particles with negligible interaction rates, the free-streaming length is related to the particle mass $m_{\rm DM}$ by \citep{bardeen}.
\begin{eqnarray}
R_{f,n}&=&7.4\times10^{-6}\left(\frac{m_{\rm DM}}{\rm 1\,MeV}\right)^{-4/3}\left(\frac{\Omega_{\rm DM}}{0.258}\right)^{1/3}\nonumber\\
&\times&\left(\frac{h}{0.719}\right)^{5/3}h^{-1}\,{\rm Mpc}.
\end{eqnarray}

\noindent However, as we will show below, the interaction rates for self-annihilating LDM in the models considered here are non-negligible. In this case, the free-streaming length is given by \citep{Boehm&Schaeffer}
\begin{eqnarray}
R_{f,i}&=&0.3\left(\frac{\Gamma_{\rm dec., DM}}{6\times10^{-24}\,{\rm s}^{-1}(1+z_{\rm dec.})^3}\right)^{1/2}\\
&\times&\left(\frac{1\,{\rm MeV}}{m_{\rm DM}}\right)^{1/2}\,{\rm Mpc},
\end{eqnarray}
\noindent where $\Gamma_{\rm dec., DM}$ is the WDM self-annihilation rate at the decoupling redshift $z_{\rm dec.}$ given by
\begin{equation}
\Gamma_{\rm dec., DM}=\frac{1}{2}\frac{\rho_{c, 0}\Omega_{{\rm DM},0}}{m_{\rm DM}}\langle\sigma_{\rm ann.}\upsilon\rangle_{\rm dec.}(1+z_{\rm dec.})^3,
\end{equation}
\noindent and $\langle\sigma_{\rm ann}\upsilon\rangle_{\rm dec.}$ is the thermally-averaged product of the WDM annihilation cross section and relative speed of two annihilating WDM particles, evaluated at the same time. In order to obtain the thermal relic density observed today, one requires $\langle\sigma_{\rm ann.}\upsilon\rangle_{\rm dec.}\simeq10^{-26}\,$cm$^3$\,s$^{-1}$. 

For $m_{\rm DM}=3\,$MeV we obtain $R_{f,n}=2.4\,$pc and $R_{f,i}=98\,$pc, while for $m_{\rm DM}=20\,$MeV, we obtain $R_{f,n}=0.19\,$pc and $R_{f,i}=15\,$pc. Hence, in both cases the co-moving free-streaming length set by WDM interactions is at least an order of magnitude larger than that when interactions are completely negligible, and consequently we must use the former in our determination of the cut-off scale in the WDM power spectrum.

We follow the treatment by \citet{avilareese} and define a characteristic free-streaming wavenumber $k_f$ such that $T_{\rm WDM}(k_f)\simeq0.5$, leading to $k_f\simeq0.46/R_f$. This wavenumber is then related to a characteristic filtering mass $M_f$ by
\begin{equation}
M_f=\frac{4\pi}{3}{\bar \rho_{\rm WDM}}\left(\frac{\lambda_f}{2}\right)^3,
\label{M_f}
\end{equation}
\noindent where $\lambda_f=2\pi/k_f=13.6R_f$. In this paper we invoke the approximation $M_{\rm min.}\sim M_f$, where here $M_{\rm min.}$ is the minimum mass of a LDM halo, and equal to approximately $46\,M_{\odot}$ and $0.16\,M_{\odot}$ for $m_{\rm DM}=3\,$MeV and $20\,$MeV respectively. Since the mass within a given co-moving volume is constant as the Universe expands, the result $(\ref{M_f})$ is independent of redshift.

Below, we perform a series of detailed calculations illustrating the enhancement of the annihilation rate relative to that obtained with a completely smooth Universe, known as the {\it clumping factor}.


\section{Calculation of the clumping factor}
\label{sec:DM_ann_rate}

\noindent We assume a standard homogeneous, isotropic Universe with a flat spatial geometry. Let $R(M, z)$ be the average annihilation rate within a generic DM halo of mass $M$ located at redshift $z$. Even for large $M$, this source can be regarded as an unresolved point-source and we assume this throughout, for all halos considered. The rate of annihilations per unit volume at a given redshift is then equal to 
\begin{equation}
\Gamma(z)=(1+z)^3\int\limits_{M_{\rm min.}}^{M_{\rm max.}}{\rm d}M\frac{{\rm d}n}{{\rm d}M}(M,z)R(M, z),
\label{rate_z}
\end{equation}
\noindent where we have introduced the unconditional halo mass function, d$n/$d$M$, i.e. the co-moving number density of virialised halos with mass $M$ located at redshift $z$, (the factor $(1+z)^3$ converts this from co-moving to proper density). The integral spans over the mass range $M>M_{\rm min.}$, where $M_{\rm min.}$ can be as small as $\sim10^{-12}M_{\odot}$, due to kinetic decoupling in the case of CDM \citep{Profumo2006}, and approximated by the filtering mass (\ref{M_f}) in the case of WDM. 

Three ingredients are required in order to calculate the annihilation rate (\ref{rate_z}). Firstly, we need to specify the annihilation cross section of our DM candidates (in our case neutralinos or LDM). Secondly, we need to specify the DM density profile of a generic halo of mass $M$ at redshift $z$. Finally, we need an estimate of the distribution of halos, i.e. an estimate of the halo mass function ${\rm d}n(M,z)/{\rm d}M$. 


\subsection{The halo mass function}
\noindent Press-Schechter theory \citep{press_schechter} postulates that the cosmological mass function of DM halos can be expressed in the universal form
\begin{equation}
\frac{\mathrm{d}n}{\mathrm{d}M}=\frac{\bar{\rho_0}}{M^2}\nu f(\nu) \frac{\mathrm{d} \log (\nu)}{\mathrm{d}\log(M)},
\end{equation}
\noindent where $\bar{\rho_0}$ is the average co-moving DM density, ${\bar \rho_0}=\rho_c\Omega_M$, and $\rho_c$ is the present critical density of the Universe. The parameter $\nu=\delta_{\rm sc}/\sigma(M)$ is defined as the ratio of the critical overdensity required for spherical collapse at redshift $z$ extrapolated using linear theory to present time, and $\sigma(M)$ is the r.m.s. of primordial density fluctuations when smoothed on a scale which contains mass $M$, again extrapolated using linear theory to present time. The form of $\delta_{\rm sc}(z)$ can be found in \citet{Tegmark2005}. $\sigma(M)$ is related to the power spectrum $P(k)$ of the linear density field extrapolated to the present time by
\begin{equation}
\sigma^2(M)=\int {\rm d}^3k W^2(kR) P(k),
\end{equation}
\noindent where $W$ is the top-hat window function at the length scale $R=\left(3M/4\pi{\bar \rho}\right)^{1/3}$ and ${\bar \rho}$ is the mean matter density. We utilise the analytical approximation specified in \citet{Tegmark2006}, relevant in the linear regime long after the relevant fluctuation modes have entered the horizon, when all modes grow at the same rate, which means that $\sigma(M)$ can be factored as a product of two functions, one solely dependent on redshift $z$ and the other solely dependent on the comoving spatial scale $R$. We normalise $P$ and $\sigma$ by computing $\sigma$ at $R=8 h^{-1}$\,Mpc and setting the result equal to the cosmological parameter $\sigma_8$ as measured by WMAP, $\sigma_8=0.796\pm0.036$ 
\citep{Dunkley:2008ie}.

The first-crossing distribution $f(\nu)$ has the following analytical fit \citep{Sheth_Tormen} to the N-body simulation results from the Virgo consortium \citep{virgo}
\begin{equation}
\nu f(\nu)=A\left[1+(a\nu)^{-p} \right]\left(\frac{a\nu}{2\pi}\right)^{1/2}\exp\left(-\frac{a\nu}{2}\right),
\end{equation}
\noindent where $a\simeq 0.7$, $p=0.3$, and $A$ is determined by the requirement that all mass lies within a given halo, i.e. $\int {\rm d}\nu f(\nu)=1$ or equivalently $\int {\rm d}M M {\rm d}n/{\rm d}M = {\bar \rho_0}$.


\subsection{The density profile of dark matter halos}
\noindent Since the rate of DM annihilation scales with density squared, it depends sensitively on the density profile of each halo. We consider three universal density profiles to model the smooth distribution of DM within each halo (substructure will be dealt with later in this section). 
Firstly, we consider the popular profile proposed by \citet*{nfw1996, nfw1997} (NFW), 
which corresponds to $\alpha=1$, $\beta=3$ and $\gamma=1$ in Eq.(\ref{rho}). Secondly, we consider a profile with a significantly larger slope, specifically the one proposed by \citet{moore},
corresponding to $\alpha=1.5$, $\beta=3$ and $\gamma=1.5$. Both of these profiles have the same functional form and are both singular towards the Galactic centre (in fact, the slope of the Moore profile must necessarily be truncated for $r<r_{\rm min.}$, where $r_{\rm min.}\sim0$ - see below, otherwise the integral of density squared will diverge).
However, there have been indications that cuspy profiles are inconsistent with observations, specifically regarding the rotation curves of small-scale galaxies \citep{flores1994,moore1994,weldrake2003,donato2004,gentile2007}, which are more likely to be consistent with density profiles possessing flattened cores similar to that which may be achieved with WDM \citep{hogan2000,colin2008}. Therefore, we lastly consider the Burkert density profile \citep{burkert}:
\begin{equation}
\rho(r)=\frac{\rho_s}{ \left[1+(r/r_s)\right]\left[ 1+(r/r_s)^{2}\right]},
\label{Burkert}
\end{equation}
\noindent which has been shown to be fairly consistent with the rotation curves of a large number of spiral galaxies \citep{salucci2000}.


\subsection{Concentration-mass relation for dark matter halos}

\noindent Here we introduce the virial concentration parameter $c_{\rm vir.}$, defined by
$c_{\rm vir.}=r_{\rm vir.}/r_s$,
where $r_s$ is the scale radius defined above and $r_{\rm vir.}$ is the virial radius of the halo. The latter is defined as the radius encapsulating the virial mass $M$ of the halo within which the average density is equal to the overdensity $\Delta_{\rm vir.}$ times the average cosmological density ${\bar \rho(z)}$ at that redshift
\begin{equation}
M=\frac{4\pi}{3}\Delta_{\rm vir.}{\bar \rho(z)}r_{\rm vir.}^3.
\label{M}
\end{equation}
For $\Delta_{\rm vir.}$, we use the approximation provided in \citet{Tegmark2006}, namely
$\Delta_{\rm vir.}\simeq 18\pi^2 + 52.8x^{0.7}+16x$,
where $x(z)=\Omega_{\Lambda}(z)/\Omega_{\rm M}(z)$, ($\Delta_{\rm vir.}\simeq311$ at $z=0$ for $\Omega_{\rm M}=0.3$ and $\Omega_{\Lambda}=0.7$). This is  accurate to within 4\% of the exact numerical calculation at relevant times.

There has been evidence from simulations revealing a strong correlation between the halo mass $M$ and its corresponding concentration $c_{\rm vir.}$, with larger concentrations in smaller mass halos, which is consistent with the idea of bottom-up hierarchical structure formation with smaller halos collapsing at earlier times when the average density of the Universe was much greater \citep{nfw1996, nfw1997}. 
This relationship was later re-affirmed by \citet{B2001} (B2001 hereafter) using a sample of simulated halos in the mass range $10^{11}\lesssim M/h^{-1}M_{\odot}\lesssim10^{14}$, who proposed a toy model to describe this behaviour, which is popular in the relevant literature: on average, a collapse redshift $z_c$ is assigned to each halo of mass $M$ through the relation $M_{*}=FM$, where at a redshift $z$ the typical collapsing mass $M_{*}(z)$ is defined implicitly by the relation $\sigma(M_{*}(z))=\delta_{\rm sc}(z)$ and is postulated to be a fixed fraction $F$ of $M$, which, following \cite{Wechsler:2001cs}, we set equal to 0.015. The density of the Universe at redshift $z_c$ is then associated with a characteristic density of the halo at redshift $z$. Therefore, here we use the average concentration-mass relation obtained using the above method, which is given by
\begin{equation}
c_{\rm vir.}(M,z)=K\frac{1+z_c}{1+z}=\frac{c_{\rm vir.}(M, z=0)}{1+z}
\label{c-m}
\end{equation}
where $K\simeq5$, for $\Omega_{\Lambda}=0.742$, $\Omega_{\rm M}=0.258$, $h=0.719$ and $\sigma_8=0.796$ \citep{Dunkley:2008ie}

Since this relation has been derived for halos with a minimum mass of $\sim10^{11} M_{\odot}$, the extrapolation to very small values of the mass, down to the mass associated with the DM free streaming length (that we take to be as small as $\sim10^{-12} M_\odot$), could be unreliable, since small mass halos become increasingly concentrated. For this reason, following \cite{ullio2002}, we introduce a cut-off mass $M_\mathrm{cut}$ such that $c_{\rm vir.}(M,z)=c_{\rm vir.}(M_\mathrm{cut},z)$ for $M<M_\mathrm{cut}$. In the following, we will either take $M_\mathrm{cut}$ equal to the mass of the smallest DM halos (i.e. no cut-off) or equal to $10^6 M_\odot$, which is the typical (mass) resolution of current numerical simulations of Galaxy-sized DM halos.


\subsection{Clumping factor for smooth halos}
\label{sec:clumpsm}

\noindent We are now able to calculate the clumping factor $C(z)$ attributed to extragalactic halos with smooth DM density profiles and concentrations given by Eq.(\ref{c-m}).  We start by calculating the annihilation rate $R(M, z)$ within a DM halo of mass $M$ located at redshift $z$ given by
\begin{equation}
R(M, z)=\frac{1}{2}\frac{\langle\sigma_{\rm ann.}\upsilon\rangle}{m_{\rm DM}^2}\int\limits^{r_{\rm vir.(M, z)}}_{r=0}\rho^2(r)4\pi r^2{\rm d}r.
\label{eq:R}
\end{equation}
The integral in (\ref{eq:R}) can be expressed in analytical form for the NFW and Moore profiles; we present the relevant formulas in Appendix \ref{app:clump}. In the case of the Moore profile, however, in order for the integral over density squared to be finite, the density must be truncated below a radius $r_{\min.}$. 

To obtain a value for $r_{\rm min.}$ we assume that, within some minimum distance from the center of the halo, most of the neutralino DM has self-annihilated, leaving a flattened density core. 
The size of the core is roughly determined by the condition that within it the time-scale for DM annihilation, $t_{\rm ann.}\sim(n_{\rm \chi}\langle\sigma_{\rm ann.}\upsilon\rangle)^{-1}$, should be smaller than the average time-scale $t_{\rm in.}$ for the replenishment of the core owing to the infall of DM from larger radii. Then  $r_{\rm min.}$ will be defined as the radius where $t_{\rm ann.} \simeq t_{\rm in.}$. We do not try to estimate $t_{\rm in.}$; instead, since we must have $t_{\rm in.} \ll t_h$, where $t_h\sim10^{17}$\,s  is the Hubble time, we have that within the core
$n_{\rm \chi}\langle\sigma_{\rm ann.}\upsilon\rangle \gg {t_h}^{-1}$,
 and since the density decreases monotonically with increasing radius we can obtain a conservative upper limit for $r_{\rm{min.}}$ from the condition $n_{\rm \chi}\langle\sigma_{\rm ann.}\upsilon\rangle \simeq t_h^{-1}$. Then, we adopt the conservative criterion $(\rho_{\rm \chi}r_{\rm min.}/m_{\chi})\langle\sigma_{\rm ann.}\upsilon\rangle\sim t_h^{-1}$, with canonical values of the neutralino mass and annihilation cross section of $m_{\chi}\sim100\,$GeV and $\langle\sigma_{\rm ann.}\upsilon\rangle\sim10^{-26}\,$cm$^3$\,s$^{-1}$. This sets an upper limit for $x_{\rm min.}$ which is $\sim10^{-8}$ for the Galactic halo at present day, which is consistent with similar approximations by other authors (see, e.g., \citet{taylor2003}).

Then,  it follows from Eq.(\ref{rate_z}) that the contribution to the DM annihilation rate per unit volume, $\Gamma_{\rm halos}(z)$, by halos located at redshift $z$ is
\begin{eqnarray}
\Gamma_{\rm halos}(z)&=&\frac{1}{2}\frac{\langle\sigma_{\rm ann.}\upsilon\rangle}{m_{\rm DM}^2}(1+z)^3\nonumber\\[0.5cm]
&&\times\int\limits_{M=M_{\rm min.}}^{M_{\rm max.}}{\rm d}M\frac{{\rm d}n}{{\rm d}M}(M,z)\int\limits^{r_{\rm vir.(M, z)}}_{r=0}\rho^2(r)4\pi r^2{\rm d}r.\nonumber\\
\label{eq:Gamma_halos}
\end{eqnarray}
The corresponding rate of DM annihilation per unit volume contributed by the smooth background density at redshift $z$ is given by
\begin{equation}
\Gamma_{\rm smooth}(z)=\frac{1}{2}\frac{\langle\sigma_{\rm ann.}\upsilon\rangle}{m_{\rm DM}^2}{\bar \rho_{\rm DM}}^2(z),
\label{eq:Gamma_smooth}
\end{equation}
where $\rho_{\rm DM}(z)=\rho_{c,0}\Omega_{\rm DM,0}(1+z)^3$. Therefore, we define the {\it clumping factor} for smooth halos,  $C_{\rm halo}(z)$, as
\begin{eqnarray}
C_{\rm halo}(z)&\equiv&1+\frac{\Gamma_{\rm halo}(z)}{\Gamma_{\rm smooth}(z)} =\nonumber\\
&=&1+\frac{(1+z)^3}{{\bar \rho_{\rm DM}^2}(z)}\nonumber\\
&&\times\int\limits_{M=M_{\rm min.}}^{M_{\rm max.}}{\rm d}M\frac{{\rm d}n}{{\rm d}M}(M,z)\int\limits^{r_{\rm vir.(M, z)}}_{r=0}\rho^2(r)4\pi r^2{\rm d}r,\nonumber\\
\label{eq:C_factor}
\end{eqnarray}
so that $C_{\rm halo}(z)\rightarrow1$ for a completely smooth universe.

\begin{figure}[t!]
	\begin{center}
	\includegraphics[width=0.9\linewidth,keepaspectratio,clip]{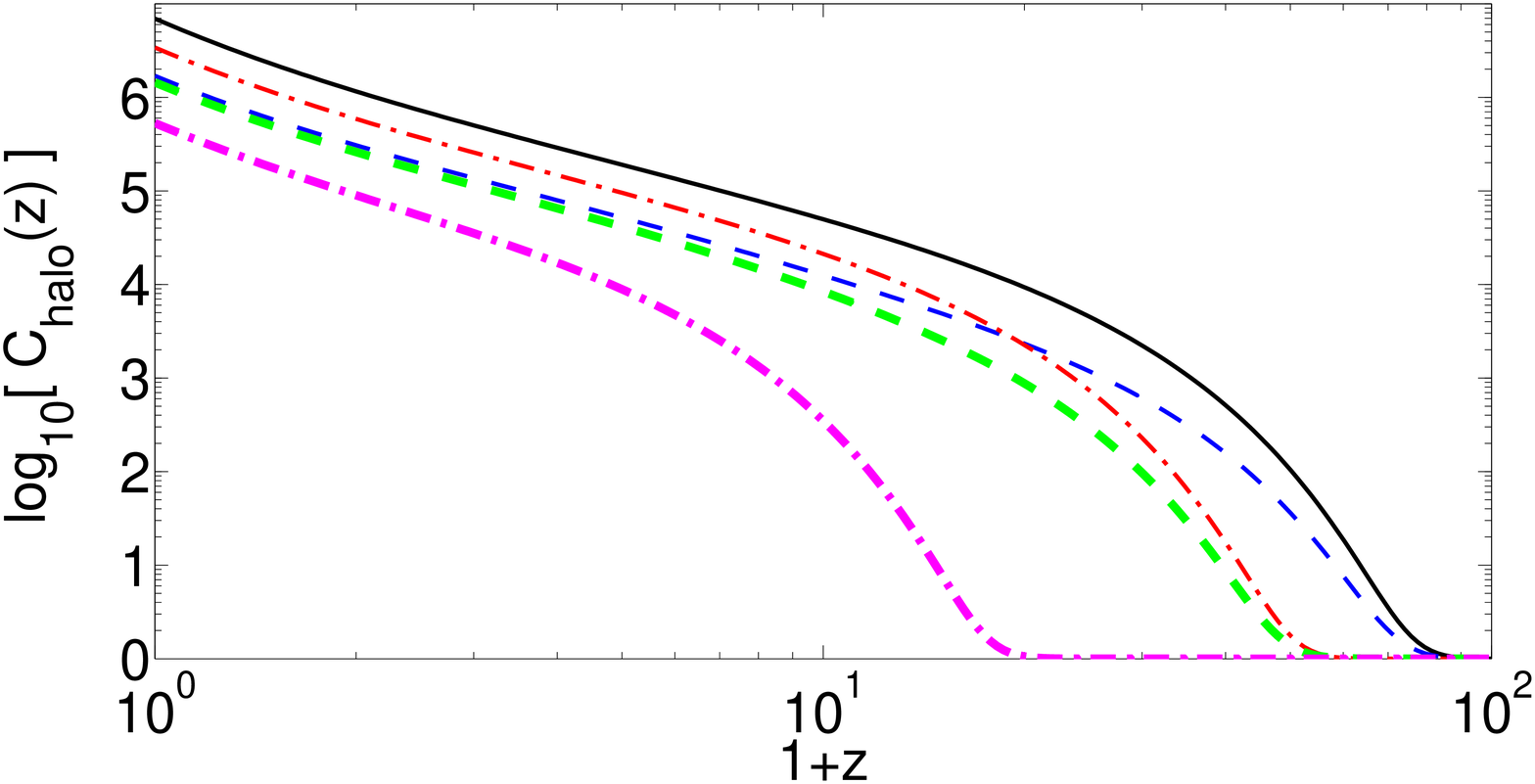}
	\includegraphics[width=0.9\linewidth,keepaspectratio,clip]{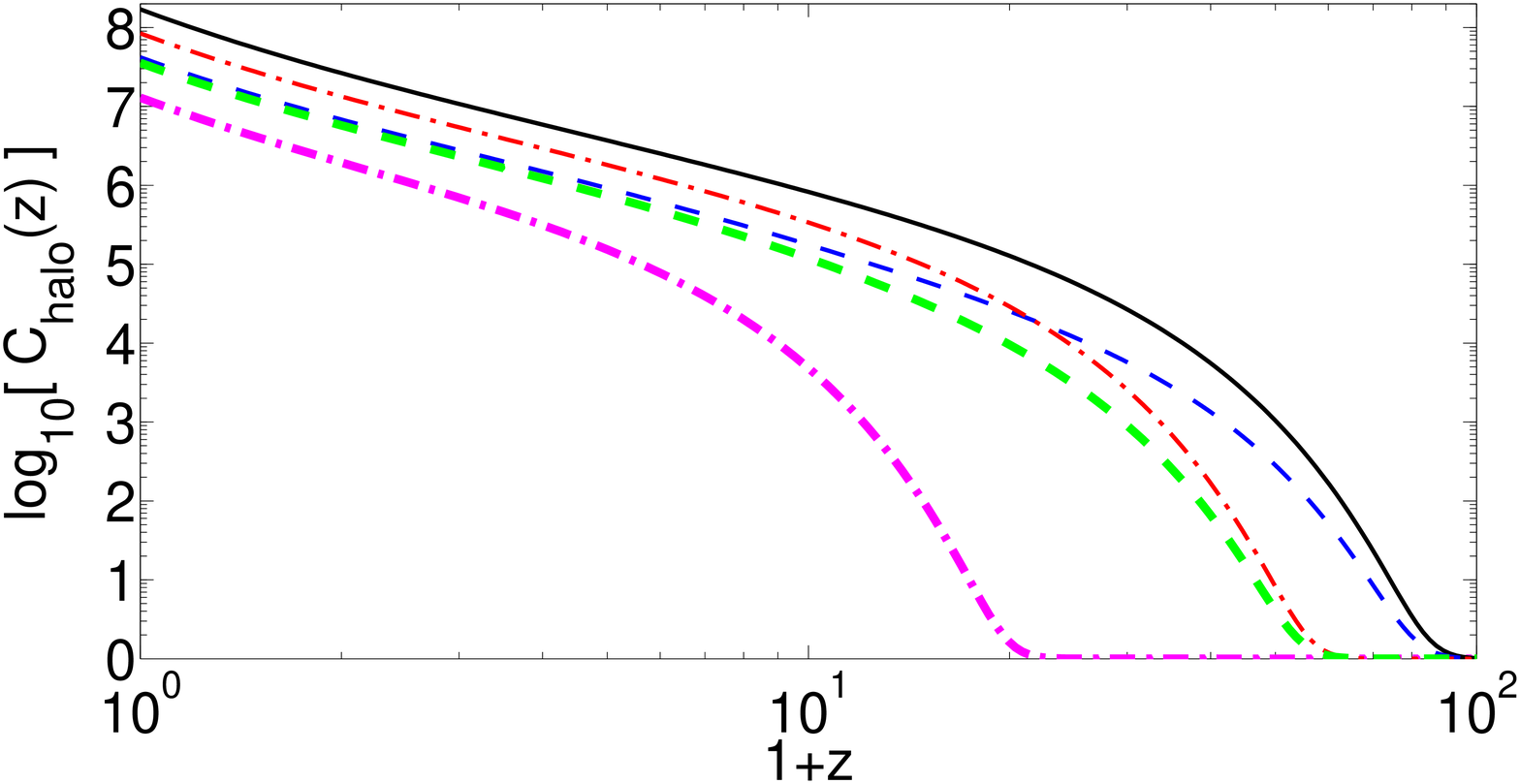}
	\includegraphics[width=0.9\linewidth,keepaspectratio,clip]{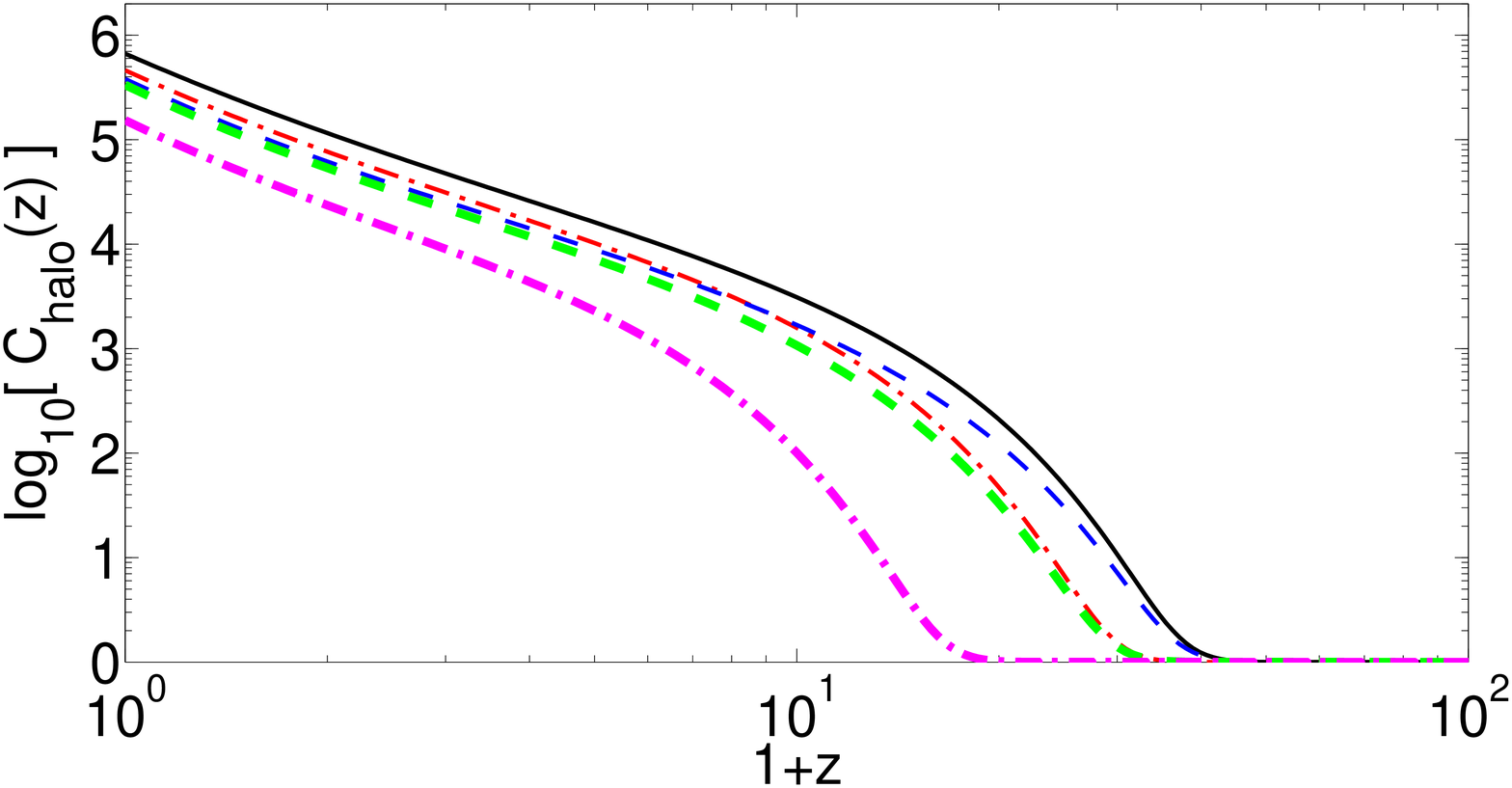}
	\caption{Clumping factor as a function of redshift for DM halos with mass $M>M_{\rm min.}$ with smooth NFW (upper panel), Moore (central panel) and Burkert (bottom panel) DM density profiles with a $c_{\rm vir.}-M$ relation described by Eq.(\ref{c-m}), truncated at a halo mass $M_{\rm cut}$. The displayed curves correspond to values of $(M_{\rm min.}/M_{\odot}, M_{\rm cut}/M_{\odot})$ of $(10^{-12},10^{-12})$ (thin black solid), $(10^{-12},10^{6})$ (thin blue dashed),  $(10^{-4},10^{-4})$ (thin red dot-dashed), $(10^{-4},10^{6})$ (thick green dashed) and $(10^{6},10^{6})$ (thick magenta dot-dashed) for the NFW and Moore profiles, and equal to $(0.16, 0.16)$ (thin black solid), $(0.16, 10^6)$ (thin blue dashed), $(46,46)$ (thin red dot-dashed), $(46, 10^6)$ (thick green dashed) and $(10^6, 10^6)$ (thick magenta dot-dashed) for the Burkert profile.}
		\label{fig:fig1}
         \end{center}
\end{figure}

In Fig.\,\ref{fig:fig1} we display plots of $C_{\rm halo}(z)$ as a function of $z$ for halos with NFW profiles (top panel), Moore profiles (central panel) and Burkert profiles (bottom panel).  Halos with cuspy density profiles, such as the NFW and Moore profiles, are typical of CDM halos for which the minimum mass cut-off scale in the matter power spectrum is determined by collisional damping and free streaming in the early Universe. For WIMP DM the value of $M_{\rm min.}/M_{\odot}$ can range from $10^{-12}$ to $10^{-4}$ for typical kinetic decoupling temperatures. 
Hence in the upper panels of Fig.\,\ref{fig:fig1} we illustrate the effect on the clumping factor for values of $M_{\rm min.}/M_{\odot}$ of $10^{-12}, 10^{-4}$ and $10^6$, where, as mentioned above, the latter value is the typical minimum mass of resolved subhalos in numerical simulations of Galactic halos. We also demonstrate the influence of truncating the Bullock {\em et al.} concentration-mass relation (referred to as the ``B2001 relation'' hereafter) below a mass of $10^6 M_{\odot}$, as well as using the relation when extrapolated to $M_{\rm min.}$. 

In the bottom panel of Fig.\,\ref{fig:fig1} we show the clumping factor for halos with flattened cores like the ones possibly formed by WDM. In particular, we plot $C_{\rm halo}$ for minimum halo masses $M_{\rm min.}\simeq46\,M_{\odot}$ and $0.16M_{\odot}$, corresponding to the values of the damping mass (\ref{M_f}), obtained using $m_{\rm WDM}=3\,$MeV and  $m_{\rm WDM}=20\,$MeV respectively. We also again illustrate the effect of using relation Eq.(\ref{c-m}) when extrapolated to $M_{\rm min.}$ or truncated at $10^6 M_{\odot}$. The  selected values of $m_{\rm WDM}$ correspond to the respective upper limits on the LDM particle mass from constraints on inner bremsstrahlung gamma ray flux from the galactic centre \citep{beacom2005} (although see \citet{boehm2006}), and from in-flight annihilation \citep{beacom2006} between positrons from LDM annihilation and electrons in the interstellar medium.

An analysis of Fig.\,\ref{fig:fig1} reveals some interesting trends. Firstly, Moore profiles tend to yield larger clumping factors than NFW profiles, which in turn yield larger clumping factors than Burkert profiles. 
This is clearly related to the relative cuspiness of the three profiles and the fact that DM annihilations are enhanced in higher-density regions. In general, we have that $C_\mathrm{halo}$ at $z=10$ is between $10^4$ and $10^6$ for Moore profiles, $10^3$ and $10^5$ for NFW profiles, and $10^2$ and $10^4$ for Burkert profiles. 
Secondly, we observe that the smaller the value of $M_\mathrm{min.}$ the sooner the clumping factor starts to deviate from unity and the larger the clumping factor is at present day. This is due to the contribution in the integral in Eq.(\ref{eq:Gamma_halos}) of the smaller halos, that form earlier, and are thus denser, than larger halos. It is however worth stressing that the mass function and the concentration parameters have not been well measured for these extremely small, high-z halos. Finally, when the Bullock et al. relation is truncated at a value $M_\mathrm{cut}=10^6\,M_\odot>M_\mathrm{min.}$, the clumping factor is smaller. In particular, this roughly amounts to an order of magnitude difference at $z=10$ for the NFW and Moore profiles with $M_\mathrm{min.}=10^{-12} M_\odot$, and, as can be expected, the difference is smaller for larger values of $M_\mathrm{min.}$ and in the case of Burkert profiles.


\subsection{Clumping factor for halos possessing sub-halos and sub-sub halos}
\label{subsub}

\noindent Thus far we have considered the amplification of the DM annihilation rate for isolated halos with smooth density profiles. However, as already mentioned, N-body simulations  indicate that a significant proportion of the smaller progenitors giving rise to larger mass halos survive the merging processes and the tidal forces exerted upon them during their orbital motion within halos. In particular, the Via Lactea II $\Lambda$CDM simulations of Galactic halos presented in \citet*{diemand2008} and in \citet*{KDM} (KDM hereafter), revealed a second generation of surviving substructures within halos (designated as ``sub-subhalos''). Further, these simulations suggest that the mass distribution of sub-subhalos within their host subhalo is approximately the same as the mass distribution of subhalos within their host halo\footnote{Once again, we remind the reader that there are results from the more recent Aquarius simulations \cite{Aq1, Aq2} that are in contention with these results (see \S\,\ref{sec:Discussion}).}.

Since the DM annihilation rate scales with density squared, these subhalos and sub-subhalos could provide significant enhancement to the annihilation rate, even for modest substructure mass fractions, within halos/subhalos. For halos of mass $M$ these have been suggested to be as much as 10\% for subhalo masses $M_{\rm s}$ in the range $10^{-5}<M_{\rm s}/M<10^{-2}$ \citep{diemand2008} (which approximately corresponds to a constant mass fraction per subhalo mass decade of 3\%, owing to the fact that the subhalo mass function has a slope of approximately 2 - see below). However, owing to the fact that substructures invariably form earlier than their host halos, and that tidal disruption is unlikely to effect the inner density profiles of structures (i.e. where the majority of the enhancement originates), the concentration of substructures may be significantly greater than that of their host halos. The simulation results recently presented in KDM are consistent with the ratio $N_c=c_{\rm vir.}^{\rm halo}/c_{\rm vir.}^{\rm subhalo}\simeq3$ for subhalos located at solar radii within galactic halos\footnote{
Although $N_c$ demonstrates a slight galactocentric radial dependence, the authors of KDM claim that the effect on the overall annihilation rate is negligible.},
whilst the numerical simulations of Bullock {\it et al.} show that, on average, $N_c\simeq1.5$ for halos of mass $M\sim5\times10^{11}M_{\odot}$ (B2001).

Here we calculate the contribution to the clumping factor by halos possessing substructures with a self-similar mass distribution. Consider a DM halo of mass $M$ with a subhalo mass distribution function given by
\begin{equation}
\frac{{\rm d}N(M)}{{\rm d}M_{\rm s}}\propto M_{\rm s}^{-\beta},
\end{equation}
\noindent where the index $\beta$ is assumed to be time-independent and approximately equal to 2, i.e. equal mass per decade in subhalos (KDM). Whilst we adopt a minimum subhalo mass equal to the minimum halo mass, $M_{\rm min.}$, for which we utilise several values as discussed above, we utilise an upper limit on $M_{\rm s}$ of $10^{-2}M$, where $M$ is the mass of the host halo, a choice motivated by recent numerical simulations (see, e.g., \cite{diemand2008}).

There are indications that $\beta$ may slightly deviate from this value, particularly for WDM substructures, for which \citet{knebe2008} claim that $\beta$ may be as small as 1.6. 
\begin{figure}[t!]
	\begin{center}
	\includegraphics[width=0.9\linewidth,keepaspectratio,clip]{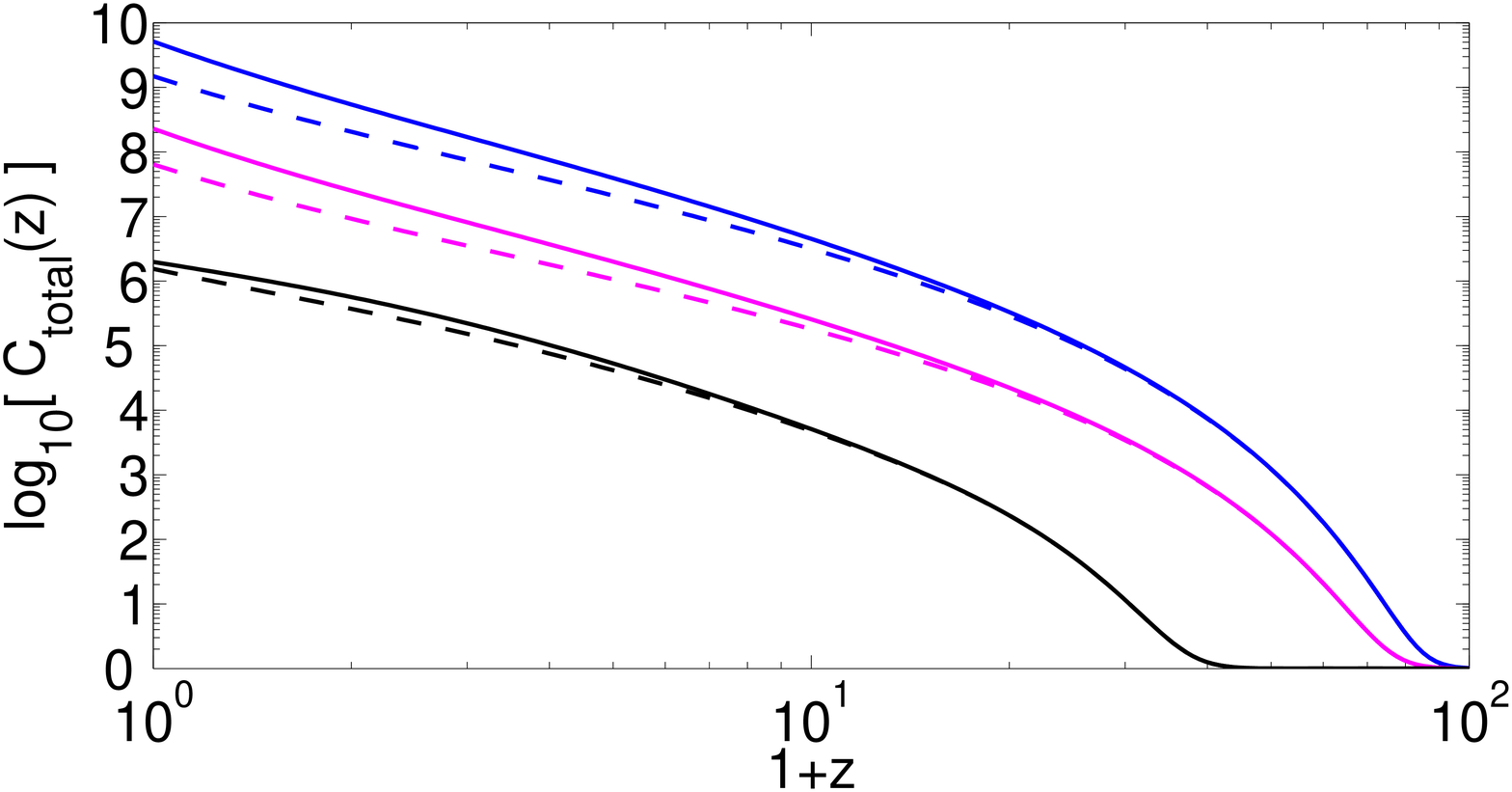}
	\caption{Total clumping factor from halos and subhalos $C_{\rm total.}(z)=1+C_{\rm halos}(z)+C_{\rm subhalos}(z)$, for different values of the substructure mass function index $\beta$. We show the values of $C_{\rm total}$ for structures with NFW (purple) and Moore (blue) density profiles for $\beta=2$ and 1.8 (solid and dashed curves respectively) and $\beta=2$ and 1.6 for the Burkert profile (black solid and dashed curves respectively). We take 
$M_{\rm cut} = M_{\rm min.} = 10^{-12}\,M_{\odot}$ for NFW and Moore profiles and $ 0.16\,M_{\odot}$ for Burkert profiles.  In all cases $F_{\rm sub.}=3\%$, $N_{c}=3$.}
	\label{fig:fig_beta}
         \end{center}
\end{figure}
However, as shown in Fig.\,\ref{fig:fig_beta}, the effect on the clumping factor by varying $\beta$ slightly from 2 is small at the times of interest. We therefore adopt the value $\beta=2$ for both CDM and WDM. Consequently, each subhalo mass decade contributes a constant fraction $F_{\rm sub.}$ of the halo mass.

Adopting a course of reasoning analogous to that used to derive Eq.\,(\ref{eq:R}), the rate of DM annihilations within a similar halo, solely due to the subhalos within it, possessing smooth density profiles $\rho(r)$, is then given by
\begin{eqnarray}
R_{\rm sub.}(M, z)&=&\frac{\langle\sigma_{\rm ann.}\upsilon\rangle}{2m_{\rm DM}^2}\int\limits^{10^{-2}M}_{M_{\rm s} = M_{\rm min.}}{\rm d}M_{\rm s}\frac{{\rm d}N(M,F_{\rm sub.})}{{\rm d}M_{\rm s}}\nonumber\\
&\times&\int\limits^{r_{\rm vir.}(z, M_{\rm s})}_{r=0}\rho^2(r, c_{\rm vir.}^{\rm sub.}[M_{\rm s}, z])4\pi r^2 {\rm d}r\nonumber\\
&=&\frac{\langle\sigma_{\rm ann.}\upsilon\rangle}{2m_{\rm DM}^2}A(M, F_{\rm sub.})\int\limits^{10^{-2}M}_{M_{\rm s}=M_{\rm min.}}{\rm d}M_{\rm s} M_{\rm s}^{-\beta}\nonumber\\
&\times&\int\limits^{r_{\rm vir.}(z, M_s)}_{r=0}\rho^2(r, c_{\rm vir.}^{\rm sub.}[M_s, z])4\pi r^2{\rm d}r,\nonumber\\
\end{eqnarray}
where $A$ is the appropriate normalisation of d$N/$d$M_{\rm s}$.
The subhalo scale density can be obtained from the expressions (\ref{rho_s_NFW}) or (\ref{rho_s_moore}) for NFW and Moore profiles respectively, with the substitutions $c_{\rm vir.}\rightarrow c_{\rm vir.}^{\rm sub.}$ and $M\rightarrow M_{\rm s}$. 
Then integrating this contribution over all halos at redshift $z$ we obtain the annihilation rate for all subhalos residing within such halos
\begin{eqnarray}
\Gamma_{\rm subhalos}(z)&=&(1+z)^3\nonumber\\
&\times&\int\limits_{M=M_{\rm min.}}^{M_{\rm max.}}{\rm d}M\frac{{\rm d}n(M,z)}{{\rm d}M} R_{\rm sub.}(M, z, F_{\rm sub.})\nonumber\\
&=&\frac{\langle\sigma_{\rm ann.}\upsilon\rangle}{2m_{\rm DM}^2}(1+z)^3\nonumber\\
&\times&\int\limits_{M=M_{\rm min.}}^{M_{\rm max.}}{\rm d}M\frac{{\rm d}n(M,z)}{{\rm d}M}A(M, F_{\rm sub.})\nonumber\\
&\times&\int\limits_{M_{\rm s}=M_{\rm min.}}^{10^{-2}M}{\rm d}M_{\rm s} M_{\rm s}^{-\beta}\nonumber\\
&\times&\int\limits^{r_{\rm vir}(z, M_{\rm s})}_{r=0}\rho^2(r, c_{\rm vir}^{\rm sub}[M_{\rm s}, z])4\pi r^2{\rm d}r,\nonumber\\
\end{eqnarray}
and following Eq.\,(\ref{eq:C_factor}), we obtain the associated subhalo clumping factor
\begin{eqnarray}
\lefteqn{C_{\rm subhalos}=1+\frac{\Gamma_{\rm subhalos}(z)}{\Gamma_{\rm smooth}(z)}}&&\nonumber\\
&=&1+\frac{(1+z)^3}{{\bar \rho^2_{\rm DM}}(z)}\int\limits_{M=M_{\rm min.}}^{M_{\rm max.}}{\rm d}M\frac{{\rm d}n(M,z)}{{\rm d}M}A(M, F_{\rm sub.})\nonumber\\
&\times&\int\limits_{M_{\rm s}=M_{\rm min.}}^{10^{-2}M}{\rm d}M_{\rm s} M_{\rm s}^{-\beta}\nonumber\\
&&\times\int\limits^{r_{\rm vir.}(z, M_{\rm s})}_{r=0}\rho^2(r, c_{\rm vir.}^{\rm sub.}[M_{\rm s}, z])4\pi r^2{\rm d}r.\nonumber\\
\label{eq:C_sub}
\end{eqnarray}

However as mentioned above, each subhalo is likely to itself host substructures with mass function approximately equal to
\begin{equation}
\frac{{\rm d}N}{{\rm d}M_{\rm ss}}=A(M_{\rm s}, F_{\rm ss}, \beta)M_{\rm s}^{-\beta_{\rm ss}}.
\end{equation}
Owing to the near self-similar nature of the mass distribution of substructures within halos, we take the values of the index $\beta_{\rm ss}$ and the sub-subhalo mass fraction per mass decade $F_{\rm ss}$ to be equal to $\beta$ and $F_{\rm sub.}$ respectively. Hence, following the above treatment for halos and their constituent subhalos, the clumping factor for all sub-subhalos with virial concentration $c_{\rm vir.}^{\rm ss}$residing within subhalos, themselves residing within halos located at redshift $z$, is given by
\begin{eqnarray}
C_{\rm sub-subhalos} &=&1+\frac{(1+z)^3}{{\bar \rho^2_{\rm DM}}(z)}\int\limits_{M=M_{\rm min.}}^{M_{\rm max.}}{\rm d}M\frac{{\rm d}n(M,z)}{{\rm d}M}\nonumber\\
&\times&A(M_{\rm s}, F_{\rm s})\int\limits_{M_{\rm s} = M_{\rm min.}}^{10^{-2}M}{\rm d}M_{\rm s} M_{\rm s}^{-\beta}\nonumber\\
&\times&A(M_{\rm ss}, F_{\rm ss})\int\limits_{M_{\rm ss} = M_{\rm min.}}^{10^{-2}M_{\rm s}}{\rm d}M_{\rm ss} M_{\rm ss}^{-\beta_{\rm ss}}\nonumber\\
&\times&\int\limits^{r_{\rm vir.}(z, M_{\rm s})}_{r=0}\rho^2(r, c_{\rm vir.}^{\rm ss}[M_{\rm ss}, z])4\pi r^2{\rm d}r.\nonumber\\
\label{eq:C_subsub}
\end{eqnarray}
where, analogous for subhalos, for a given host subhalo of mass $M_{\rm s}$ and minimum permitted mass $M_{\rm min.}$, we allow for sub-subhalo masses in the range $M_{\rm min.}\le M_{\rm ss}\le10^{-2}M_{\rm s}$.

Finally, using Eqs.(\ref{eq:C_factor}), (\ref{eq:C_sub}) and (\ref{eq:C_subsub}), we obtain the total clumping factor for all structures at redshift $z$
\begin{eqnarray}
C_{\rm total}&=&1+(C_{\rm halo}(z)-1)\nonumber\\
&+&(C_{\rm subhalos}(z)-1)\nonumber\\
&+&(C_{\rm sub-subhalos}(z)-1),
\end{eqnarray}
where it should be understood that the normalisation of expressions $C_{\rm halo}$ and $C_{\rm subhalos}$ is modified to take into account the fact that a specified percentage of the mass of each halo and subhalo is provided by smaller substructures.

\begin{figure}[!]
	\begin{center}
	\includegraphics[width=0.9\linewidth,keepaspectratio,clip]{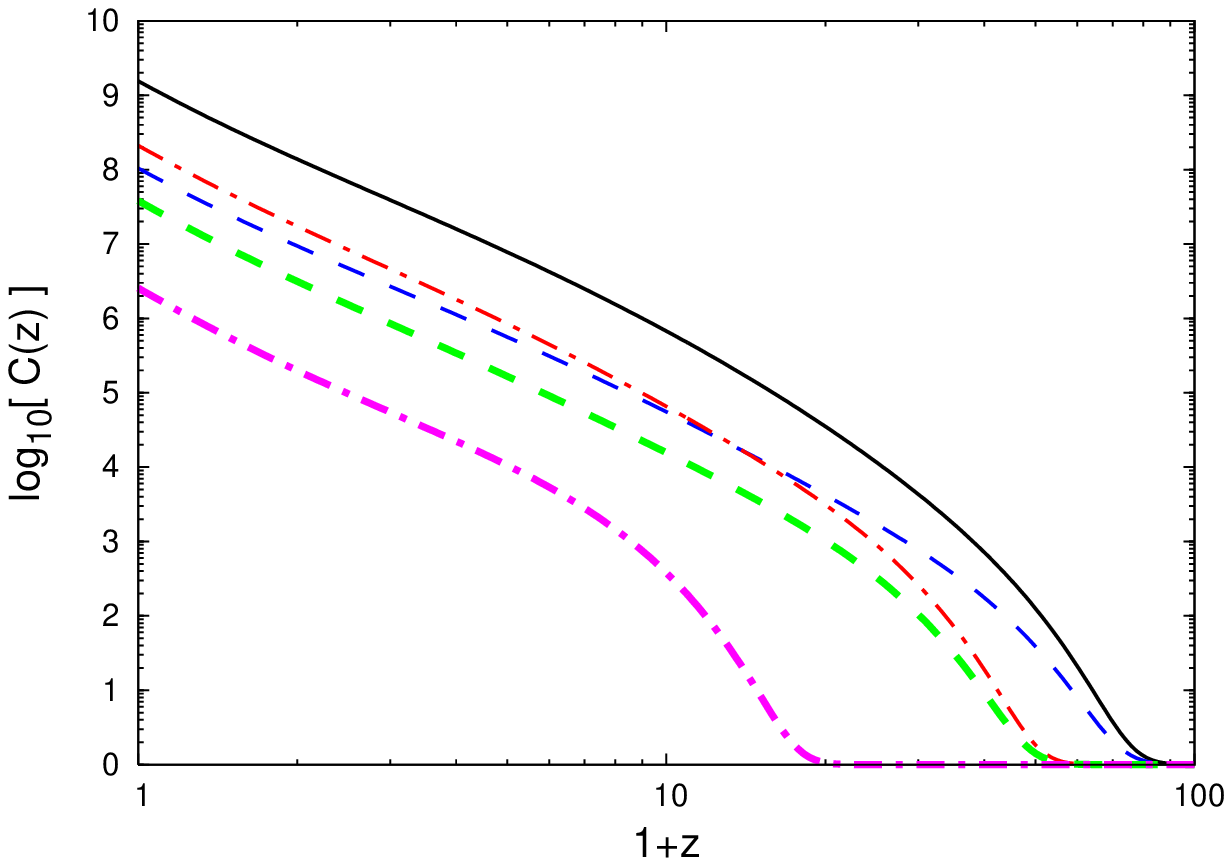}
	\includegraphics[width=0.9\linewidth,keepaspectratio,clip]{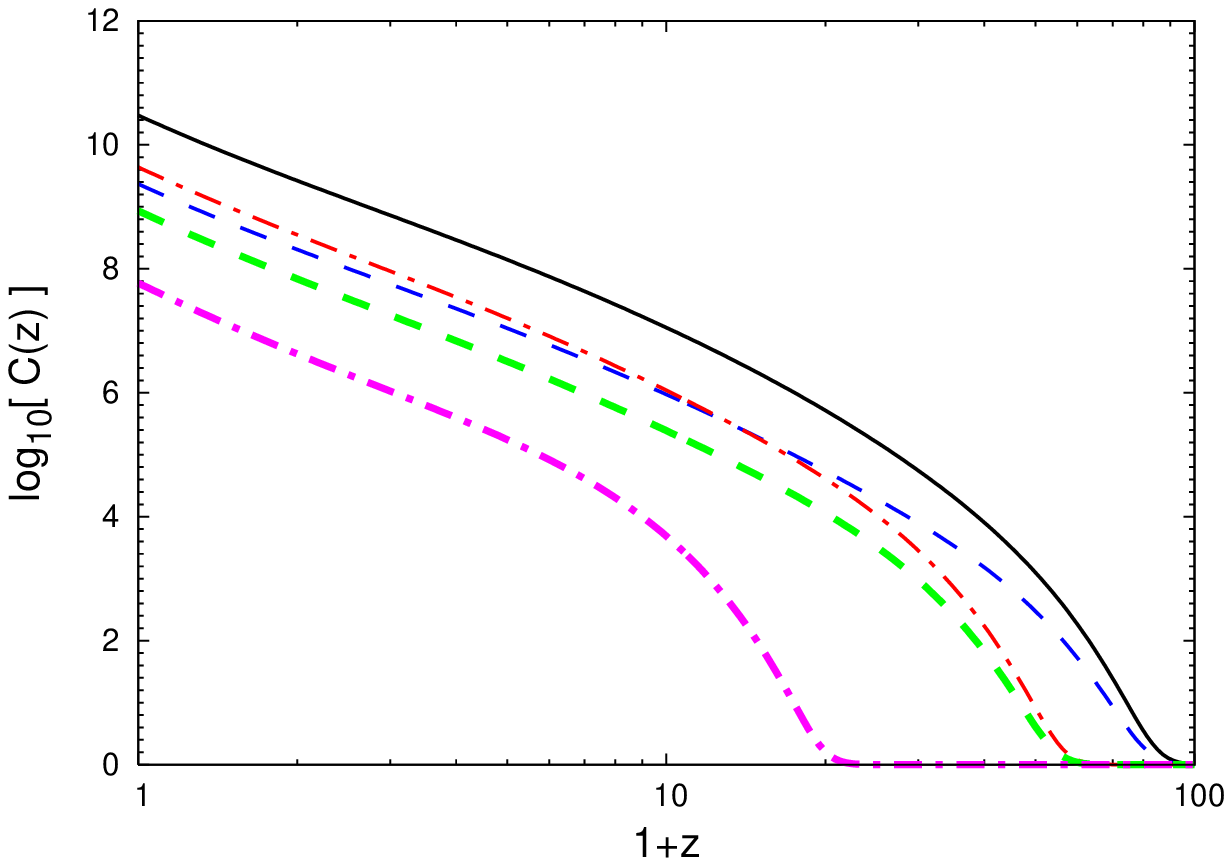}
	\includegraphics[width=0.9\linewidth,keepaspectratio,clip]{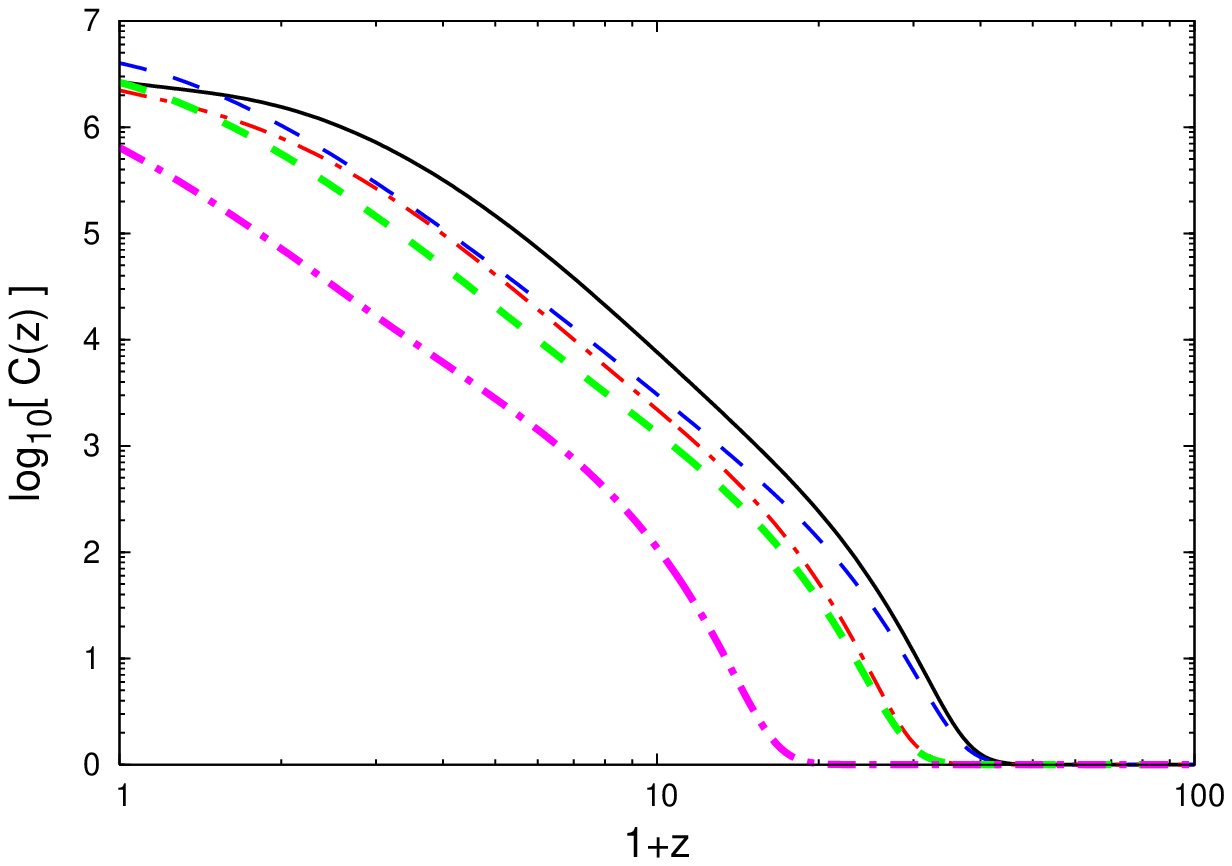}
	\caption{Plots of $C_{\rm total}(z)$ for structures with NFW (top panel), Moore (central panel) and Burkert (bottom panel) density profiles with subhalo and sub-subhalo mass fractions per decade   $F_{\rm sub.}$ and $F_{\rm ss}$ of 0.03, and a relative concentration ratio $N_c$ of 3.0. The different curves correspond to different values of $(M_{\rm min.}/M_{\odot}, M_{\rm cut}/M_{\odot})$, as in Fig.\,\ref{fig:fig1}.}
	\label{fig:fig2c}
         \end{center}
\end{figure}

In Fig.\,\ref{fig:fig2c} we show the total clumping factor as a function of $z$ for halos with NFW profiles (left panel), Moore profiles (central panel) and Burkert profiles (right panel). We find the same trends as in the case of smooth halos. However, the presence of substructures boosts the clumping factor, more effectively so for cuspier (i.e. Moore and NFW) profiles and for smaller values of $(M_\mathrm{min.},\,M_\mathrm{cut})$. In particular, we find that $C_\mathrm{halo}$ at $z=10$ is in the range between $10^4$ and $10^8$ for Moore profiles,  
between $10^3$ and $10^{6}$ for NFW profiles, and between $10^2$ and $\sim10^4$ for Burkert profiles.

From the recursive structure of Eq.(\ref{eq:C_subsub}), one can easily observe how to extend the present scenario to include higher generations of substructures, but since there is no evidence for such structures we omit them in this study. 
Moreover, we have found that the relative contribution of halos, subhalos and sub-subhalos to $C_{\rm total}(z)$ is increasingly smaller at the redshifts of interest for realistic values of the concentration ratio $N_c$ and substructure fraction $F_{\rm sub}$ such that further generations of substructures, if they exist, are unlikely to increase $C_{\rm total}(z)$ by more than a few percent.

\section{Energy absorbed fraction}
\label{sec:Absorbed_Fraction}
\noindent A key quantity entering our computations is the fraction $\fabs$ of energy produced in each DM annihilation that is effectively absorbed by the IGM. In fact, taking $\fabs=1$ would be quite a poor approximation as sometimes just a very small fraction of the total energy actually goes into the heating/ionisation of the IGM. We describe in detail the method used to compute $\fabs$ in Appendix \ref{sec:app-fabs}; here we describe the annihilation spectra used for the different particle physics models that we adopt, provide some qualitative arguments to gain a physical insight on the mechanisms that lead to the absorption of particles, and finally, show the results obtained when the full method is invoked.


\subsection{Dark matter annihilation spectra}\label{sec:DM_ann}
\noindent Neutralino DM can annihilate directly into either 
a fermion pair or weak gauge bosons. Since the cross section
for annihilation to fermion pairs is proportional to 
the square of the final state fermion mass, this process
will be dominated by heavy final states, namely $b\bar b$, $\tau^-\tau^+$ and
$t\bar t$ (if kinematically allowed), while direct annihilation
into electron-positron pairs will be strongly suppressed. 
Hence we need only consider the following 
annihilation modes: $\chi\chi\to W^+W^-$, $\chi\chi\to ZZ$, 
$\chi\chi\to b\bar b$, $\chi\chi\to\tau^+\tau^-$ and $\chi\chi\to t \bar t$.
Both the gauge bosons and the fermion pairs
produced in neutralino annihilations will initiate a cascade that
will eventually lead to a continuum of photons, neutrinos,
electron/positrons pairs and protons in the final states, extending to energies
much smaller than the rest mass of the DM particle.
Here we utilise PYTHIA\footnote{
http://home.thep.lu.se/~torbjorn/Pythia.html
} \citep{Sjostrand:2000wi} to calculate these spectra.

The actual spectrum produced by the annihilations will depend 
on the branching ratios of the various channels; this in 
turn will be determined by the gaugino and higgsino fractions
of the neutralino. In the following, we will consider 
four representative supersymmetric scenarios, in a similar way
to what was done by \citet{Hooper:2003ad}.
First, we consider a 50\,GeV neutralino with an annihilation branching ratio
of 0.96 to $b\bar b$ and of 0.04 to $\tau^+ \tau^-$ (designated as model 1). 
Such a particle could be gaugino-like or higgsino-like, since for masses below the gauge boson masses, these modes dominate for 
either case. 
Second, we consider two cases for a 150\,GeV neutralino: One (designated as model 2) which annihilates as described in model 1, and  another (designated as model 3) which annihillates entirely to gauge bosons ($W^+W^-$ or $ZZ$). Such neutralinos are typically gaugino-like and higgsino-like respectively.  
Finally, we consider heavy, 600\,GeV neutralinos, which annihilate to $b\bar{b}$ with a ratio 
of 0.87 and to $\tau^+\tau^-$ or $t^+t^-$ the remaining time (designated as model 4).
Although these models do not fully encompass the extensive 
parameter space available to neutralinos at present, they do describe effective MSSM benchmarks. Furthermore, the relevant 
results for neutralinos with a mixture of the properties of those above can be inferred by interpolating 
between those presented. 

\begin{center}
\begin{figure*}
\includegraphics[width=0.45\textwidth,keepaspectratio,clip]{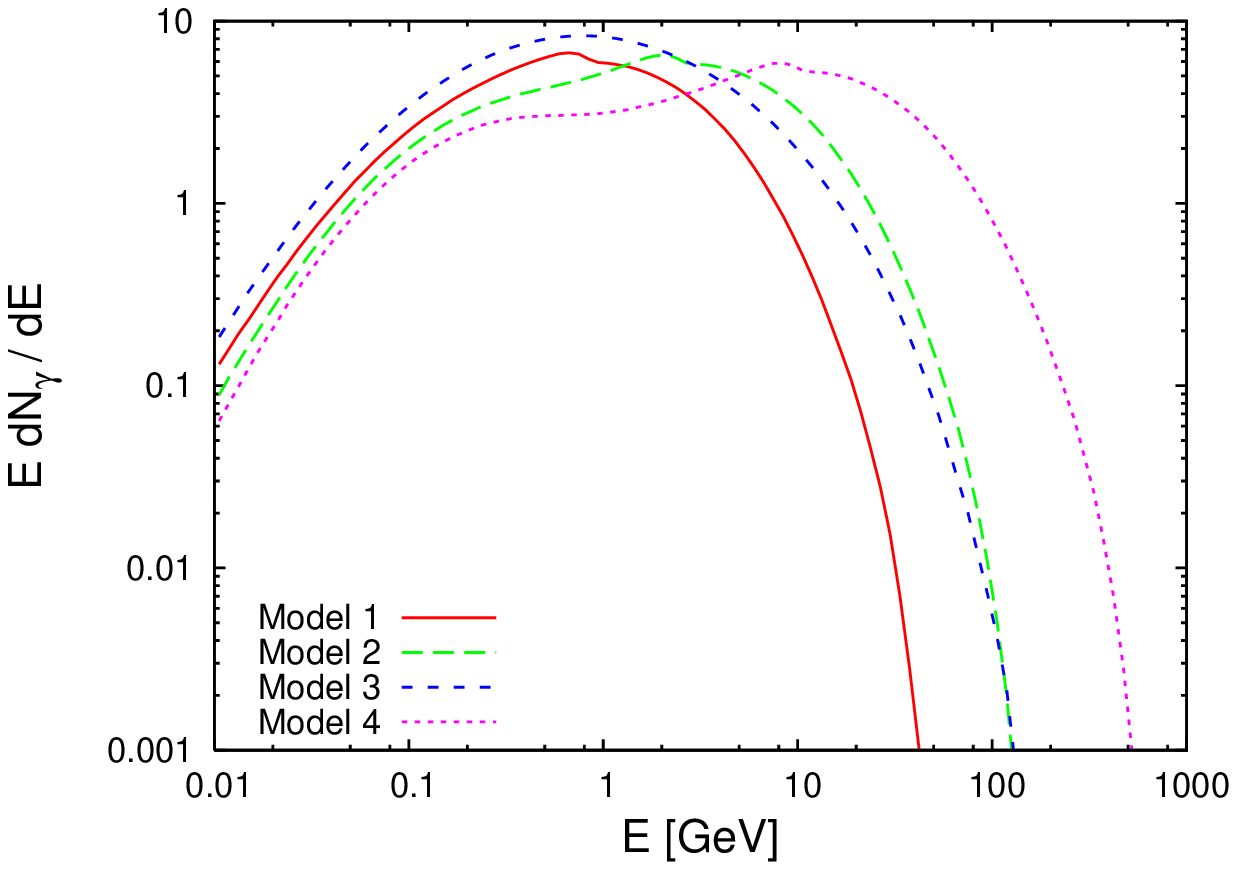}
\includegraphics[width=0.45\textwidth,keepaspectratio,clip]{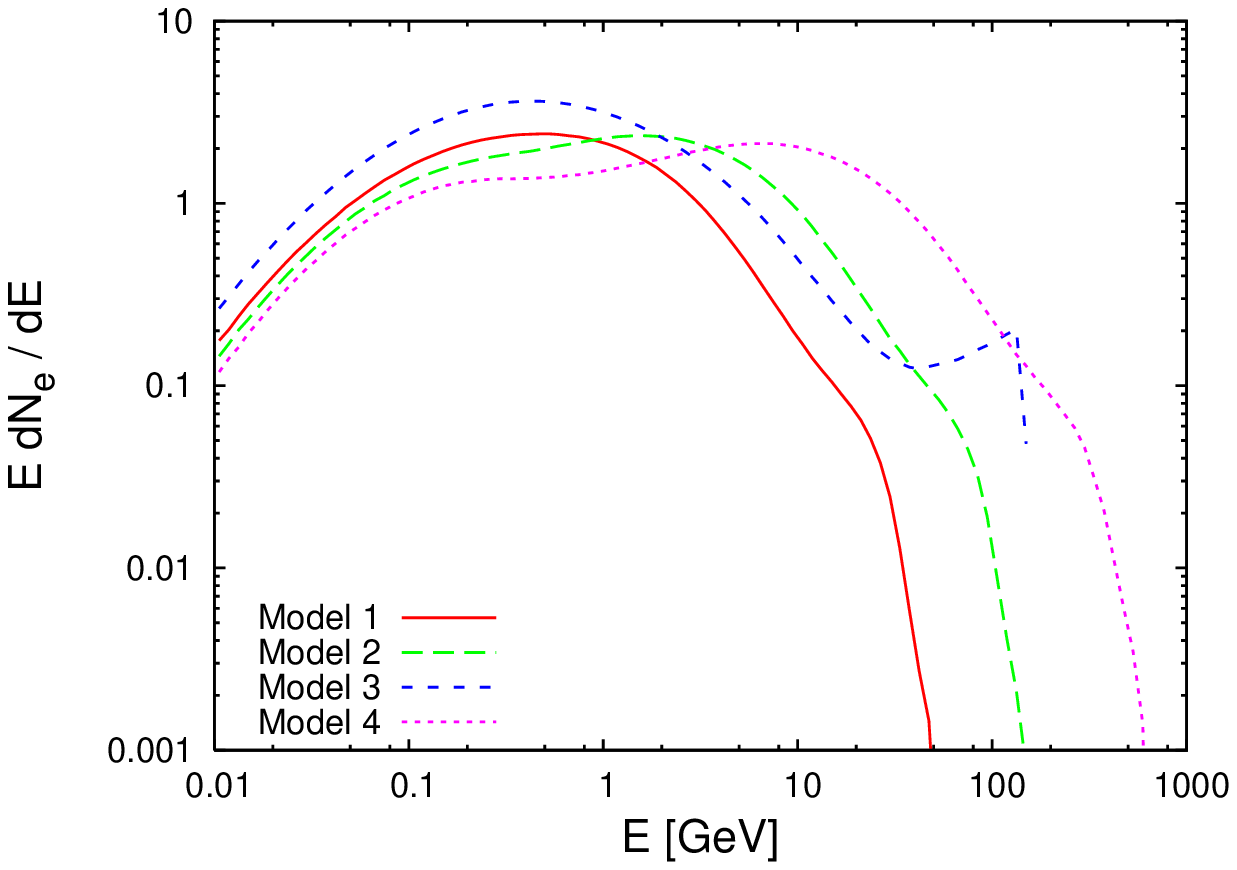}
\caption{Photon (left) and electron (right) energy spectra $E\,{\rm d}N/{\rm d}E$ resulting from a single neutralino annihilation, for our four benchmark models.}
\label{fig:spec_g}
\end{figure*}
\end{center}

In Fig.\,\ref{fig:spec_g} we show the spectrum of photons and electrons produced in a single annihilation for our four neutralino models. As we shall describe in more detail in Appendix \ref{sec:app-fabs}, in the numerical computation of $\fabs$ we will be
 make the approximation that the annihilation spectra are monochromatic and peaked at the average energy. In Table\,\ref{tab:av-en} we show the average photon and electron energy
 for the four models described above, together with the average number of particles produced in each annihilation.

In addition to neutralinos, we also consider light (MeV) DM candidates, which annihilate directly into electron-positron pairs.
This will result in a monochromatic spectrum with $E_\epm=m_\mathrm{DM}c^2$. We consider two different
LDM candidates with masses $m_\mathrm{DM} = 3$ and $20\,\MeV$ respectively. For completeness, we present the ``average'' energy and number of electrons (which is equal to the number of positrons) produced in
each annihilation in Table~\ref{tab:av-en}.
\begin{table}
\begin{center}
\begin{tabular}{lcccc}
\hline
Model &  $\bar N_\gamma$ & $\bar E_\gamma$ & $\bar N_{e^\pm}$ & $\bar E_{e^\pm}$  \\
\hline
\hline
SUSY-1 & 21.6 	&  1.3\,GeV  & 9.2  & 1.1\,GeV\\
SUSY-2 & 24.7 	& 3.5\,GeV   & 10.6 & 2.8\,GeV\\
SUSY-3 & 31.5		& 2.1\,GeV   & 14.6 & 2.6\,GeV\\
SUSY-4 & 25.8 	& 12\,GeV    & 11.3 & 9.8\,GeV\\
LDM 3\,MeV 		& - & - & 1 & 3\,MeV\\
LDM 20\,MeV & - & - & 1 & 20\,MeV  \\
\hline
\end{tabular}
\caption{Average number and energy of the photons and electrons (positrons) produced in a single DM annihilation, for the different models considered.}
\label{tab:av-en}
\end{center}
\end{table}


\subsection{Interaction of the annihilation products with the IGM}
\label{sec:int}

\noindent In this section we discuss the different processes by which the annihilation products of our DM candidates inject energy into the IGM. 
We will be only concerned with the interaction of photons and electron-positron pairs
with the IGM. Protons are very penetrating and thus do not transfer energy to the 
IGM; neutrino interactions
are so weak that they are also unable to transfer energy to the IGM,
so that the annihilation energy that ends up into protons and neutrinos is effectively
lost for the purpose of heating/ionising the IGM.

We will compute the transparency and opacity windows for photons and $e^+e^-$ pairs in order to gain a qualitative insight to the
regions in the $(E,z)$ plane where energy injection is expected to be efficient or not.
However, for the actual calculations of the absorbed energy fraction, the energy transfer between the annihilation products and the 
IGM is treated in more detail as explained in Appendix \ref{sec:app-fabs}.


\subsubsection{Photons}
\noindent As far as photons are concerned, we are mostly interested in the absorption of $\gamma$-ray 
photons. The absorption processes of x-ray and $\gamma$-ray photons at cosmological distances
were discussed by \citet{ZS89}. In principle, the possible energy loss mechanisms
for photons are: photoionisation of atoms; Compton scattering on electrons; pair production on atoms;
pair production of free electrons or nuclei; scattering on CMB photons; pair production on CMB photons.
The total rate for fractional energy loss, $\phi_\gamma(z,\,E)$, i.e., the fraction of the photon energy that
is lost in a unit time, is given by a sum over the contributions of the individual processes:
\begin{equation}
\phi_\gamma(z,\,E) = - \frac{1}{E}\frac{dE}{dt} = \sum_i \phi_{\gamma,i}
\end{equation} 
where the index $i$ runs over the different processes listed above.
However, for $z\lesssim 1500$, and in the range of energies we are interested (namely
$E\lesssim 10$~GeV), the relevant processes are photoionization, Compton scattering and pair production
on atoms or free electrons and nuclei. The effectiveness of these processes depends upon, other than on the photon energy, the density of the Universe at the redshift of interest. Roughly speaking, we can say that photoionization is effective
for energies below $\sim10$\,keV, while pair production is the dominant absorption mechanism at $z\gtrsim1000$ for $100\,\MeV\lesssim E\lesssim 10\,\GeV$.
Compton scattering is effective only for $z \gtrsim 100$, in a range of photon energies roughly
centered around $\sim 1$ MeV;   at $z=500$, the region where absorption
is dominated by Compton scattering extends roughly from 10\,keV to 30\,MeV.
The other processes, namely scattering on CMB photons and photon-photon pair production,
can be safely neglected for our purposes since they are only relevant either at
large redshifts or for very large ($E\gtrsim 100\,\GeV$) energies.

The rate for fractional energy loss by photoionisation $\phgion$ is given by
\begin{equation}
\phgion(z,E)=\frac{\sigma_{\mathrm{He+H}}(E)}{16}n_b(z) c,
\label{eq:photoion}
\end{equation}
where $n_b(z)$ is the number density of baryons at redshift $z$, and
$\sigma_{\He+\Hy}$ is the absorption cross section per helium atom
(hence the factor of 16 in Eq.(\ref{eq:photoion}), since $\nHe = \nb/16$),
given by
\begin{equation}
\sigma_{\He+\Hy}(E)=5.1\times 10^{-20}\left(\frac{E}{250\,\eV}\right)^{-p}\cm^2,
\end{equation}
where $p=3.3$ for $E>250\,\eV$, $p=2.65$ for $25\,\eV\le E\le250\,\eV$.

The fractional energy loss rate by Compton scattering is
\begin{equation}
\phgcom(z,E)=\sigma_T\, \epsilon\, g(\epsilon) n_e(z) c,
\label{eq:compton}
\end{equation}
where $\sigma_T$ is the Thomson cross section, $\epsilon=E/m_ec^2$ is the photon energy
in units of the electron mass, $n_e \simeq 0.88\,\nb$ is the total electron density at redshift $z$ (including
both free and bound electrons), and $g(\eps)$ is
\begin{multline}
g(\eps)=\frac{3}{8}\Bigg[\frac{(\eps-3)(\eps+1)}{\eps^4}\ln(1+2\eps)\\
+\frac{2\left(3+17\eps+31\eps^2+17\eps^3-10\eps^4/3\right)}{\eps^3(1+2\eps)^3}\Bigg].
\end{multline}
The corresponding term for pair production over atoms is given by
\begin{equation}
\phgpp(z,E)=0.63\times \alpha \sigma_T n_e(z) c \ln\left(\frac{513\eps}{\eps+825}\right),
\label{eq:pair-prod}
\end{equation}
while the one for pair production over ionized matter is
\begin{equation}
\phgpp(z,E)=0.8\times \alpha \sigma_T n_e(z) c \left(\ln 2\eps - \frac{109}{42} \right),
\label{eq:pair-prod2}
\end{equation}
where $\alpha$ is the fine structure constant.

A simple rule of thumb to assess the efficiency of the above energy loss mechanisms 
is to compare the rate $\phi_\gamma$ with the expansion rate, as
given by the value of the Hubble constant $H(z)$. When $\phi_\gamma\gg H(z)$,
the photon loses all of its energy on a time scale small compared to the cosmological time, so that
the energy loss mechanisms are very effective and the universe is opaque to
its propagation. It can then be assumed that all the photon energy is instantly
lost, and, in the case of photoionisation and Compton scattering, instantly deposited into the IGM 
(in the case of pair production, one should take into account the subsequent interaction of the pair with the IGM - see Appendix \ref{sec:app-fabs} for details). 
In the opposite regime, $\phi_\gamma\ll H(z)$, the 
photon loses a significant fraction of its energy on a time scale larger than 
the cosmological time, and the Universe 
is effectively transparent to the photon propagation.

\begin{center}
\begin{figure}
\includegraphics[width=\linewidth,keepaspectratio,clip]{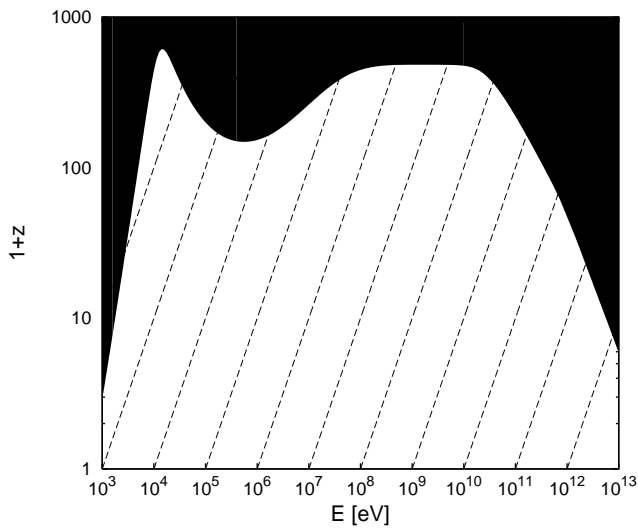}
\caption{Photon transparency window. In the black region, the photon loses
all of its energy through interaction with particles in the IGM and CMB photons.
In the white region, the photon can propagate freely. The dashed lines represent
photon trajectories.}
\label{fig:transp}
\end{figure}
\end{center}

Following Chen \& Kamionkowski \cite{chen04}, in Fig.\,\ref{fig:transp} we show the photon transparency window in the 
$(E,\,z)$ plane. For illustrative purposes, we consider redshifts
as large as $z=1000$ and energies up to 10\,TeV in the figure, so that, in addition
to the three processes for which we have listed explicitly the 
energy loss rates, we have
also included the scattering and pair production over CMB photons
in the total rate $\phi_\gamma$.
In the filled region, $\phi_\gamma > H(z)$, 
corresponding to the optically thick regime. In the white region,
$\phi_\gamma < H(z)$, 
corresponding to the optically thin regime.

Although the transparency window can be useful as a preliminary tool, in order, for example, to assess which processes are important at a given redshift and energy range, it has some limitations nevertheless. First of all, it does not allow us to treat properly the regime $\phi_\gamma \simeq H(z)$, i.e. the regime where the energy loss happens on cosmological time scales. In this case, the approximation of an instantaneous energy deposition is not appropriate, since part of the energy can be deposited at a redshift lower than the redshift of emission. Secondly, even if in the regime $\phi_\gamma \gg H(z)$ one can safely conclude that all the energy of the annihilation products is instantaneously lost, this does not mean that it is instantaneously transferred (or even transferred at all) into the IGM: in some cases the interactions of the annihilation products generate secondary particles, like the $e^+e^-$ generated by the pair production of photons on atoms or nuclei, whose propagation has to be followed as well. For these reasons we follow (with some small modifications) the approach of \citep{Ripamonti:2006gq} to calculate $\fabs$; the detailed calculations and the results for $\fabs$ are described in Appendix \ref{sec:app-fabs}.

In any case we can gain some qualitative insight by looking at the Figs.\,\ref{fig:spec_g} and \ref{fig:transp}. For the supersymmetric models considered, the average
energy of the photons produced in each annihilation is of order of a few GeV, and
their energy is at most a few hundred GeV (in the case of our heaviest candidate, the 600\,GeV neutralino of model 4, only
$\sim 1\%$ of the total energy produced in the annihilation is released in the form of photons with energy $E_\gamma>200$\,GeV). 
As it can be seen from Fig.\,\ref{fig:transp} above, these photons lie in the middle of the transparency window: their energy is too low for pair production, as already noticed, but on the other hand it is too high for photoionisation
(or Compton scattering at $z>100$) to be effective. These photons will propagate freely and
their energy will decrease due to cosmological expansion. Although it is in principle possible that,
due to cosmological redshifting, a photon produced in the transparency
window at a given time will be absorbed later, we see from Fig.\,\ref{fig:transp} that this is practically never
the case. In conclusion, we expect that the absorbed fraction for photons at $z<1000$ will be very small, and that the
photons produced in neutralino annihilations will instead show up in the diffuse gamma-ray background.


\subsubsection{Electron-positron pairs}

\noindent Electrons and positrons can lose energy through
collisional ionisation of atoms or through
inverse Compton scattering off of CMB photons. In addition, positrons
can annihilate with thermal electrons. Other energy loss mechanisms,
like synchrotron radiation loss, can be safely neglected.

The rate of energy loss through collisional ionisation is given by
\begin{multline}
\pheion(E,z) = \frac{v}{E}\frac{2\pi e^4}{m_e v^2}\\
	\times\Bigg\{Z_\Hy \nH\bigg[\ln\left(\frac{m_e v^2\gamma^2 T_\mathrm{max,\Hy}}{2 I^2_\Hy}\right)+{\cal D}(\gamma)\bigg]\\
+Z_\He \nHe\bigg[\ln\left(\frac{m_e v^2\gamma^2 T_\mathrm{max,\He}}{2 I^2_\He}\right)+{\cal D}(\gamma)\bigg]\Bigg\},
\end{multline}
where $v$ is the electron velocity, $\gamma=E/m_ec^2$ is the electron Lorentz factor, $I_\Hy=13.59\,\eV$ and 
$I_\He=24.6\,\eV$ are the hydrogen and helium ionisation thresholds respectively, $Z_\Hy$ and $Z_\He$ are the hydrogen
and helium atomic numbers respectively, the function ${\cal D}(\gamma)$ is given by
\begin{equation}
{\cal D}(\gamma)=\frac{1}{\gamma^2}-\left(\frac{2}{\gamma}-\frac{1}{\gamma^2}\right)\ln{2}+\frac{1}{8}\left(1-\frac{1}{\gamma}\right)^2,
\end{equation}
and $T_{\mathrm{max.},\Hy}$ and $T_{\mathrm{max.},\He}$ are the maximum energy transfers in a single collision
\begin{align}
T_{\mathrm{max.},\Hy}&=\frac{2\gamma^2 m_\Hy^2 m_e v^2}{m_e^2+m_\Hy^2+2\gamma m_e m_\Hy},\\
T_{\mathrm{max.},\He}&=\frac{2\gamma^2 m_\He^2 m_e v^2}{m_e^2+m_\He^2+2\gamma m_e m_\He}.
\end{align}
The fractional energy loss rate through inverse Compton scattering is given by
\begin{equation}
\phecom(z,E) = \frac{4}{3}\frac{\sigma_T U_{\CMB}(z)}{{m_e}}\,\frac{\gamma^2-1}{\gamma},
\label{eq:invcom}
\end{equation}
where $U_{\CMB}(z)$ is the CMB energy density at redshift $z$.

In the case of inverse Compton losses, we must take into account that the electrons and positrons do not transfer their energy directly into the IGM; instead, they accelerate the CMB photons they interact with, boosting their energy by a factor $\sim \gamma^2$ . These up-scattered photons can either be absorbed by the IGM or escape, depending on their energy. 
As explained above, at redshifts below a few hundred, the only relevant photon absorption processes are photoionisation, Compton scattering and  pair production; however, a simple calculation shows that the electrons and positrons produced in the  annihilation of the DM candidates considered here are not energetic enough to boost the CMB photons above the threshold for pair production. We can also safely neglect Compton scattering, since it is only efficient for $z>100$ and in a small energy region around 1\,MeV. Therefore we need only consider photoionisation as the secondary process leading to the absorption of the photons produced by inverse Compton scattering of electrons and positrons. The method that we use to estimate the energy injected in the IGM by the up-scattered photons is described in detail in Appendix \ref{sec:app-fabs}. Here we just show the results concerning the opacity window of electrons and positrons.

\begin{center}
\begin{figure}
\includegraphics[width=\linewidth,keepaspectratio,clip]{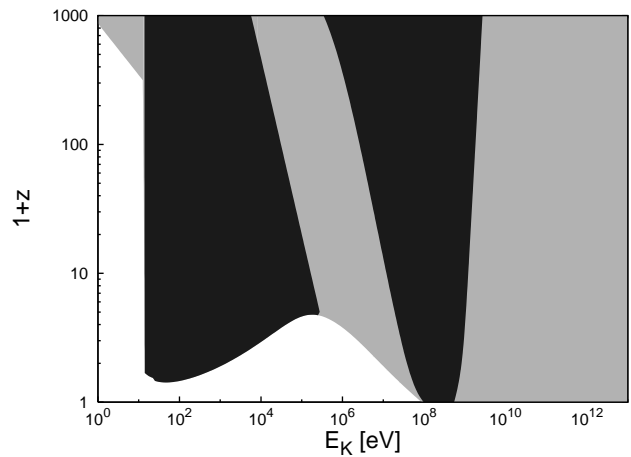}
\caption{Transparency window for electrons. In the white region, electrons propagate freely. In the grey regions,
electrons transfer all of their energy to the CMB photons, but these are subsequently lost, so that no energy is injected in the IGM. In the black regions, all the electron energy is efficiently injected in the IGM. See text for discussion.}
\label{fig:el-trans}
\end{figure}
\end{center}

The behaviour of electron-positron pairs with respect to the energy transfer to the IGM is summarised in  Fig.\,\ref{fig:el-trans}. In the white region, the total energy loss rate is smaller than the expansion rate: $\pheion+\phecom < H$, so that the Universe
is transparent to the propagation of electrons. In the grey regions, the electrons and positrons interact by inverse Compton scattering, but the resulting photons fall in the photon transparency window. This means that the Universe is opaque to the propagation of electrons, but nevertheless their energy is not transferred to the IGM.
Finally, in the black regions the electron energy is efficiently transferred to the IGM. In particular, the region on the left correspond to the case in which 
collisional ionisation is the dominant process; the region on the right 
is where inverse Compton is the dominant interaction, and the up-scattered CMB photons fall into the photon absorption window.

\section{the 21\,cm Background}
\label{sec:21cm_Background}


\subsection{CMB-kinetic temperature coupling}
\label{sec:CMB-T_K coupling}

\noindent In this section we briefly review the basic physics behind the 21\,cm signal. For a more in-depth discussion, we refer the reader to Refs. \cite{Madau:1996cs,Furlanetto:2006jb,Barkana:2006ep} and references therein.

The emission or absorption of the 21\,cm line signal emanating from neutral gas is associated with the transition between the $n=1$ triplet and singlet hyperfine levels of hydrogen. The transition rate is governed by the spin temperature, $T_s$, defined as
\begin{equation}
\frac{n_1}{n_0}=3\,{\rm exp}\left(-\frac{T_*}{T_S}\right),
\end{equation} 
where $n_0$ and $n_1$ are the respective number densities of hydrogen atoms in the singlet and triplet states, and $T_*=0.068$K is the equivalent temperature corresponding to the transition energy. 


In the presence of the CMB radiation field, the spin temperature of the neutral hydrogen gas rapidly tends towards the CMB temperature $T_{\rm CMB}\simeq2.725(1+z)$K. In order for neutral hydrogen gas to produce a detectable signal in the 21\,cm background, be it in absorption or emission, that is distinguishable from that generated from the CMB, the kinetic temperature $T_K$ of the gas must decouple from $T_{\rm CMB}$.

In a Universe containing stable, non-annihilating DM, the spin temperature and the kinetic temperature of HI gas are coupled to $T_{\rm CMB}$ until $z\simeq200$ \cite{Peebles:1994xt}. At $30\lesssim z \lesssim 200$, prior to the formation of non-linear baryonic structures, the IGM cools adiabatically, i.e. $T_K\propto(1+z)^2$, compared to $T_{\rm CMB}\propto(1+z)$. During this epoch, spin-exchange collisions between hydrogen atoms, protons and electrons are efficient at coupling $T_K$ and $T_S$ of the gas, and consequently an absorption at wavelength $\lambda=21(1+z)$\,cm can be observed until approximately $z\simeq70$. At later times cosmological expansion reduces the frequency of these collisions significantly, to the extent where $T_S$ re-couples with $T_{\rm CMB}$, diminishing the 21\,cm absorption signal.

However, in a Universe containing annihilating/decaying DM which injects  appreciable energy into the IGM, the thermal history of the gas may be significantly altered to the extent where the corresponding changes in the evolution of the 21\,cm signal are detectable by current and future radio experiments. Of particular importance is the high sensitivity of these changes with respect to the nature of the DM, making them a powerful tool for constraining the properties of potential DM candidates.

There are two mechanisms which can decouple $T_S$ from $T_{\rm CMB}$: firstly, the aforementioned spin-exchange collisions between neutral atoms, electrons and protons \cite{purcell56}, which are effective at $z\ge70$ before the Hubble expansion has rarefied the gas in the IGM, and secondly, scattering by Lyman-$\alpha$ radiation (known as the ``Wouthuysen-Field'' effect, also known as ``Lyman-$\alpha$ pumping'' \citep{wouthuysen, field, hirata}), which couples $T_S$ to $T_K$ via the mixing of the $n=1$ hyperfine states through intermediate transitions to the $2p$ state.
 
In the quasi-static approximation for the population of the hyperfine levels in question, and in the absence of radio sources, the HI spin temperature is a weighted mean involving $T_K$ and $T_{\rm CMB}$, 
\begin{equation}
T_S=\frac{T_{\rm CMB}+yT_K}{1+y}, 
\end{equation}
\noindent The coupling coefficient $y$ can be written as
\begin{equation}
y=y_{\alpha}+y_C,
\end{equation}
\noindent where $y_{\alpha}$ is the term associated with Lyman-${\alpha}$ pumping, given by
\begin{equation}
y_{\alpha}=\frac{P_{10}T_*}{A_{10}T_K},
\end{equation}
\noindent whilst $y_C$ is associated with the de-excitation of the HI hyperfine triplet state due to collisions with neutral atoms, electrons and protons, collectively written as
\begin{equation}
y_C=\frac{T_*}{A_{10}T_K}(C_{\rm H}+C_e+C_p).
\end{equation}
In the above equations $A_{10}=2.85\times10^{-15}$\,s is the rate of spontaneous photon emission, $P_{10}$ is the de-excitation rate of the hyperfine triplet state due to Lyman-$\alpha$ scattering, and $C_{\rm H}, C_e$ and $C_p$ are the de-excitation rates associated with collisions of hydrogen atoms with other hydrogen atoms, electrons and protons respectively.
We write $P_{10}=(16\pi J_{\alpha}\sigma_{\alpha})/{(27 h_p \nu_\alpha)}$, where $J_{\alpha}$ is the background intensity of Lyman-${\alpha}$ photons, $\sigma_{\alpha}$ is the Lyman-${\alpha}$  photon absorption cross section for neutral hydrogen and $h_p$ is Planck's constant.  We neglect the small corrections to the above expressions proposed by \citet{hirata}.
The H-H collision term can be written as $C_{\rm H}=k_{10}n_{\rm HI}$, where $k_{10}$ is the effective single-atom collision rate coefficient for which we adopt the fit: $k_{10}=3.1\times10^{-11}T_{K}^{0.357}\exp(-32\,{\rm K}/T_K)\,$cm$^3$\,s$^{-1}$ proposed by \citet{kuhlen2006}, which is accurate to within $0.5\%$ in the range $10<T_K<10^3\,$K. 
For the e-H collision term, $C_e=n_e\gamma_e$, we have used the following fit\footnote{Updated rates can be found in \cite{Furlanetto:2006su}.} proposed by \citet{liszt}: $\log(\gamma_e/{\rm cm}^3\,{\rm s}^{-1})=-9.607+0.5\,\log(T_K/{1\,\rm K})\exp\left\{-\left[\log(T_K/1{\,\rm K})\right]^{4.5}/1800\right\}$ for $1<T_K<10^4$\,K, $\log(\gamma_e/{\rm cm}^3\,{\rm s}^{-1})=-9.607+0.5\,\log(T_K/1\,{\rm K})$ for $T_K<1$\,K \citep{smith}, and $\gamma_e(T_K>10^4\,{\rm K})=\gamma_e(10^4\,{\rm K})$. We ignore de-excitations involving collisions with protons since they are typically much weaker than those involving electrons at the same temperature, although it has been shown that they can be relevant at low temperatures \cite{Furlanetto:2007te}.


\subsection{Modifications to IGM thermal evolution in the presence of DM}

\noindent In this section we describe the modifications to the standard equations describing the  thermal and ionisation history of the IGM when we incorporate the potentially significant energy deposition of the products of annihilating DM.

We parameterize the effect of DM annihilation by the rate of energy injection given by Eq.(\ref{eq:ann_rate}). This energy is then used to excite and ionise the hydrogen and helium atoms in the IGM. We will not enter here into the detail of the partition of energy between hydrogen and helium, but instead assume that it is divided proportionally to the respective number densities. This means that a fraction $1/(1+f_\He)$ of the absorbed energy will go to hydrogen, while a fraction $f_\He/(1+f_\He)$ will go to helium, $f_\He$ being the helium to hydrogen number ratio. Then we need to know how the energy is partitioned between the different processes.
The relative fractions $\chi_i, \chi_h$ and $\chi_{e}$ of the energy absorbed which is diverted towards respectively ionising, heating and exciting hydrogen and helium atoms were calculated by \citet{shull_vsteenberg}. Their results can be approximated by 
\citep{chen04}
 \begin{eqnarray}
\chi_{i,j}(z)&\sim&\frac{[1-x_{j}(z)]}{3},\\
\chi_{h,j}(z)&\sim&\frac{[1+2x_{j}(z)]}{3},\\
\chi_{e,j}(z)&\sim&\frac{[1-x_{j}(z)]}{3},
\end{eqnarray} 
where $x_{j}(z)$ is the ionisation fraction of the relevant nuclear species $j$ (i.e. $j$=H or He for hydrogen or helium nuclei respectively), defined as
\begin{equation}
x_j=\frac{n_{j^+}}{n_j},
\end{equation}
where $n_{j^+}$ is the number density of ionised nuclei of the species $j$. 
We can also define a total ionisation efficiency $\chi_i\equiv (\chi_{i,\Hy}+f_\He\chi_{i,\He})/(1+f_\He)$, and similar quantities for heating and excitation. 

Following \citep{padm05}, we compute the ionisation and thermal history of the IGM, when incorporating our chosen species of DM, using the publicly available code RECFAST \citep*{seager1999, seager2000}, modifying the standard evolution equations for the ionisation fractions of hydrogen and helium nuclei, as well as the evolution equation for the kinetic temperature as follows:
\begin{eqnarray}
-\delta\left(\frac{{\rm d}x[{\rm H}]}{{\rm d}z}\right)&=&\frac{{\dot\epsilon}}{I_{H}}\frac{\chi_{i, {\rm H}}}{(1+f_{\rm He})}\frac{1}{H(z)(1+z)},\\
-\delta\left(\frac{{\rm d}x[{\rm He}]}{{\rm d}z}\right)&=&\frac{{\dot\epsilon}}{I_{{\rm He}}}\frac{\chi_{i, {\rm He}}}{(1+f_{\rm He})}\frac{1}{H(z)(1+z)},\\
-\delta\left(\frac{{\rm d}T_k}{{\rm d}z}\right)&=&\frac{2\dot\epsilon}{3k_{\rm B}}\frac{(\chi_{h, {\rm H}}+f_{\rm He}\chi_{h, {\rm He}})}{(1+f_{\rm He})H(z)(1+z)}.
\end{eqnarray}
A further equation needed to calculate the 21\,cm signal is that describing the evolution of the Lyman-$\alpha$ background intensity $J_{\alpha}$, which can couple the spin and kinetic temperatures of the H-atoms in the IGM via the Wouthuysen-Field effect. H-atoms, excited by collisions with fast photoelectrons subsequently produce a cascade of line photons, including Lyman-$\alpha$ photons which are then likely to be re-absorbed by the optically-thick IGM. We utilise the approximation adopted by \citet{furlanetto} that approximately half of the total energy which is diverted to excite hydrogen is used to produce Lyman-$\alpha$ photons\footnote{The authors of \cite{Pritchard:2006sq} actually find that $\chi_\alpha$ is somewhat larger than the value used here, $\chi_\alpha\simeq 0.79\chi_{e,\Hy}/2$. However we do not think this would alter our results significantly; in any case it would result in a larger deviation of the brightness temperature, so our results can be considered as more conservative.}, i.e. $\chi_{\alpha}\sim\chi_{e,\Hy}/2$.  

Following the treatment by \citet{valdes} we obtain (however, note the additional correction factor of $\nu_\alpha$ in front of the expression)
\begin{equation}
J_{\alpha}=\frac{n_{\rm H}^2hc\nu_\alpha}{4\pi H(z)}\left[x_ex_{e, {\rm H}}\alpha_{2^2P}^{\rm eff.}+x_{e, {\rm H}}x_{\rm HI}\gamma_{\rm eH}+\frac{\chi_\alpha{\dot\epsilon}}{n_{\rm H}h\nu_{\alpha}}\right],
\end{equation}
where the first two terms are the contributions associated with the collisional excitation involving electrons discussed above, and the last term is the contribution from DM. Also, $\alpha_{2^2P}^{\rm eff.}$ is the effective recombination coefficient to the $2^2P$ level \citep{pengelly}, and $\gamma_{e{\rm H}}\simeq2.2\times10^{-8}\exp\left[-11.84/(T/10^4\,{\rm K})\right]$cm$^3$\,s$^{-1}$ is the collisional excitation rate of HI atoms involving electrons \citep{shull_vsteenberg}.

The quantity most intimately associated with observations of the cosmological 21\,cm signal is the differential brightness temperature deviation, $\delta T_b$, between the 21\,cm signal and the CMB, approximately given by \citep{field1958, field, ciardi}
\begin{eqnarray}
\delta T_b&\simeq&26~{\rm mK}~x_{\rm HI}\left(1-\frac{T_{\rm CMB}}{T_S}\right)\left(\frac{\Omega_bh^2}{0.02}\right)\nonumber\\
&\times&\left[\left(\frac{1+z}{10}\right)\left(\frac{0.3}{\Omega_{\rm M}}\right)\right]^{1/2},
\label{eq:dT_b}
\end{eqnarray}
\noindent where $x_{\rm HI}=1-x_{\rm H}$ is the average fraction of neutral hydrogen in the patch of sky being observed.


\section{Results}
\label{sec:Results}

\noindent In the following section, we illustrate our predictions for the effects on the thermal history of the IGM caused by the additional energy injected into it by annihilating neutralino CDM and LDM, when including the enhancement effects from DM structures. 

Since, as we have seen, this enhancement can be very large, boosting the DM annihilation rate by several orders of magnitude, we want to be sure that this does not contradict other observations. Consequently, we perform two tests on each
of the clumping factors investigated, before taking into consideration its effect on the 21\,cm brightness temperature. First of all, we check that the huge injection of energy into the IGM does not lead to premature re-ionisation. We use for this purpose our modified version of the RECFAST code described above, and discard all clumping factors for which the ionised fraction $x_e>0.01$ at $z=14$. We choose this value of the redshift because it is close to the
$3\sigma$ upper limit to $z_\mathrm{reion.}$ coming from WMAP7 observations \cite{Komatsu:2010fb}. Secondly, we
check that the diffuse photon flux produced does not exceed the observed diffuse gamma-ray and x-ray background (adopting the conservative approximation that
$\fabs \sim 0$, that for all the models we consider is quite a good approximation at the present time $z=0$, where the clumping factor reaches its maximum value). We use to this purpose the measurements of the diffuse gamma-ray background in the 1\,MeV - 100\,GeV range conducted by EGRET \citep{Sreekumar:1997un,Strong:2004ry} and COMPTEL \citep{comptel}, and those of the diffuse x-ray background in the sub-MeV range conducted by the SPI spectrometer aboard INTEGRAL \citep{integral}.

In the following, owing to the fact that the effects on the spin and brightness temperature can be very subtle, we display results only for the the most optimistic clumping factors, which we define, for a given DM model, as those which yield the largest difference in the differential brightness temperature, $\delta T_b-\delta T_{b,0}$ (see \S\,\ref{sec:Discussion}), at $z=30$ (i.e., the smaller $z$ were plausibly astrophysical effects are not important, see discussion at the end), 
while at the same time conforming to the above criteria. For reference, in Appendix \ref{sec:app-tables} we have tabulated the relevant astrophysical parameters associated with all clumping factors investigated, indicating which have been excluded on the basis of the criteria described above.


\subsection{Neutralino dark matter}

\begin{figure*}
	\begin{center}
	\includegraphics[width=0.45\linewidth,keepaspectratio,clip]{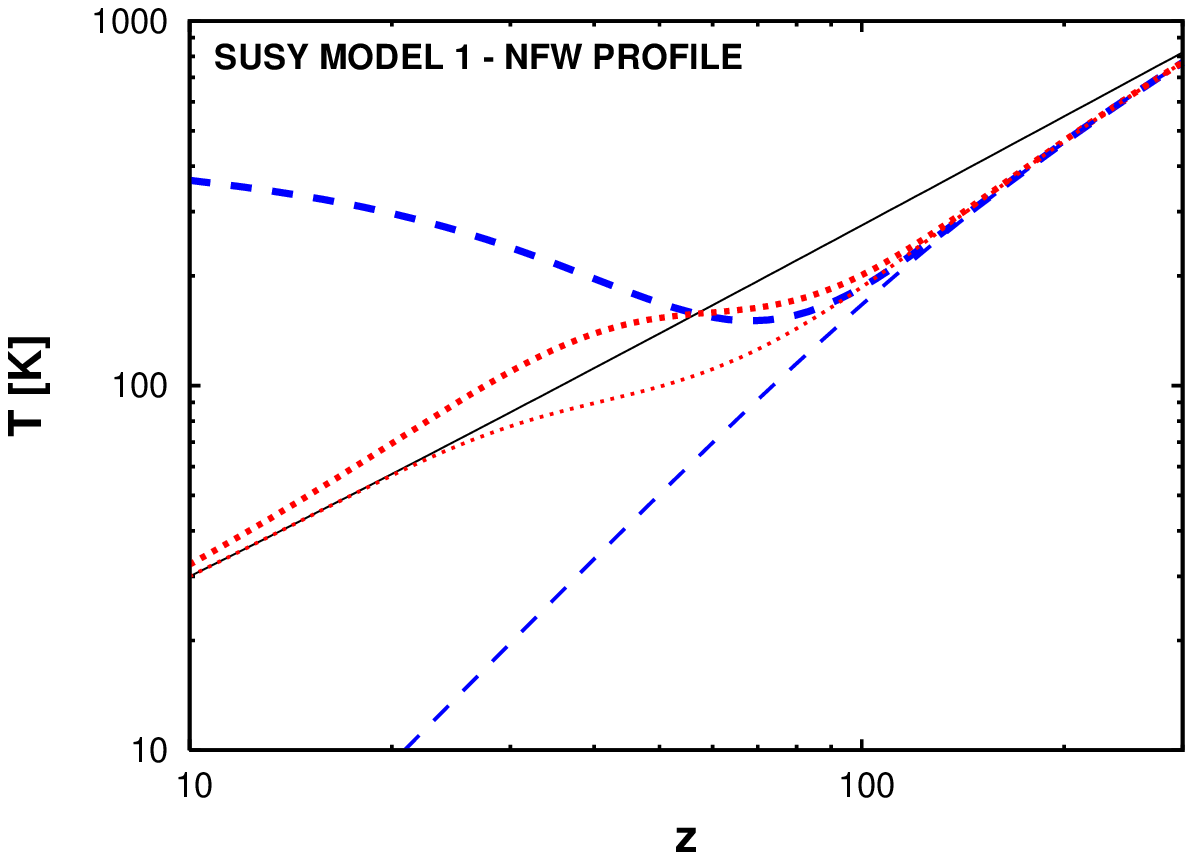}
	\includegraphics[width=0.45\linewidth,keepaspectratio,clip]{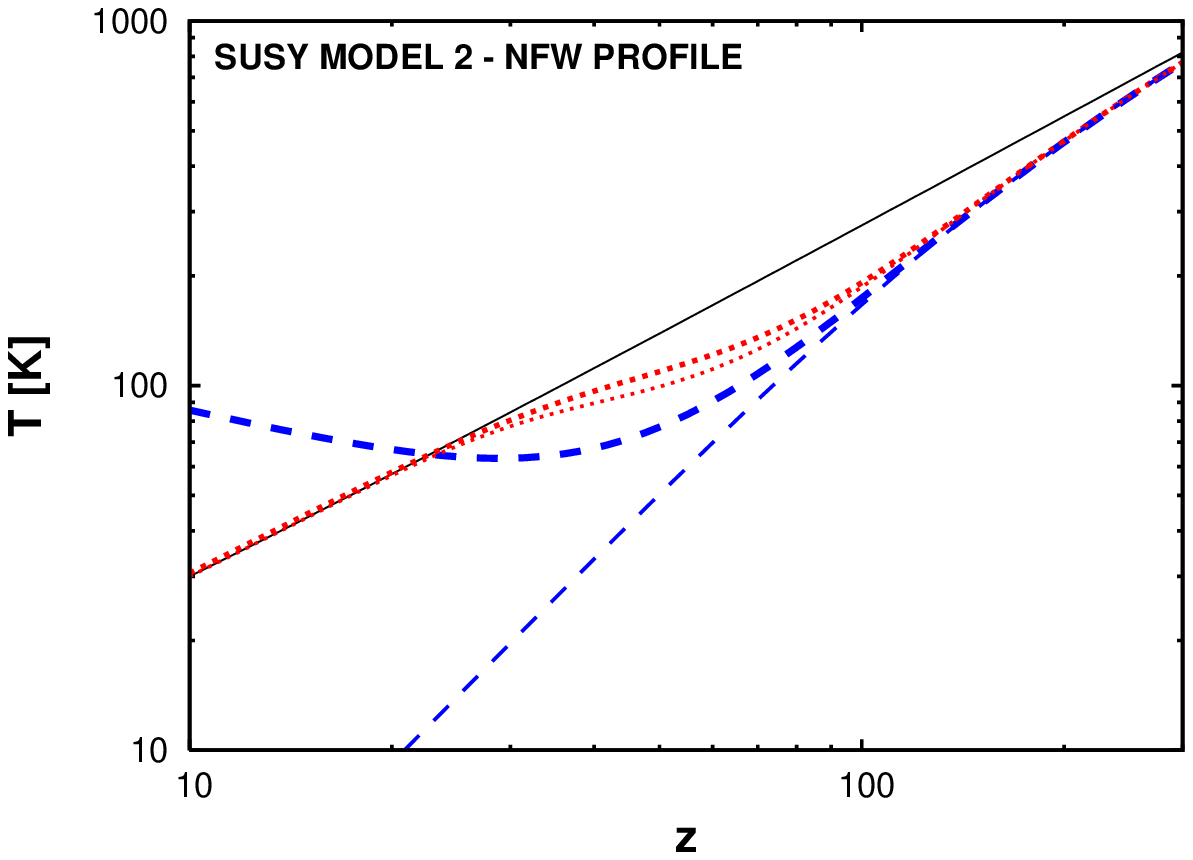}
	\includegraphics[width=0.45\linewidth,keepaspectratio,clip]{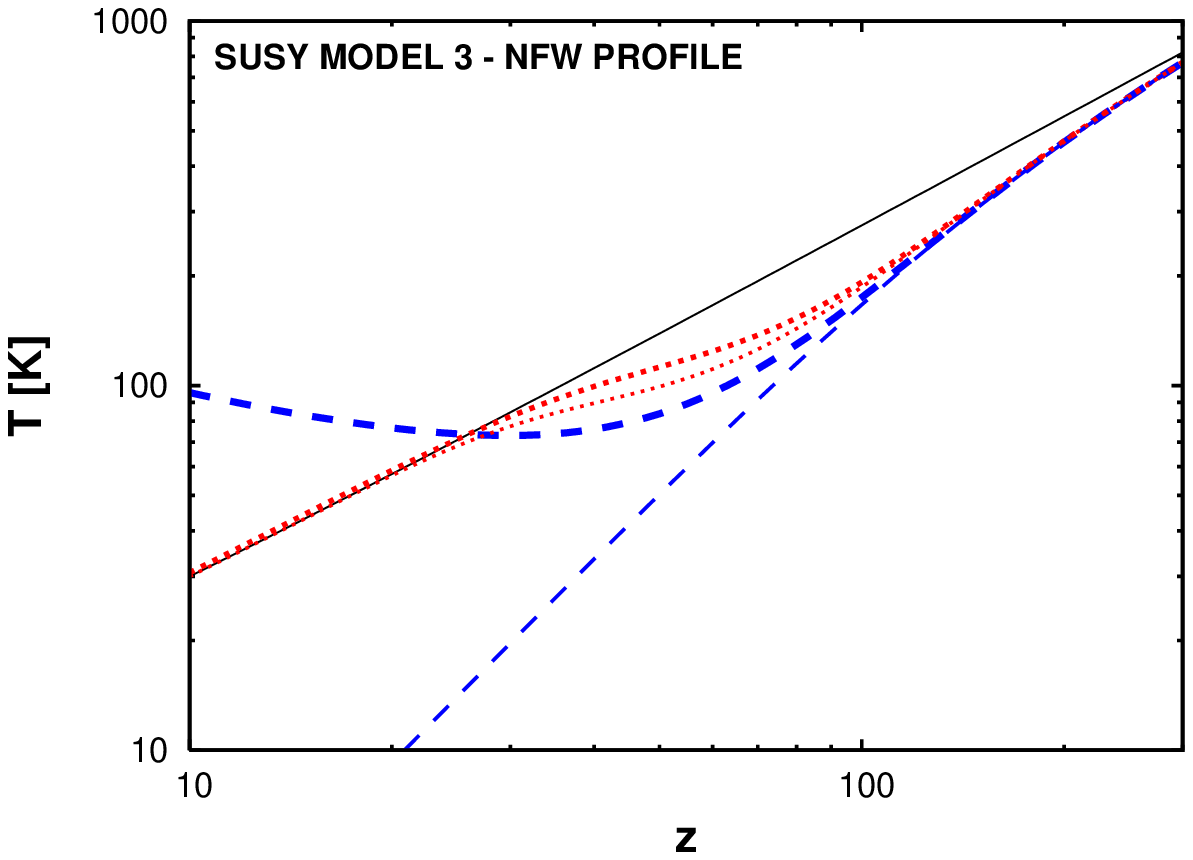}
	\includegraphics[width=0.45\linewidth,keepaspectratio,clip]{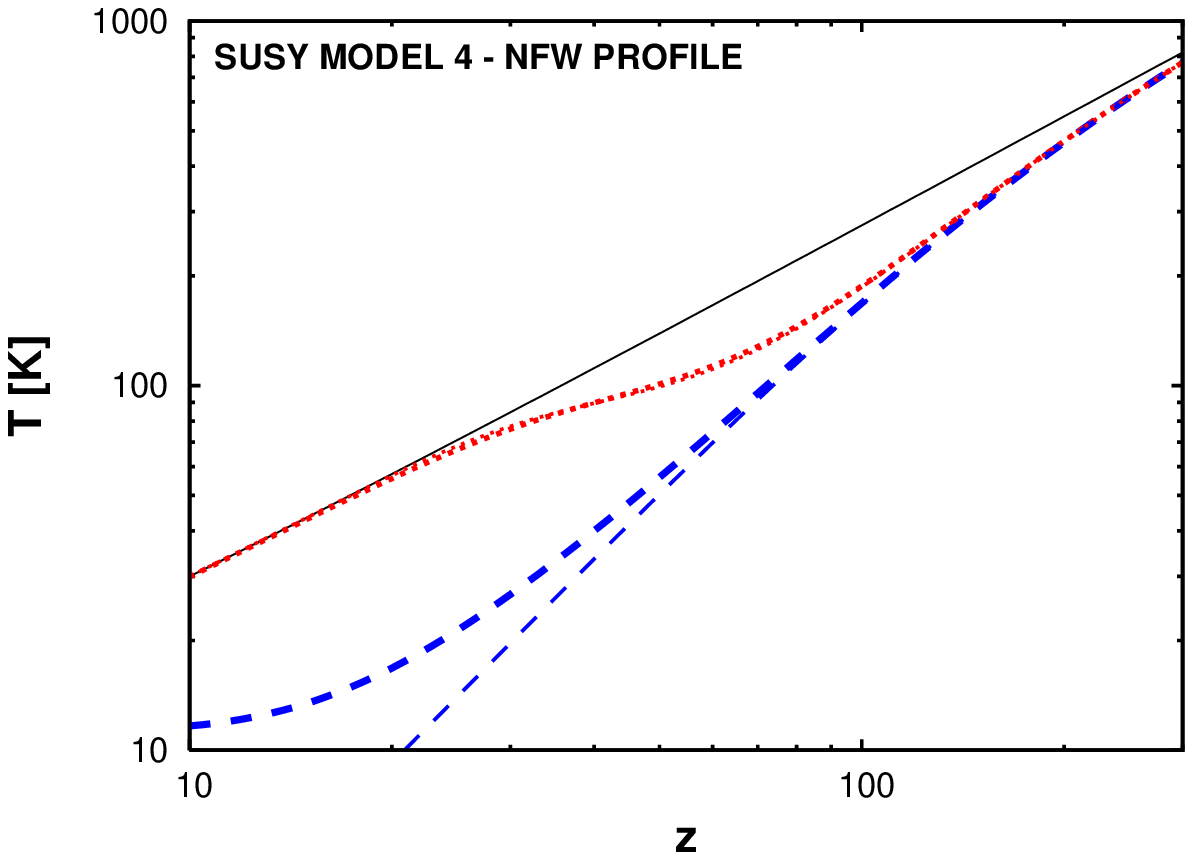}
	\caption{Evolution of the IGM kinetic (thick dashed blue curves) and spin (dotted red curves) temperatures for our four supersymmetric models. In each plot we show the results for the most optimistic clumping factors using the NFW density profile (thick curves) and, for comparison, the kinetic and spin temperatures in the absence of DM annihilations (thin curves).  The CMB temperature is also shown (black solid curve). The annihilation cross section $\langle\sigma_{\rm ann.}\upsilon\rangle=2.7\times10^{-26}$\,cm$^3$\,s$^{-1}$ in all plots. The clumping factors used are N4 for model 1, and N2 for models 2, 3, and 4.}
	\label{fig:T-plot_NFW}
         \end{center}
\end{figure*}

\begin{figure*}
	\begin{center}
	\includegraphics[width=0.45\linewidth,keepaspectratio,clip]{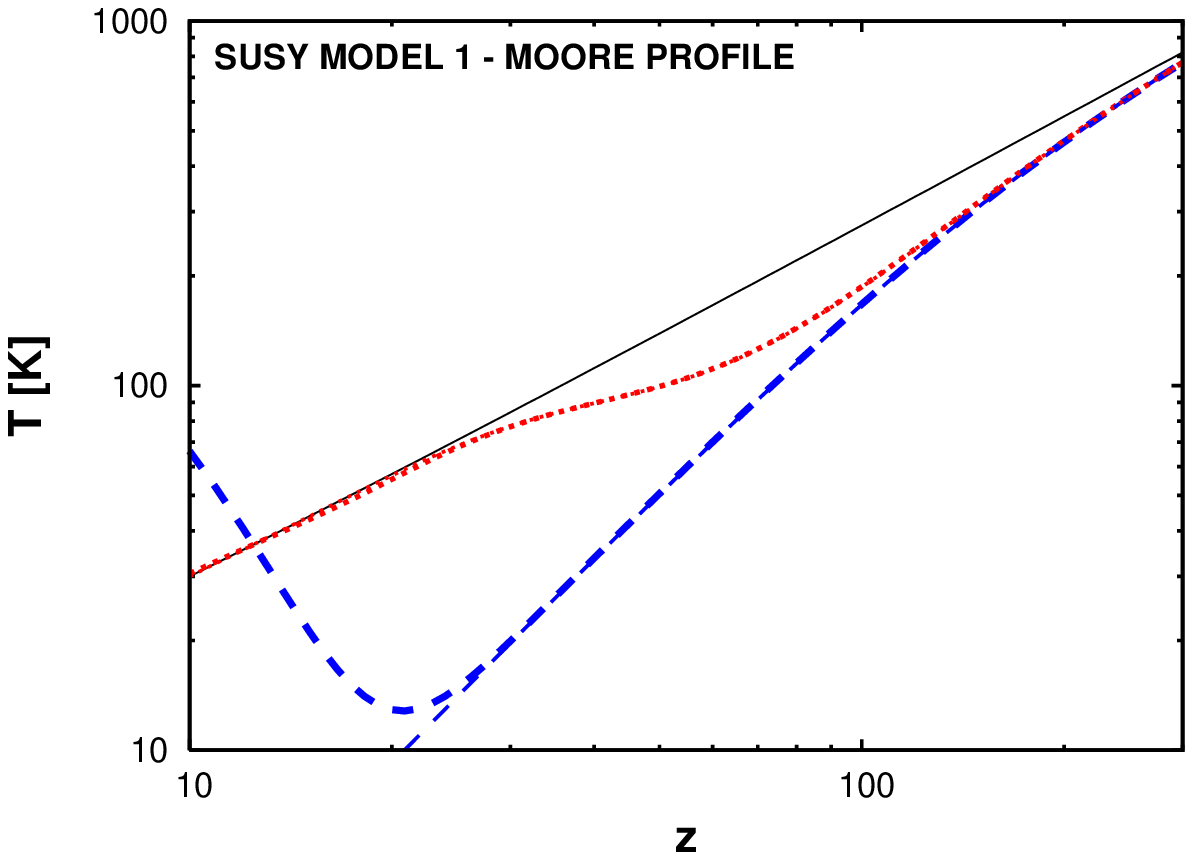}
	\includegraphics[width=0.45\linewidth,keepaspectratio,clip]{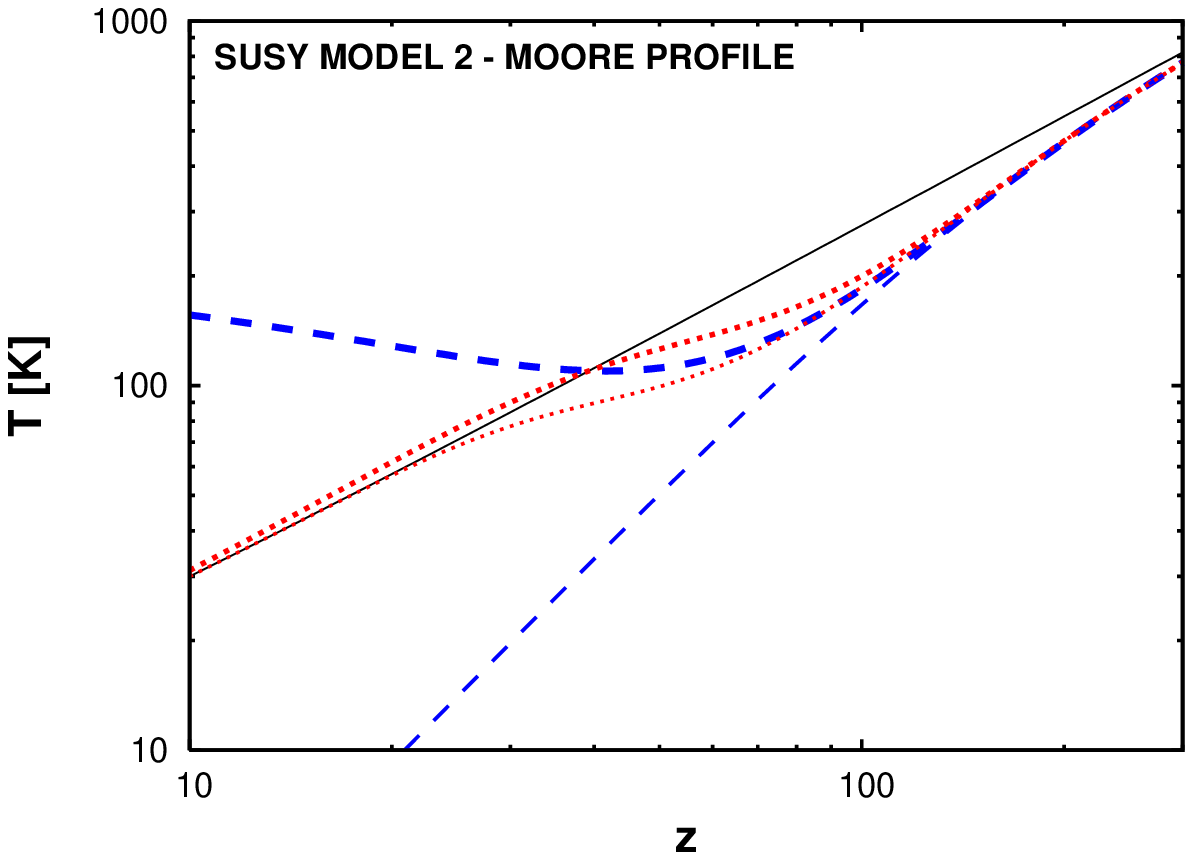}
	\includegraphics[width=0.45\linewidth,keepaspectratio,clip]{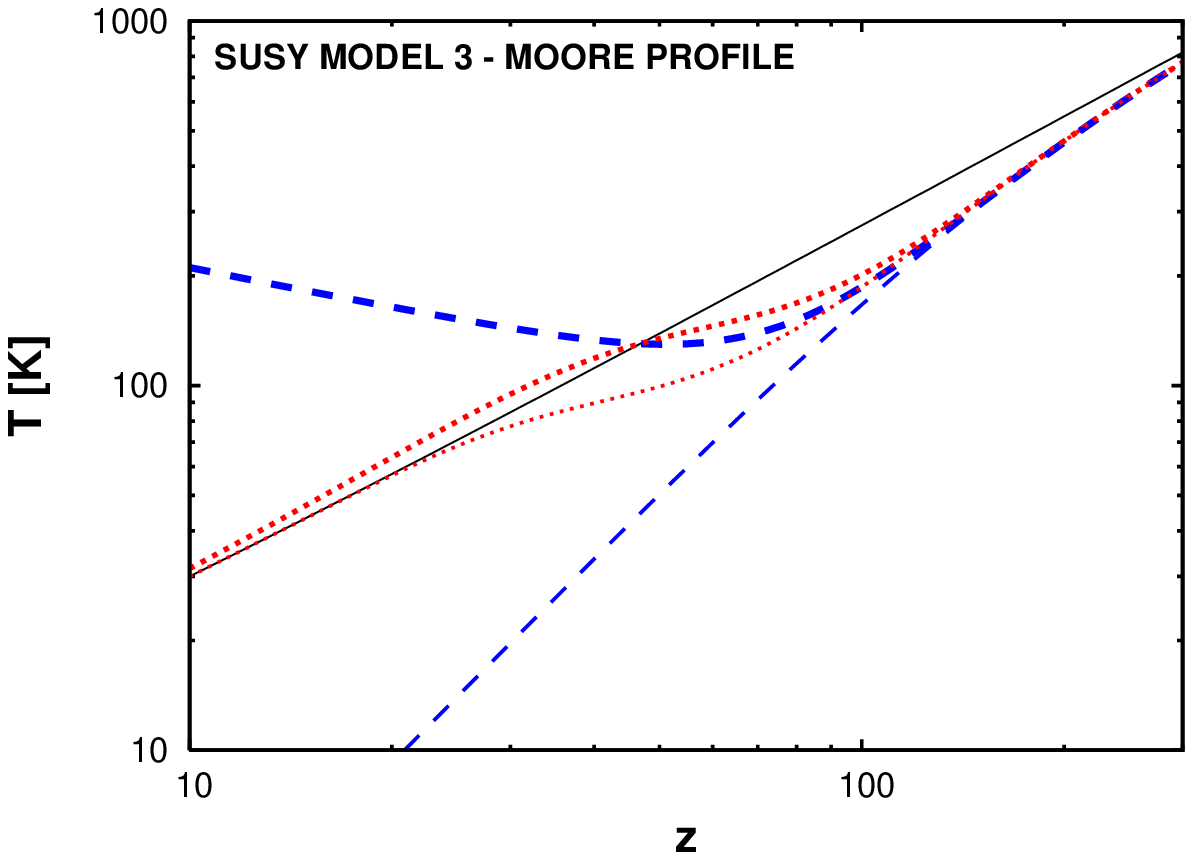}
	\includegraphics[width=0.45\linewidth,keepaspectratio,clip]{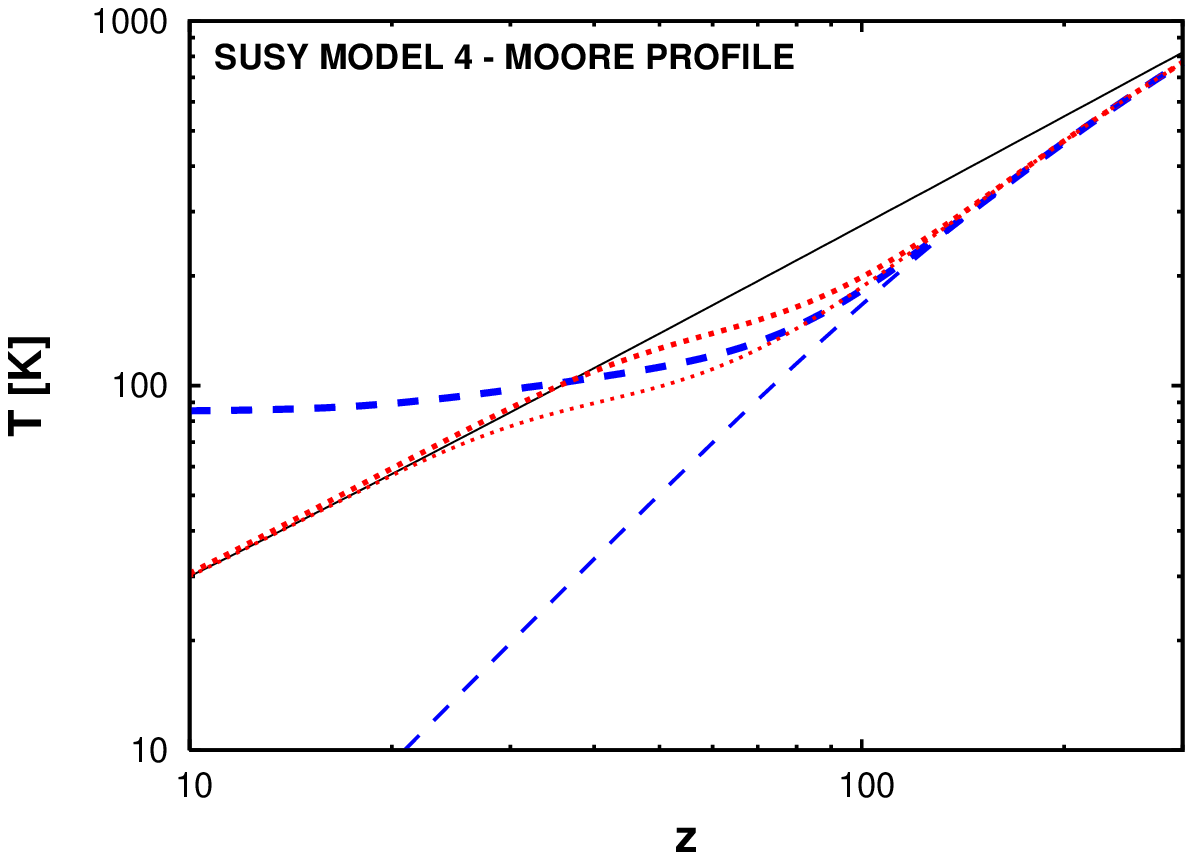}
	\caption{Evolution of the IGM kinetic (dashed blue curves) and spin (dotted red curves) temperatures for our four supersymmetric models. In each plot we show the results for the most optmistic clumping factors using the Moore density profile (thick curves) and, for comparison, the kinetic and spin temperatures in the absence of DM annihilations (thin curves).  The CMB temperature is also shown (black solid curve). The annihilation cross section $\langle\sigma_{\rm ann.}\upsilon\rangle=2.7\times10^{-26}$\,cm$^3$\,s$^{-1}$ in all plots. The clumping factors used are M18, M8, M7 and M4 for models 1, 2, 3, and 4 respectively.}
	\label{fig:T-plot_Moore}
         \end{center}
\end{figure*}

\noindent We show the results for the supersymmetric models described in \S\,\ref{sec:DM_ann} in Figs.\,\ref{fig:T-plot_NFW} and \ref{fig:T-plot_Moore} for halos with
NFW and Moore density profiles respectively. In particular, in each panel we show the evolution of $T_K$ and $T_S$ for the most optimistic clumping factors consistent with our selection criteria, as described above. For comparison, we also display the corresponding results for the ``no DM'' scenario, i.e., in the absence of annihilating DM.

We start by considering model 1, i.e., 
50\,GeV neutralinos that annihilate to $b{\bar b}$ pairs $96\%$ of the time and to $\tau^+\tau^-$ otherwise, with a canonical annihilation cross section of $\langle\sigma_{\rm ann.}\upsilon\rangle=3\times10^{-27} \mathrm{cm}^3\mathrm{\,s}^{-1}/\Omega_{\rm DM,0}h^2\simeq2.7\times10^{-26}$cm$^3$\,s$^{-1}$. 
Owing to the relatively small mass of this neutralino, the associated energy injection rate into the IGM per annihilation is large (since overall, ${\dot \epsilon_{\rm DM}}$ scales as $m_{\rm DM}^{-1}$). Consequently, the majority of the clumping factors calculated using the Moore profile are excluded based on our criterion involving the diffuse radiation background. The most optimistic clumping factors consistent with our selection criteria are N4 and M18, for the NFW and Moore profiles resepectively.
It is known that when the enhancement inside structures is neglected (i.e. when $C=1$), the energy injection from neutralino annihilation is insufficient to significantly alter the evolution of the IGM kinetic temperature. It can then be expected that significant heating of the IGM by annihilations can only start once the clumping factor is deviates significantly from unity. This corresponds to $z\simeq25$ for M18 and $z\simeq85$ for N4, corresponding to the time when the least massive DM structures start to form in these scenarios. The function $\fabs$ for model 1 neutralinos is almost constant during the period $10<z<90$, therefore we expect the evolution of the clumping factor to almost completely determine the evolution of the deviations from the ``no DM'' scenario. This is illustrated somewhat by the rapid elevation in $T_K$ corresponding to the rapid increase in the M18 clumping factor at $z\simeq20$, compared to that associated with N4, which maintains a more uniform increase in $\log(T_K)$, reflecting the almost constant value of $d\log(C)/d\log(1+z)$ at this time.

Next, we consider DM composed of neutralinos described by model 2, that is, 150\,GeV gaugino-dominated neutralinos that annihilate $96\%$ to $b{\bar b}$ and $4\%$ to $\tau^+\tau^-$. 
As for model 1 neutralinos, the majority of the clumping factors calculated using the Moore profile are excluded owing to the overproduction of the diffuse radiation background with M8 being the most optimistic clumping factor to survive this constraint. For structures with NFW profiles, nearly all clumping factors are permitted, with N2 being the most optimistic. 
The clumping factors M8 and N2 are very similar in their evolution, both significantly exceeding unity at approximately $z=85$, although M8 always exceeds N2 for the times $100\gtrsim z \gtrsim2$. Hence, we expect the evolution of $T_K$ and $T_S$ in both cases to be similar, with larger deviations from the ``no DM'' scenario expected for the M8 scenario. 
Unlike model 1, $f_{\rm abs.}$ for model 2 neutralinos decreases by nearly a factor of 3 during this period. The decrease in $f_{\rm abs.}$ largely mitigates the increasing heating rate resulting from the formation of DM structures, leading to the rather flatter evolution of $T_K$ that is observed. This can be compared with the earlier results for model 1 neutralinos, which display constantly increasing $T_K$'s (for $C\gg1)$, owing to the constant value of $f_{\rm abs.}$ during such times.

Next we consider DM composed of neutralinos described by model 3, that is, 150\,GeV higgsino-dominated neutralinos that annihilate $58\%$ to $W^+W^-$ and $42\%$ to $ZZ$. 
Despite the fact that the gaugino fractions of the neutralinos described by models 2 and 3 are significantly different, their spectra of injected electrons and photons, and thus the absorbed fraction $f_abs$ are quite similar.
Hence we expect that the permitted clumping factors will also be quite similar, and in fact, this is the case: the most optmistic clumping factors are N2 and M7 for the NFW and Moore profiles, respectively. This results in the evolution of $T_K$ to be very similar to that predicted for model 2, and all the considerations made above apply.

Finally, we consider DM composed of neutralinos described by model 4, that is, 600\,GeV gaugino-dominated neutralinos that annihilate $87\%$ to $b{\bar b}$ and $13\%$ to $\tau^+\tau^-$. 
The relatively low energy injection rate per annihilation (arising from the large neutralino mass) allows for correspondingly larger clumping factors that satisfy our diffuse background constraints. In fact, the most optimistic clumping factors that are allowed are N2 and M4.
Both M4 and N2 have similar patterns of evolution owing to the similar values of the parameters associated with each model (see Table\,\ref{tab:SUSY_Moore}). However, despite this, M4 is always much larger than N2 (as can be expected when comparing clumping factors for structures possessing Moore and NFW profiles with similar structural parameters), with a maximum difference of approximately one order of magnitude at $z\simeq30$. Also, M4 increases slightly more quickly than N2 for times $20<z<90$, explaining the correspondingly larger increase in $T_K$ associated with the M4 model despite the larger substructure mass fraction associated with N2. This explains why the displayed M4 result for $T_K$ increases so rapidly compared to that for N2 during these times.
The associated $\fabs$ function for model 4 decreases significantly over the period $100>z>10$ (approximately 0.05 to 0.006), owing to the significantly larger energies of annihilation products produced compared to the lighter neutralinos of models 1, 2 and 3. Further, unlike the other three models, $f_{\rm abs.}$ here has no minimum within the times of interest. This, in addition to the steeply decreasing nature of $\fabs$ explains the distinct lack of a rise in $T_K$ at later times, unlike that observed in models 1, 2 and 3. However, the similar values in $d\log(C)/d\log(1+z)$ for $z<20$ for both M4 and N2 result in similar rates of decrease in $\log(T_K)$ at such times.


\subsection{Light dark matter}
\noindent In this section we consider the effects on the kinetic and spin temperature of the IGM caused by LDM particles that annihilate entirely to monochromatic $e^+e^-$ pairs.

\begin{figure*}[!ht]
	\begin{center}
	\includegraphics[width=0.45\linewidth,keepaspectratio,clip]{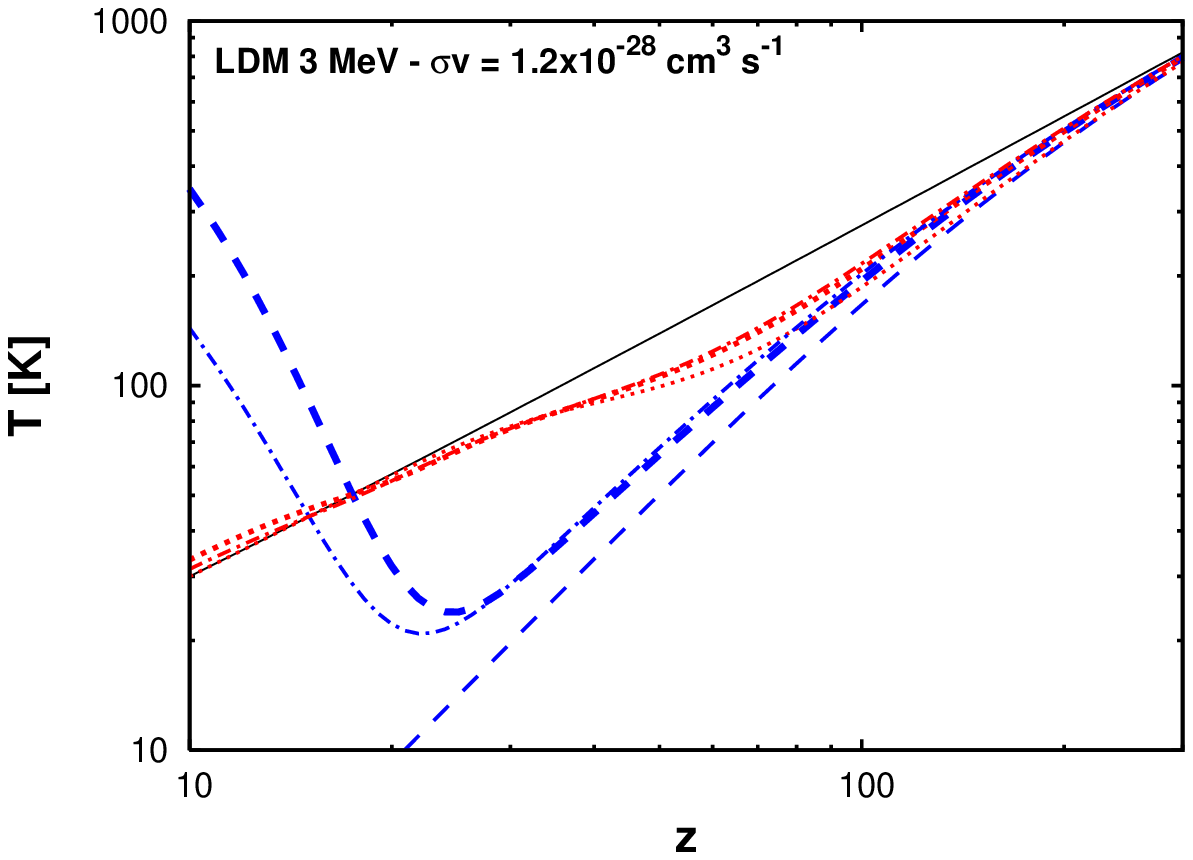}
	\includegraphics[width=0.45\linewidth,keepaspectratio,clip]{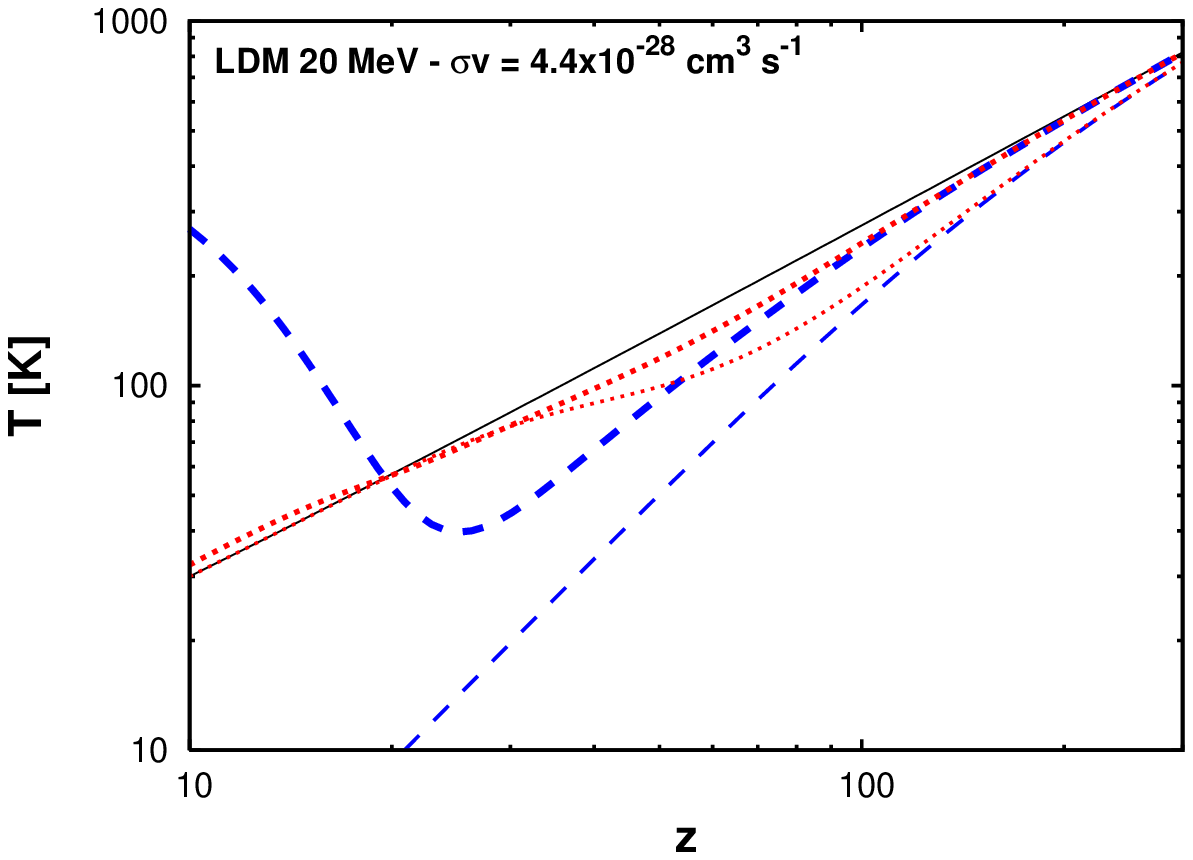}
	\includegraphics[width=0.45\linewidth,keepaspectratio,clip]{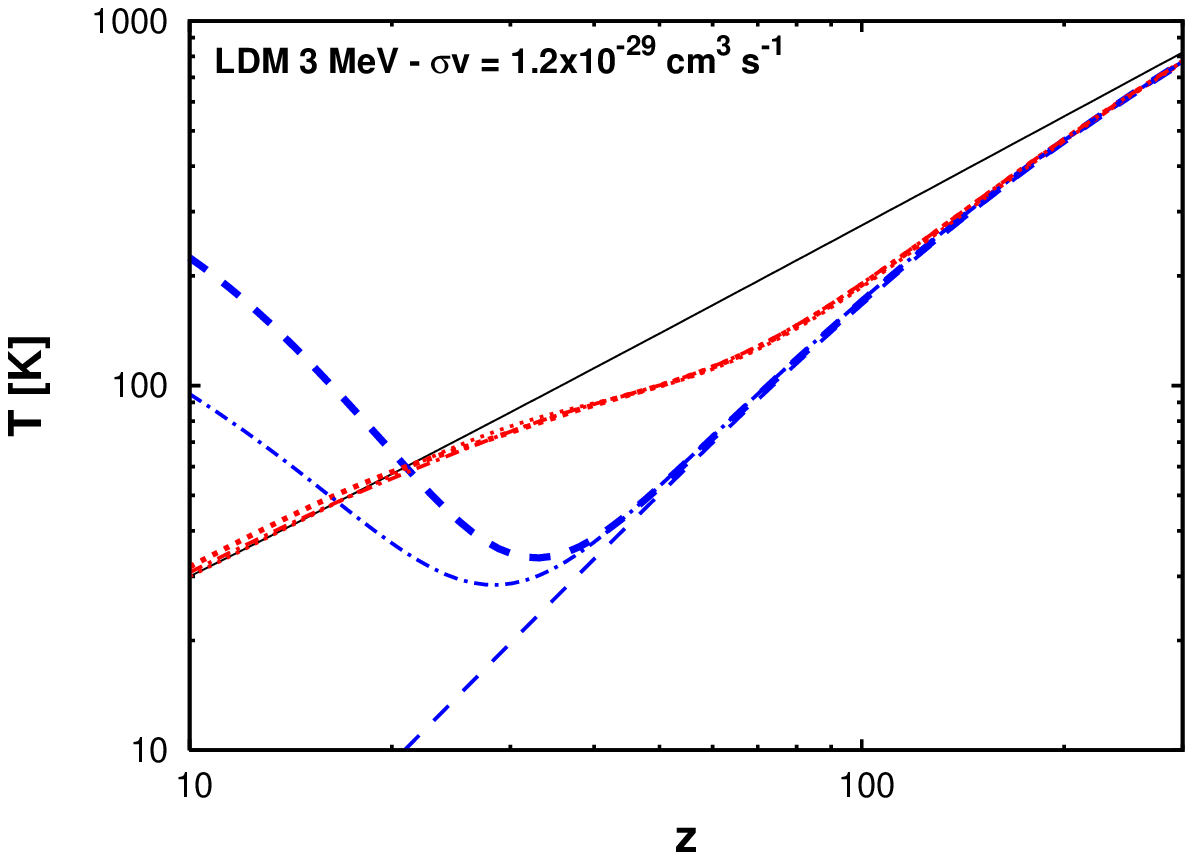}
	\includegraphics[width=0.45\linewidth,keepaspectratio,clip]{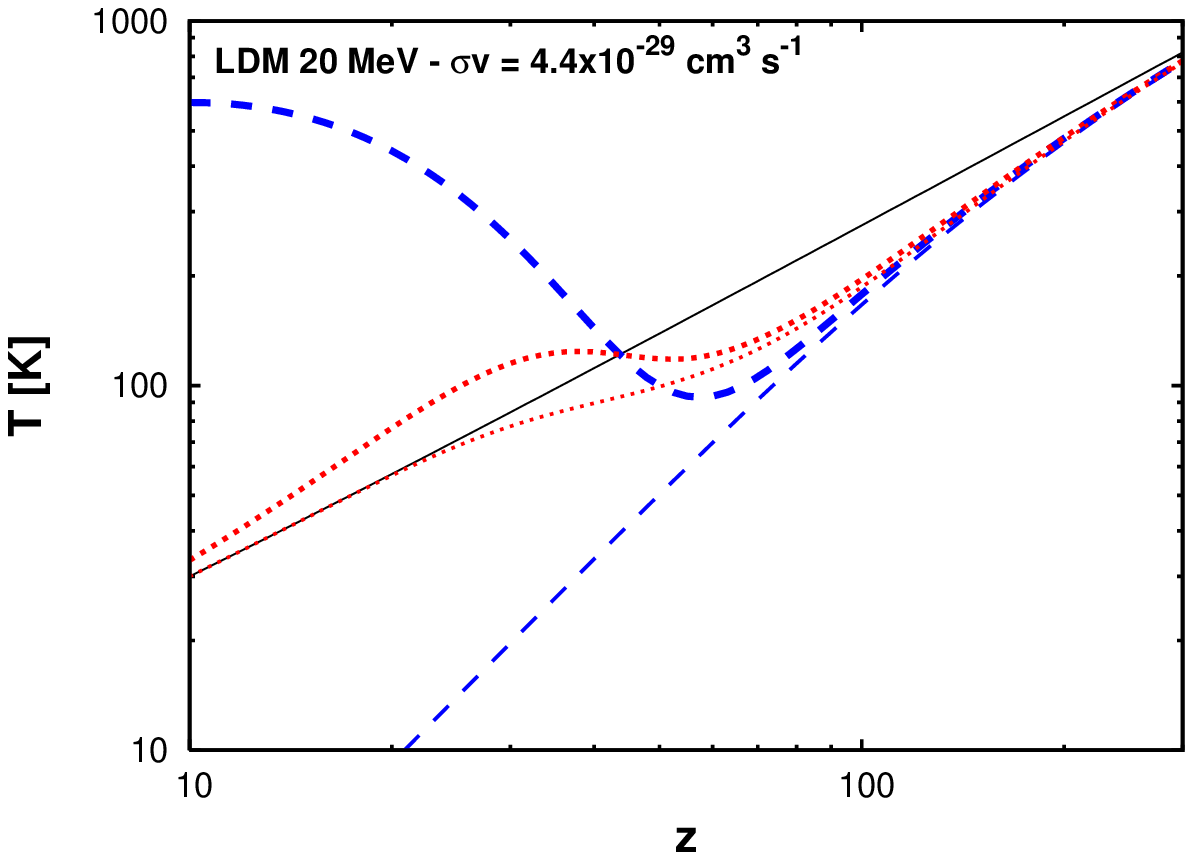}
	\caption{Evolution of the IGM kinetic (dashed blue curves) and spin (dotted red curves) temperatures for annihilating LDM particles with a mass of 3\,MeV (left panels) and 20\,MeV (right panels). In each plot we show the results for the most optmistic clumping factors using the Burkert density profile (thick curves) and, for comparison, the kinetic and spin temperatures in the absence of DM annihilations (thin curves). The CMB temperature is also shown (black solid curve). In the case of 3\,MeV LDM, we also show the results for $T_K$ and $T_S$, obtained using the fitting formula for $\fabs$ of \cite{Ripamonti:2006gq} (dot-dashed thin red curves). The annihilation cross section used is given by $\langle\sigma_{\rm ann.}\upsilon\rangle=1.2\times10^{-28}$\,cm$^3$\,s$^{-1}$ and $\langle\sigma_{\rm ann.}\upsilon\rangle=4.4\times10^{-28}$\,cm$^3$\,s$^{-1}$ in the upper left and right plots respectively, and a factor of 10 smaller in the corresponding lower plots. The clumping factors used are B19 in the two upper panels, B15 in the lower left panel, and B3 in the lower right panel.}
	\label{fig:T-plot_LDM}
         \end{center}
\end{figure*}

In the following, we calculate results utilising values of the LDM annihilation cross section $\langle\sigma_{\rm ann.}\upsilon\rangle$
based on constraints derived from the predicted effects of LDM annihilations on the CMB presented in \cite{Zhang:2006, Ripamonti:2006gq} as
\begin{equation}
\langle\sigma_{\rm ann.}\upsilon\rangle\le2.2\times10^{-29}\,{\rm cm}^3\,{\rm s}^{-1}\,f_{\rm abs.}^{-1}\left(\frac{m_{\rm LDM}}{1\,{\rm MeV}}\right).
\label{eq:LDMconstraint}
\end{equation}
We follow the conservative treatment in \cite{Ripamonti:2006gq} and substitute a value of $f_{\rm abs.}$ approximately equal to its maximum value, $f_{\rm abs.}^{\rm max.}$, into Eq.(\ref{eq:LDMconstraint}), in order to determine our conservative estimate for $\langle\sigma_{\rm ann.}\upsilon\rangle$. For comparison, we also calculate results for a value of $\langle\sigma_{\rm ann.}\upsilon\rangle$ one order of magnitude smaller than this limiting value. We show our results for $T_K$ and $T_S$ in Fig.\,\ref{fig:T-plot_LDM}.

Firstly we consider 3\,MeV LDM particles. 
In the left panels of Fig.\,\ref{fig:T-plot_LDM} we display the effects of DM composed of these particles on the evolution of $T_K$ and $T_S$, when using annihilation cross sections $\langle\sigma_{\rm ann.}\upsilon\rangle$ equal to $1.2\times10^{-28}\,$cm$^3$\,s$^{-1}$ (upper left panel) and $1.2\times10^{-29}\,$cm$^3$\,s$^{-1}$ (lower left panel). We display results calculated using the B19 (upper left panel) and B15 (lower left panel) clumping factors, calculated for structures possessing Burkert density profiles, deduced to be the most optimistic models consistent with our constraints involving the diffuse radiation background when using values of $\langle\sigma\upsilon\rangle$ equal to $1.2\times10^{-28}\,$cm$^3$s$^{-1}$ and $1.2\times10^{-29}\,$cm$^3$s$^{-1}$ respectively. As usual, we compare these results for $T_K$ and $T_S$ to the ``no DM'' scenario. In the 3\,MeV case, we also compare our results to those obtained using the formula for $\fabs$ provided by \citet{Ripamonti:2006gq} (see Appendix \ref{sec:app-fabs} for details).

The clumping factors B19 and B15 possess significant differences in their evolution owing to the different values of the minimum halo mass associated with them ($10^6\,M_{\odot}$ for B19 and $46\,M_{\odot}$ for B15). 
This results in the time at which B19 starts to significantly exceed unity occurring much more recently ($z\simeq20$) than for B15 ($z\simeq35$). 
As can be observed from Fig.\,\ref{fig:T-plot_LDM}, these times closely correspond with the respective minima in $T_K$ immediately before its rapid increase. 
However, unlike neutralinos, LDM can still significantly heat the IGM at times when $C\sim1$, provided that $\langle\sigma_{\rm ann.}\upsilon\rangle$ is large enough. 
This can again be observed in Fig.\,\ref{fig:T-plot_LDM}, where in the upper left panel significant deviations in $T_K$ relative to the ``no DM'' scenario occur for $z>20$, whereas such deviations are negligible for $\langle\sigma_{\rm ann.}\upsilon\rangle=1.2\times10^{-29}\,$cm$^3$\,s$^{-1}$ at times $z>35$, as can be seen from the lower left panel.
Further, at recent times, B19 increases much more rapidly than B15, resulting in the correspondingly larger value of $d\log(T_K)/d\log(z)$ that is observed.

We also show the effect of using the function $\fabs$ as calculated in Ref. \cite{Ripamonti:2006gq}. At the respective times when the clumping factors B19 and B15 are much greater than unity, both the absorbed fraction calculated in this paper and that of  Ref. \cite{Ripamonti:2006gq} are monotonically increasing, with the former roughly twice larger than the latter (see upper panel of Fig.\,\ref{fig:fabs_ldm}). This is illustrated in the left panels of Fig.\,\ref{fig:T-plot_LDM} by the larger values $T_K$ associated with our $\fabs$ relative to those associated with the $\fabs$ of Ref. \cite{Ripamonti:2006gq}.



Next, we consider 20\,MeV LDM particles. 
In the right panels of Fig.\,\ref{fig:T-plot_LDM} we display the effects of DM composed of these particles on the evolution of $T_K$ and $T_S$  when using annihilation cross sections $\langle\sigma_{\rm ann.}\upsilon\rangle$ equal to $4.4\times10^{-28}\,$cm$^3$\,s$^{-1}$ (upper right panel) and $4.4\times10^{-29}\,$cm$^3$\,s$^{-1}$ (lower right panel). 
We display results calculated using the B19 and B3 clumping factors, calculated for structures possessing Burkert density profiles, deduced to be the most optimistic models consistent with our constraints involving the diffuse radiation background when using values of $\langle\sigma_{\rm ann.}\upsilon\rangle$ equal to $4.4\times10^{-28}\,$cm$^3$\,s$^{-1}$ and $4.4\times10^{-29}\,$cm$^3$\,s$^{-1}$ respectively. In all cases, we utilise the function $f_{\rm abs.}$
calculated according to the procedure described in Appendix \ref{sec:app-fabs}.
Once again, we compare these results for $T_K$ and $T_S$ to those when DM is absent.
The function $f_{\rm abs.}$ starts to deviate from unity at $z\sim1000$, but the decrease is quite slow until $z\sim30-50$; then it becomes more rapid until it reaches 0.2 at $z\simeq10$ (see lower panel of Fig.\,\ref{fig:fabs_ldm}). This accounts for the slight decrease in d$T_K$/d$z$ observed during the period $z<30$.
Hence, at times $z>30$, the evolution of the LDM heating rate is dominated by that of the clumping factor B19 (upper right panel) or B3 (lower right panel). 
There are significant differences in the evolution of these two clumping factors.
In particular, the respective minima in $T_K$ closely correspond to the times at which the clumping factor $C$ starts to become much greater than unity. 
However, there appears to be significant heating by LDM at times when $C\sim1$, indicated in the right panels of Fig.\,\ref{fig:T-plot_LDM} by the non-negligible deviations from the ``no DM'' model, especially when using $\langle\sigma_{\rm ann.}\upsilon\rangle=4.4\times10^{-28}\,$cm$^3$\,s$^{-1}$ (upper right panel) at least up to $z=300$, and up to $z\simeq150$ when using $\langle\sigma\upsilon\rangle=4.4\times10^{-29}\,$cm$^3$s\,$^{-1}$ (lower right panel).


\subsection{The 21\,cm global signature}
\noindent Using the above results for the evolution of the spin temperature, $T_S$, in this section we present corresponding results for the differential brightness temperature $\delta T_b$, calculated using Eq.(\ref{eq:dT_b}), that is most readily associated with measurements of the 21\,cm background.


\subsubsection{Neutralino dark matter}

  \begin{figure}
	\begin{center}
	\includegraphics[width=\linewidth,keepaspectratio,clip]{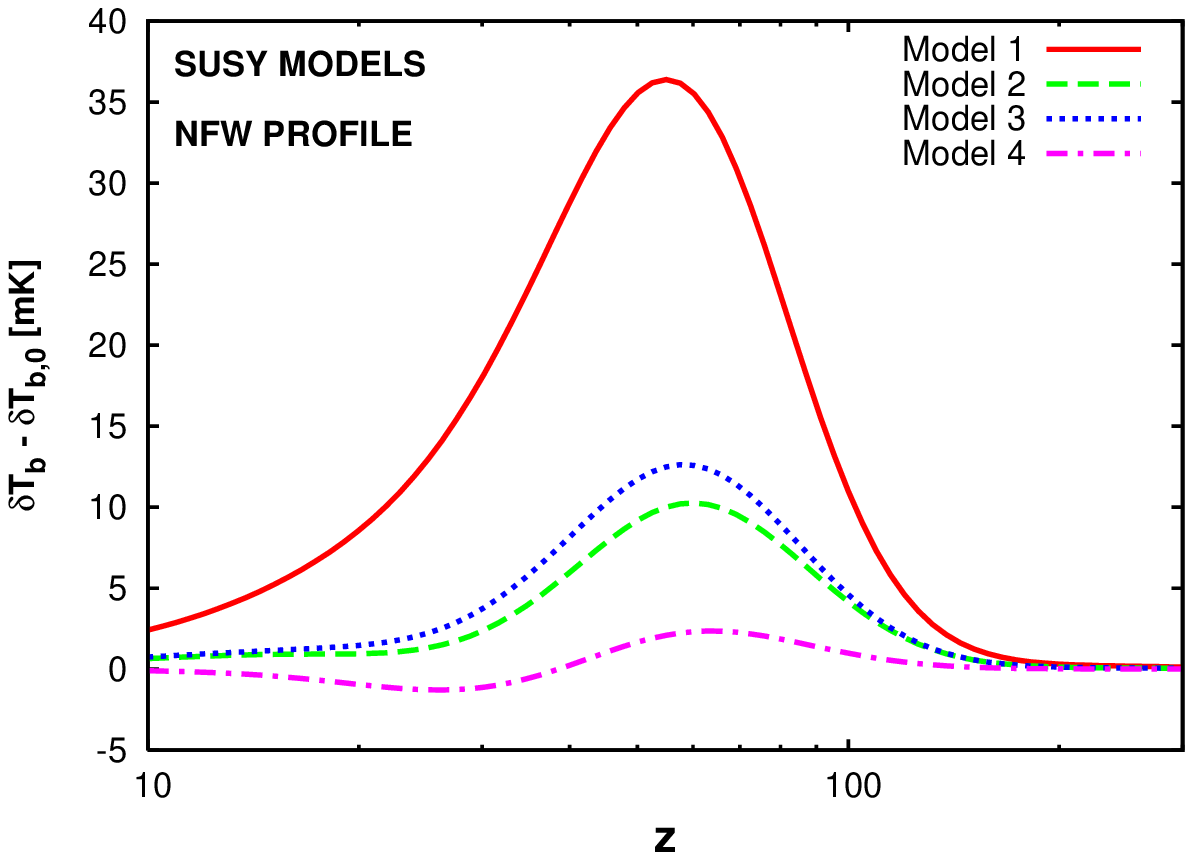}
	\includegraphics[width=\linewidth,keepaspectratio,clip]{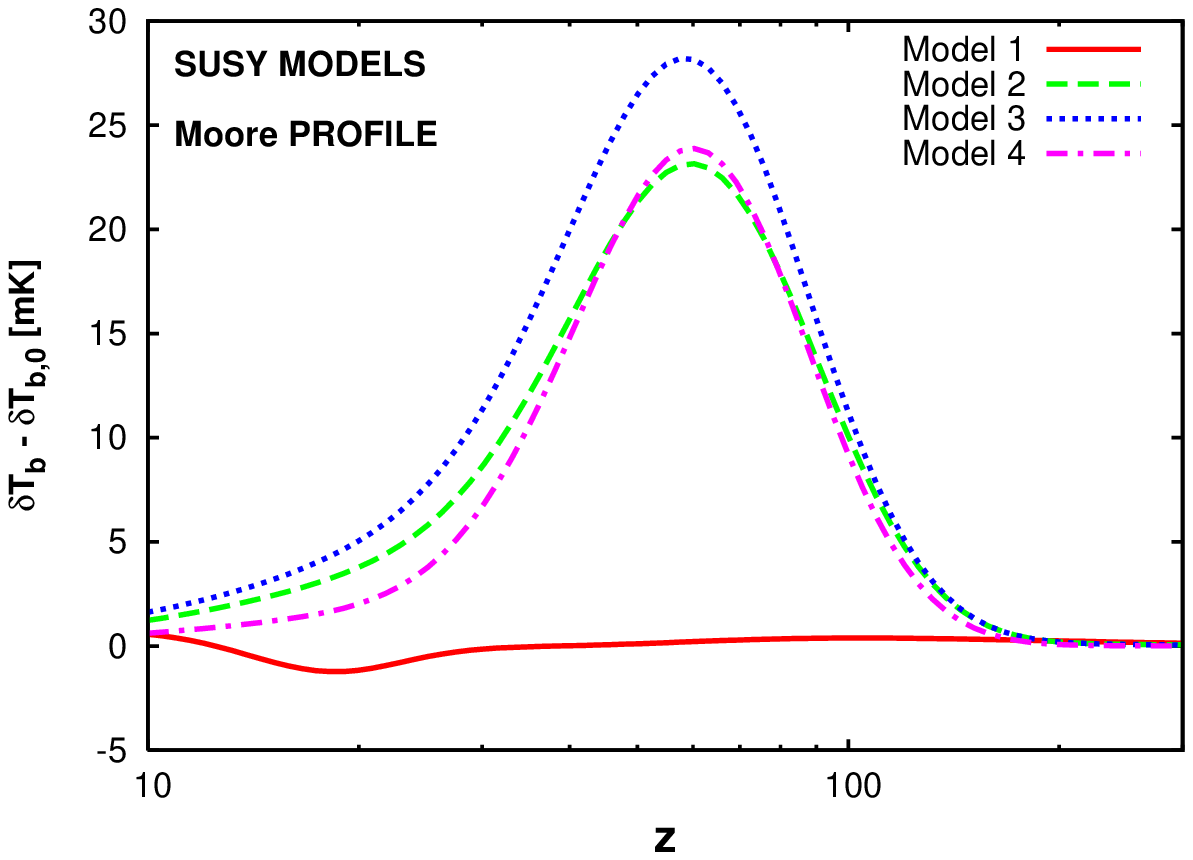}
	\caption{
	Evolution of the 21\,cm differential brightness temperature $\delta T_b$, relative to $\delta T_{b,0}$ calculated for the ``no DM'' model, for the
	four supersymmetric models considered in the text, in the case of NFW (upper panel) and Moore (lower panel) profiles.
	For each model and density profile we display results using the most optmistic clumping factors compatible with our selection criteria based on the reionization redshift 	and on the gamma-ray background (top panel:  N4 for model 1, N2 for models 2, 3, 4; bottom panel: M18, M8, M7, M4 for models 1, 2, 3 and 4 respectively). 
	The annihilation cross section $\langle\sigma\upsilon\rangle=2.7\times10^{-26}\,$cm$^3$s$^{-1}$ for all models.
	}
	\label{fig:dTb_susy}
         \end{center}
\end{figure}


\noindent As in \S\,\ref{sec:Results}, we firstly consider neutralino DM.
In Fig.\,\ref{fig:dTb_susy} we display our predictions for the evolution of $\delta T_b$ in the presence of neutralino DM, relative to that calculated for the ``no DM'' scenario, $\delta T_{b,0}$, for our four benchmark SUSY models.  In each case, we utilise an annihilation cross section $\langle\sigma\upsilon\rangle=2.7\times10^{-26}\,$cm$^3$\,s$^{-1}$. 
For each DM model we display results when using the most optimistic clumping factors associated with Moore (lower panel) and NFW (upper panel) density profiles. For the Moore profiles, these are the clumping factors M18, M8, M7, M4 for models 1, 2, 3 and 4 respectively; for the NFW profile, they are the N4 clumping factors for model 1, and N2 for models 2, 3 and 4. For comparison, we have also calculated the differential brightness temperature for the least optimistic clumping factors, and also in the absence of structures (i.e. $C(z)=1$). We do not show the results, but we have found that in both cases the deviations from the ``no DM'' behaviour are smaller than 1\,mK at all redshifts greater than 10.

We observe that generally the evolution of $\delta T_b$ using the most optimistic clumping factors presents some common features among the four models.
In most cases we find a peak in the 21 cm emission at $z\simeq60$ of up to $\sim40\,$mK.  These features emphasise the additional heating by DM at times prior to the formation of baryonic structures, when the Universe cools adiabatically, when there is a characteristic absorption feature in the ``no DM'' scenario arising from the efficient coupling (via collisions) between $T_K$ and $T_S$ at these times (see, e.g., Ref. \cite{valdes}). In the case of the NFW profile, we have that the clumping factors used for the four models are very similar (in fact, models 2, 3 and 4 use the same clumping factor N4, while model 1 uses N2), so that we expect the differences among the models to be mainly driven by the difference in the injected energy. In particular, lower-mass neutralino yield a larger signal, since overall the energy produced by annihilations scales as $m_\mathrm{DM}^{-1}$. Also, in the case of lighter neutralinos, a larger part of the energy produced is effectively absorbed by the IGM (see Figs. \ref{fig:fabs_susy_A} and \ref{fig:fabs_susy_B}). For these reasons, model 1 gives a peak of $\simeq 35$~mK at $z=60$ ($\simeq 18$~mK at $z=30$), while model 4 presents only $\sim 1$ mK deviations from the ``no dark matter'' case.

In the case of the Moore profile, the clumping factors used differ more, so that we should factor this in when interpreting the results for $\delta T_b$. In the case of model 1, the large energy injection leads to the violation of the constraints on the diffuse background for nearly all clumping factors, so that has to be compensated by relatively small values of $C(z)$. The result is that there is actually an overcompensation, so that for the clumping factor considered there is no significant heating of the IGM and the brightness temperature basically has the same evolution as in the ``no dark matter'' case. Of course, our exploration of the parameter space for the clumping factor is far from complete, so that it could well be possible that there is a ``soft spot'' in parameter space (in particular a value of the minimum mass $M_\mathrm{min}$ somewhere between $10^{-6}$ and $10^4 M_\odot$) where the clumping factor is large enough to produce sizeable differences in $\delta T_b$, without at the same time violating the constraints on the reionization redshift and on the diffuse gamma backgrund. On the other hand, the clumping factors used for models 2 and 3 are very similar, and in fact, considering also that the neutralino mass is the same in the two models, the results for the brightness temperature are very similar. The differences can be traced in the larger value of the absorbed energy fraction for model 3. Finally, in the case of model 4, the smaller energy injection with respect to models 2 and 3 is nearly completely offset by the use of a larger clumping factor. In general, in models 2, 3 and 4, we find a peak of $\simeq 25$~mK at $z=60$ ($\simeq 5--10$ mK at $z=30$) in the deviation of the differential brightness temperature with respect to the ``no dark matter'' case.


\subsubsection{Light dark matter}
\begin{figure}
	\begin{center}
	\includegraphics[width=\linewidth,keepaspectratio,clip]{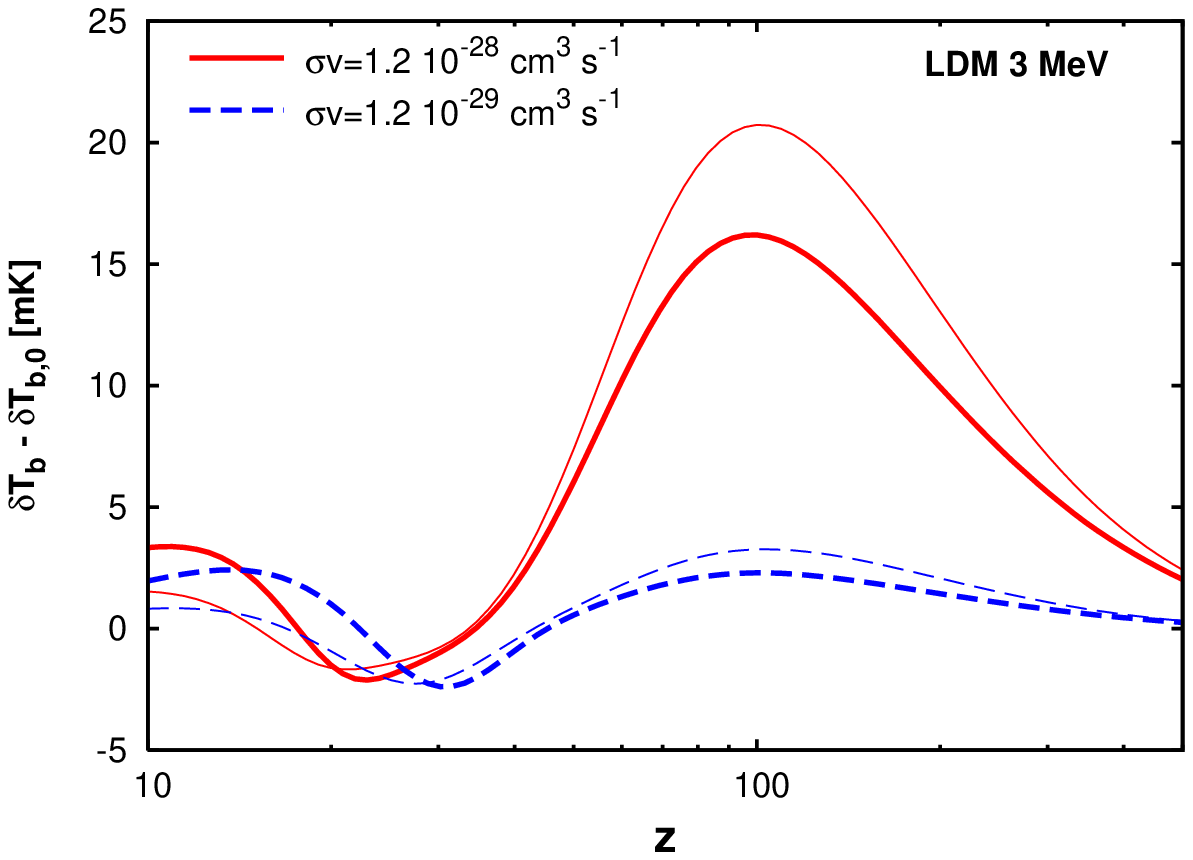}
	\includegraphics[width=\linewidth,keepaspectratio,clip]{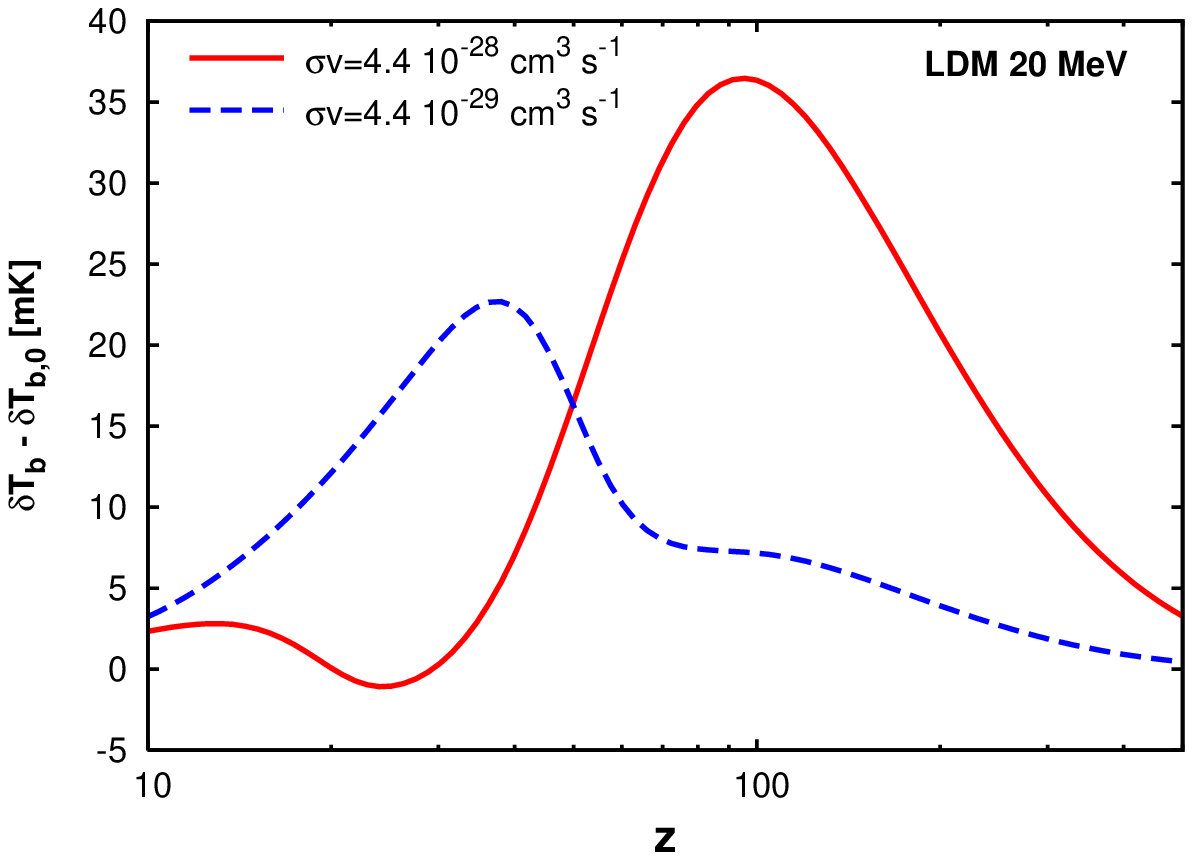}
	\caption{
	Evolution of the 21\,cm differential brightness temperature $\delta T_b$, relative to $\delta T_{b,0}$ calculated for the ``no DM'' model, for LDM
	particles with a mass of 3 MeV (upper panel) and 20 MeV (lower panel), for structures with Burkert profiles. For each value of the mass, we consider two values of the 	annihilation cross section. In the case of 3 MeV LDM, we also show the results obtained using the fitting formula for $\fabs$ of \cite{Ripamonti:2006gq} (thin curves). For each model we display results using the most optmistic clumping factors compatible with our selection criteria based on the reionization 	redshift and on the gamma-ray background [top panel:  B19 for $\langle\sigma\upsilon\rangle=1.2\times10^{-28}\cm^3\mathrm{s}^{-1}$, B15 for $\langle\sigma\upsilon\rangle=1.2\times10^{-29}\cm^3\mathrm{s}^{-1}$; bottom panel:  B19 for $\langle\sigma\upsilon\rangle=4.4\times10^{-28}\cm^3\mathrm{s}^{-1}$, B3 for $\langle\sigma\upsilon\rangle=1.2\times10^{-28}\cm^3\mathrm{s}^{-1}$]. 
	}	
	\label{fig:dTb_LDM}
         \end{center}
\end{figure}


\noindent We now consider light dark matter. 
In the top panel of Fig.\,\ref{fig:dTb_LDM} we display predictions for the evolution of $\delta T_b-\delta T_{b,0}$ in the presence of DM solely composed of 3\,MeV LDM particles. We use values of the annihilation cross section $\langle\sigma_{\rm ann.}\upsilon\rangle$ equal to $1.2\times10^{-28}\,$cm$^3$\,s$^{-1}$ (red solid curves) and $1.2\times10^{-29}\,$cm$^3$\,s$^{-1}$ (blue dashed curves). For comparison, we also show the corresponding results calculated using the function $f_{\rm abs.}$ derived in \cite{Ripamonti:2006gq} (thin curves). 
For each value of $\langle\sigma_{\rm ann.}\upsilon\rangle$ we utilise the clumping factors associated with Burkert density profile which were determined to be the most optimistic, namely, B19 for the larger cross section and B15 for the smaller.

For both values of $\langle\sigma_{\rm ann.}\upsilon\rangle$, like neutralinos, there is a characteristic peak in $\delta T_b-\delta T_{b,0}$  which, as expected, is more distinct for the larger value of $\langle\sigma_{\rm ann.}\upsilon\rangle$ because of the larger deviations in the heating rate of the IGM relative to the ``no DM'' scenario.
However, for LDM this peak occurs, in most cases, quite earlier ($z\simeq100$) than for neutralino DM (with maxima at $z\simeq60$).


Finally, in the bottom panel of Fig.\,\ref{fig:dTb_LDM} we display predictions for the evolution of $\delta T_b-\delta T_{b,0}$ in the presence of DM solely composed of 20\,MeV LDM particles. 
Our results are calculated using values of the annihilation cross section $\langle\sigma_{\rm ann.}\upsilon\rangle$ equal to $4.4\times10^{-28}\,$cm$^3$\,s$^{-1}$ (red solid curve) and $4.4\times10^{-29}\,$cm$^3$\,s$^{-1}$ (blue dashed curve).
As for 3\,MeV LDM particles, for each of these values of $\langle\sigma_{\rm ann.}\upsilon\rangle$ we utilise two most optimistic clumping factors associated with the Burkert density profile, which in this case were B19 for the larger cross section and B3 for the smaller.

For $\langle\sigma_{\rm ann.}\upsilon\rangle=4.4\times10^{-29}\,$cm$^3$\,s$^{-1}$, the most optimistic clumping factor, B3, results in a rate of IGM heating that yields a peak in $\delta T_b-\delta T_{b,0}$ of $\simeq 25$~mK at $z\simeq40$. 
This results in a deviation at $z\simeq 30$  of roughly 20\,mK.
For $\langle\sigma_{\rm ann.}\upsilon\rangle=4.4\times10^{-28}\,$cm$^3$\,s$^{-1}$,
the heating rate is sufficient to produce a significantly larger deviation of $\simeq40\,$mK at $z\simeq 100$; however, the deviation is $< 1$ mK at $z\simeq 30$.


\section{Discussion}
\label{sec:Discussion}

\noindent We have calculated predictions for the effects on the evolution of the cosmological HI 21\,cm signal during the Dark Ages for various forms of annihilating neutralino and light dark matter. 
In doing so, we fully accounted for the significant enhancements made to the annihilation rate of these DM particles arising from DM structures. 
We utilised results from the state-the-art N-body simulations to calculate the evolution of the aforementioned enhancement in the annihilation rate, referred to here as the ``clumping factor'', owing to the distribution of halos and the near self-similar distribution of increasingly smaller substructures predicted to exist within them. We did this for a diverse range of values of astrophysical parameters consistent with the uncertainties in the dynamics of the simulated halos. 
We performed detailed calculations of the absorbed fraction of the energy injected into the IGM by the annihilation products of our DM candidates.
We used the standard equations for the evolution of the kinetic and spin temperatures of the IGM with modifications to account for the additional energy injected into by the IGM from DM annihilations. 
Finally, we calculated the resulting deviations in the evolution of the differential brightness temperature $\delta T_b$ relative to a scenario where DM is absent. 

In our calculations of $\delta T_b$ we have neglected the influence of astrophysical processes affecting the 21~cm background. In fact, these effects dominate the 21~cm signal, thus obscuring the 
cosmological information, once star formation becomes important at redshifts $z \simeq 25$ \cite{Pritchard:2008da}. This means that the effect of the presence of annihilating DM at $z\lesssim 25$
(corresponding to frequencies $\nu \gtrsim 55$~Hz) will 
likely be undetectable due to the uncertainity in the astrophysical modelization.

The global 21\,cm signal is the target of several experiments, like the ``Cosmological Reionization Experiment'' (CORE) \cite{CORE} and the ``Experiment for Detecting the Global EOR Signature'' (EDGES) \cite{edges1, edges2}. Both experiments roughly operate in the frequency range from $\sim 100$ to $\sim 200$~MHz, corresponding to the  redshift range $7 \lesssim z \lesssim 14$. In fact, the single antenna experiment EDGES has already released preliminary results \cite{edges1, edges2}. This experiment attempts to separate the redshifted HI 21\,cm signal from the contribution of the Galactic and Extragalactic foregrounds at the same frequencies by taking advantage of the fact that these foregrounds are anticipated to be smooth power-law spectra. Conversely, the cosmological signal, is expected to have up to three rapid transitions in brightness temperature corresponding to the cooling and heating of the IGM, and most recently from reionisation. However, the rapidly varying (with respect to frequency) systematic and instrumental contributions to the measured power spectrum can easily be mistaken for a cosmological signal. Currently, the r.m.s. level of systematic contributions is approximately $T\sim75$\,mK, but further reductions in these systematics are anticipated in the near future \cite{JuddPC}. Unfortunately, both CORE and EDGES operate in a frequency range where, as explained above, the evolution of the ionized fraction and of the kinetic temperature is dominated by astrophysics, so that presently they cannot be useful in constraining the models considered here. However, the EDGES team plans to expand the frequency coverage of the instrument down
to 50\,MHz or lower \cite{edges1}, thus opening the possibility of exploring the regime where the 21\,cm signal is dominated by the cosmological contribution. In order to exclude at least the most optimistic models considered here,
the EDGES experiment should be able to reduce the systematics at 50\,MHz to below $\sim20$\,mK, although the contribution of the Galactic synchrotron foreground increases significantly at the lower frequencies.

Another interesting way to potentially put observational constraints on the energy injection from DM annihilation would be to consider the effects on the spatial fluctuations of the brightness temperature, as encoded by their power spectrum, ${\cal P}_{T_b}$, rather than the average signal as we have done in this paper. 
This was done, for example, in \cite{furlanetto} (although the authors neglect the clumpiness of DM). 
Such fluctuations are somewhat easier to measure than the average signal since they are less contaminated by foregrounds. 

Even more interestingly, peculiar velocities give rise to an anisotropy of the 21\,cm power spectrum that can be used to separate the cosmological signal from the (uncertain) astrophysical contribution, thus allowing, at least in principle, one to detect the effects of DM annihilation in the astrophysics-dominated regime \cite{Barkana:2004zy,McQuinn:2005hk,Pritchard:2008da}. 

The 21\,cm fluctuations are themselves the target of several experimens, such as LOFAR \cite{lofar}, MWA \cite{mwa}, PAPER \cite{paper}, 21CMA \cite{21CMA} and SKA \cite{ska}. In particular, the impending LOFAR epoch of reionization experiment is designed to observe radio fluctuations at frequencies 115-215 MHz, corresponding to the redshifted 21\,cm signal in the range $6<z<11.5$ \citep{LOFAR}. Unfortunately, the cosmological signal is contaminated by a plethora of astrophysical and non-astrophysical components - including, Galactic synchrotron emission from diffuse and localized sources \citep{shaver99}, Galactic free-free emission \citep{shaver99}, integrated emission from extragalactic sources, such as radio galaxies and clusters \citep{shaver99, dimatteo2002, dimatteo2004, oh2003, cooray2004}, ionospheric scintillation and instrumental response - the fluctuations in  which can significantly exceed the cosmological signal (see e.g.~\citep{lofar_foregrounds, lofar_foregrounds2, Liu2009}). 


We did not consider the possibility that the DM annihilation cross section is enhanced inside cold substructures due to non-perturbative quantum corrections \cite{Lattanzi:2008qa}. This so-called Sommerfeld enhancement could increase the annihilation cross section by orders of magnitude with respect to its early-Universe value. The effect of such an enhancement on the heating/ionisation history of the IGM has been studied in \cite{Cirelli:2009bb}, where it has been shown that it could similarly increase the IGM temperature by orders of magnitude.

An interesting extension of our analysis would be to consider other DM particles, such as ``Exciting DM'' (XDM) (see, e.g.,\,\citep{fink07}).  XDM can annihilate to produce two intermediate scalars $\phi$ that can subsequently decay to standard model particles. If $2m_e<m_\phi<2m_\mu$, the $\phi$ will mainly decay to an $e^+e^-$ pair, with an energy spectrum extending up to the mass of the XDM particle \citep{cholis08}.
Such particles are well motivated DM candidates and have been proposed to explain several other astronomical observations \cite{fink07, chen09, cholis08}. Like LDM, the direct production of boosted $e^{+}e^{-}$ pairs following self-annihilation gives XDM the potential to produce observable features in the global 21\,cm signal. In fact, recently it has been determined that particles with collisional long-lived excited states, and inspired by XDM models, may have observable effects on the CMB and the 21\,cm background signal \cite{fink08}. 

We would also like to acknowledge the recent simulations by the Virgo Consortium as part of its Aquarius project \cite{Aq1, Aq2}, conducted during the writing of the most recent version of this paper. These simulations investigate the properties of a Galaxy-sized DM halo and its substructure with unprecedented resolution. Whilst the results are largely consistent with those deduced from the Via Lactea simulations, there are significant differences. These include (i) four generations of resolved substructure (rather than the two in Via Lactea II), (ii) a distribution of substructures that is not self-similar to its host halo (rather than a fractal-like distribution observed in Via Lactea), (iii) a subhalo mass function with an index of -1.9 (rather than the -2.0 used in this study, although a brief discussion of the significance of such deviations on the clumping factor are made in \S\,\ref{subsub}). Whilst we acknowledge that these differences may change some of our conclusions regarding the detectability of the 21\,cm global signature, we do not pursue an investigation of these results here, but intend to incorporate them into a subsequent study \cite{21cm_Cumberbatch_new}.

Finally, we note that the effect of DM annihilations on the 21\,cm signal has been further studied in  
\cite{Natarajan:2009bm}, including the effects on the brightness temperature fluctuations, that we have not considered here. On the other hand, the authors do not include the effect of substructures in their calculations.


{\em Acknowledgements:}  DTC is supported by the Science and Technology Facilities Council. Part of this work was conducted while DTC was at CERCA Case Western Reserve University, funded by NASA grant 4200188792 and also supported in part by the US DoE, and also whilst at the University of Oxford, funded by the Science and Technology Facilities Council. 
Part of this work was conducted while ML was at the University of Oxford supported by the INFN. 
We would also like to thank Saleem Zaroubi, Rajat Thomas and the other members of the Kapteyn Institute, Groningen for their very helpful discussions regarding LOFAR sensitivities. 
We would also like to thank Judd Bowman, Aaron Chippendale and Ron Ekers for their helpful comments regarding the CoRE and EDGES experiments.



\appendix

\section{Analytical expressions for the halo annihilation rate} \label{app:clump}

\noindent In this appendix we report the analytical formulas for the annihilation rate within halos with NFW and Moore DM density profiles.
For the Burkert profile, it is not possible to express the relevant integrals in analytical form. 

Applying Eq.\,(\ref{eq:R}) to the case of a halo of mass $M$ with an NFW profile with concentration $c_{\rm vir.}(M,z)$, the annihlation rate $R(M,z)$ is easily calculated to be
\begin{eqnarray}
R(M, z)&=&\frac{1}{2}\langle\sigma_{\rm ann.}\upsilon\rangle\left(\frac{\rho_s}{m_{\rm DM}}\right)^2\frac{4\pi}{3}\left(\frac{r_{\rm vi.r}(z,M)}{c_{\rm vir.}(z,M)}\right)^3\nonumber\\
&\times&\left\{1-\frac{1}{\left[1+c_{\rm vir.}(M,z)\right]^3}\right\}.\nonumber\\
\label{R_NFW}
\end{eqnarray}
By equating (\ref{M}), for the virial mass, $M$, to the integral
\begin{eqnarray}
M&=&\int\limits^{r_{\rm vir.}}_{r=0}\rho(r, c_{\rm vir.}(M,z))4\pi r^2 {\rm d}r\nonumber\\
&=&4\pi\left(\frac{r_{\rm vir.}}{c_{\rm vir.}(M,z)}\right)^3\rho_s(M,z)\nonumber\\
&&\times\left[\log\left[1+c_{\rm vir.}(z,M)\right]-\left(\frac{c_{\rm vir.}(z,M)}{\left[1+c_{\rm vir.}(z,M)\right]}\right)\right],\nonumber\\
\label{M_int_NFW}
\end{eqnarray}
we obtain the relation for the scale density
\begin{eqnarray}
\rho_s(M, z)&=&\frac{M}{4\pi\left(\frac{r_{\rm vir.}}{c_{\rm vir.}(M,z)}\right)^3}\nonumber\\
&\times&\frac{1}{\left[\log\left[1+c_{\rm vir.}(z,M)\right]-\left(\frac{c_{\rm vir.}(z,M)}{\left[1+c_{\rm vir}(z,M)\right]}\right)\right]}.\nonumber\\
\label{rho_s_NFW}
\end{eqnarray}

For the Moore profile, in order for the integral over density squared to be finite we must truncate the density below a radius $r_{\min.}$ (see discussion in \S\,\ref{sec:clumpsm}). Defining the variable $x=rc_{\rm vir.}/r_{\rm vir.}$ with $x_{\rm min.}=r_{\rm min.}c_{\rm vir.}/r_{\rm vir.}$ we find that for a Moore profile

%
\begin{eqnarray}
R(M, z)&=&\frac{1}{2}\frac{\langle\sigma_{\rm ann.}\upsilon\rangle}{{\bar n_b}(z)}\left(\frac{\rho_s}{m_{\rm DM}}\right)^2\frac{4\pi}{3}\left(\frac{r_{\rm vir.}(z,M)}{c_{\rm vir.}(z,M)}\right)^3\nonumber\\
&\times&F_1(c_{\rm vir.}, x_{\rm min.}),\nonumber\\
\label{R_moore}
\end{eqnarray}
where
\begin{eqnarray}
F_1(c_{\rm vir.}, x_{\rm min.})&=&\frac{1}{3}\frac{1}{(1+x_{\rm min.})^2}\nonumber\\
&+&\frac{1}{1.5}\left[\log\left(\frac{ c_{\rm vir.}^{1.5}(1+x_{\rm min.}^{1.5}) }{x_{\rm min.}^{1.5}(1+c_{\rm vir.}^{1.5})}\right)\right]\nonumber\\
&+&\frac{1}{1.5}\left[\frac{1}{1+c_{\rm vir.}^{1.5}}-\frac{1}{1+x_{\rm min.}^{1.5}}\right],\nonumber\\
\label{F_1}
\end{eqnarray}
and
\begin{eqnarray}
\rho_s(M,z)&=&\frac{M}{4\pi\left(\frac{r_{\rm vir.}}{c_{\rm vir.}(M,z)}\right)^3}\nonumber\\
&\times&\left[\frac{1}{1.5}\log\left(\frac{1+c_{\rm vir.}^{1.5}}{1+x_{\rm min.}^{1.5}}\right)+\frac{1}{3}\frac{x_{\rm min.}^{1.5}}{(1+x_{\rm min.}^{1.5})}\right]^{-1}\nonumber\\
&\equiv&\frac{M}{4\pi\left(\frac{r_{\rm vir.}}{c_{\rm vir.}(M,z)}\right)^3F_2(c_{\rm vir.}, x_{\rm min.})}\nonumber\\
\label{rho_s_moore}
\end{eqnarray}
respectively.


\section{Computation of the absorbed fraction}
\label{sec:app-fabs}

\noindent Here we describe the method that we have used to compute the energy absorbed fraction $\fabs$. It is mainly based on the method first described in \cite{Ripamonti:2006gq} (henceforth referred to as RMF07).

We denote with $N(z,z')$ the number of particles (per hydrogen nucleus) produced in the annihilations at redshift $z'$, that are still present (and able to transfer energy to the IGM) at redshift $z\le z'$, and their energy spectrum with ${\rm d}N(z,E,z')/{\rm d}E$. The rate $\dot\epsilon$ of energy absorption per H nucleus at redshift $z$ is obtained by integrating over all energies and production redshifts:
\begin{equation}
\dot \epsilon = \int_z^\infty {\rm d}z'\int {\rm d}E \frac{{\rm d}N}{{\rm d}E}(z,E,z') E \Phi(z,E),
\end{equation}
where $\Phi(z,E)$ is the fractional energy transfer rate to the IGM by a particle with energy $E$ at redshift $z$. A summation over the different particle species produced in the annihilations is also implicit in $\Phi$. The upper integration limit over redshift is formally infinite, but from the numerical point of view it is enough to take a redshift large enough so that all the absorption happens locally and does not give contributions at later times. In our calculations, we have taken, as an upper integration limit, $z_\mathrm{max.}=1500$, and checked explictly that increases in $z_{\rm max.}$ do not significantly alter our results.

This expression can be simplified assuming that the energy spectrum of the particles is very peaked around the mean energy (as it is often the case) and thus approximating the energy spectrum with a Dirac delta function: ${\rm d}N/{\rm d}E(z,z',E) \propto \delta(E-\bar E(z,z'))$, where of course the peak energy $\bar E$ depends on both $z$ and $z'$. The normalization is given by the condition that $\int {\rm d}N/{\rm d}E(z,z',E) {\rm d}E = N(z,z')$. Then:
\begin{equation}
\dot \epsilon = \int_z^{\infty} {\rm d}z' N(z,z') \bar E(z,z') \Phi(z,\bar E(z,z')).
\label{eq:epsdot}
\end{equation}

Finally, the energy absorbed fraction is simply given by the ratio between the energy absorbed and the energy produced by the annihilations:
\begin{equation}
\fabs= \frac{\dot\epsilon}{\frac{1}{2}\frac{n_{\rm DM, 0}^2}{n_{\rm H, 0}} \langle \sigma_{\rm ann}\upsilon\rangle m_{\rm dm} (1+z)^3}
\end{equation}

The problem then reduces to the computation of the two functions $N(z,z')$ and $\bar E(z,z')$ and to the subsequent computation of the integral in Eq.\,(\ref{eq:epsdot}). In the following, we will describe how we computed the evolution of $N$ and $\bar E$ for different DM particles (in our case, SUSY neutralinos and LDM) and annihilation products (in our case, namely photons, electrons and positrons).

In order to compute the evolution of the two quantities $N(z,z')$ and $\bar E(z,z')$ with redshift, it is useful to divide the
possible interactions of the annihilation products between those that result in the loss of a particle, without changing their average energy, and those that, conversely, reduce the average energy without changing the number of particles. An example of the first class is the photoionisation of hydrogen atoms by photons produced in the annihilation, as the photon responsible for the ionisation is effectively removed;  an example of the second class is the Compton scattering of photons over electrons. We shall denote with $\phi_N(z,E)$ and $\phi_E(z,E)$ the interaction rates for the two kind of processes, respectively. Another thing that should be considered is that the total energy loss rate, $\phi = \phi_N + \phi_E$, is not necessarily equal to the total rate of energy transfer, $\Phi$, that appears in  Eq.\,(\ref{eq:epsdot}). The reason is that not all the energy that is lost by the particles is actually transferred to the IGM; for example, as we shall see later, high energy electrons produced in DM annihilations can lose all of their energy very rapidly by inverse Compton scattering on CMB photons, but it can be the case that the up-scattered photons do not subsequently interact with the IGM. The net result is that, although the electrons lose all of their energy, this does not end up heating or ionising the IGM. We shall take this into account using, for a given process, an efficiency factor, $\eta(z,E)\le1$, such that $\Phi = \eta \phi$.

We first consider the absorption of photons. We can write the equations describing the evolution of $N_\gamma(z,z')$ and $\bar E_\gamma(z,z')$ (for $z<z'$) as:
\begin{align}
&\frac{{\rm d}N_\gamma}{{\rm d}z}(z,z')=N_\gamma(z,z')  \frac{\phi_{N,\gamma}[z,\bar E_\gamma(z,z')] }{H(z)(1+z)} \label{eq:dNdz}
\end{align}
and
\begin{align}
&\frac{{\rm d}\bar E_\gamma}{{\rm d}z}(z,z')=\bar E_\gamma(z,z')\left( \frac{\phi_{E,\gamma}[z,\bar E_\gamma(z,z')] }{H(z)(1+z)} + \frac{1}{1+z}\right), \label{eq:dEdz}
\end{align}
where the second term in Eq.\,(\ref{eq:dEdz}) takes into account the cosmological redshifts of photons, and the factor $H(z)(1+z)$ originates from the transformation from cosmological time to redshift. The initial conditions at $z=z'$ for the above system are:
\begin{align}
&N_\gamma(z',z') = \zeta_{\gamma,1} \frac{\dot N_{\rm DM}(z')}{H(z')(1+z')}\\
&\bar E_\gamma(z',z') = \zeta_{\gamma,2} m_{\rm dm} c^2
\end{align} 
where $\dot N_{\rm DM}$ is the rate of decrease of DM particles per H nucleus, $\zeta_{\gamma,1}$ is the average number of photons produced per every annihilating DM particle (i.e., it is equal to the number of photons produced in the annihilation, divided by 2), and $\zeta_{\gamma,2}$ is the average fraction of the DM rest mass energy that goes to each photon. The values of $\zeta_{\gamma,1}$ and $\zeta_{\gamma,2}$ can be easily inferred from the values presented in Table\,\ref{tab:av-en}, while the rate of decrease of DM particles is given by 
\begin{equation}
\dot N_{\rm DM}=\frac{1}{2} \frac{n_{\rm dm}(z)}{n_H(z)}^2 \langle\sigma\upsilon\rangle
\end{equation}

The next step is to model the rates $\phi_N$ and $\phi_E$ in order to include the relevant processes in the energy and redshift ranges of interests. For the supersymmetric models we consider here, the average photon energy ranges from $\sim1\,\GeV$ in model 1 to $\sim10\,\GeV$ in model 2. From \cite{ZS89} and from the discussion in \S\,\ref{sec:int} we know that at these energies the only relevant absorption process is pair production over atoms (if $z<z_{\rm dec.}$) or over ions and free electrons (if $z>z_{\rm dec.}$); even this process is only effective for redshifts larger than a few hundred. At more recent times, the Universe is basically transparent to GeV photons. Compton scattering over electrons should also be taken into account as it can contribute to the absorbed fraction, and is in fact the main mechanism of energy transfer at redshifts $z\lesssim 100$, as we have checked explicitly by computing the absorbed fraction with and without the Compton term. 
Since pair production results in the loss of a photon, while Compton scattering just changes its energy, we take $\phi_N = \phgpp$ and $\phi_E=\phgcom$, and numerically solve Eqs.\,(\ref{eq:dNdz}) and (\ref{eq:dEdz}).

Once the time evolution of $N(z,z')$ and $\bar E(z,z')$ is known, the only ingredient required before we can proceed with the computation of the integral in Eq\,.(\ref{eq:epsdot}) is the estimation of the photon absorption rate, $\Phi_\gamma$, i.e. of the total photon energy loss rate weighted with the absorption efficiency. We can in general write
\begin{equation}
\Phi_\gamma = \sum_i \eta_i \phi_i,
\end{equation}
the sum running over all relevant processes. In the present case, all the energy lost through Compton scattering over electrons directly goes into the IGM, so we can take $\eta_\mathrm{Com.} = 1$. On the other hand, not all electron-positron pairs produced by the interaction of photons actually inject energy into the IGM. We model the subsequent absorption as follows. Every photon will produce an electron and a positron, each with energy $E_{e^\pm}\simeq E_\gamma/2$. These electrons and positrons very quickly lose all of their energy through inverse Compton scattering off of CMB photons. The up-scattered photons have an average energy $E_\gamma' =\gamma^2 k_{\rm B} T_{\rm CMB}^0 (1+z)$, where $\gamma = E_{e^\pm} / (m_e c^2)$ is the Lorentz factor of the electron or positron. Finally, following RMF07, we assume that these secondary photons are absorbed through Compton scattering with an efficiency $\eta$ given by
\begin{equation}
\eta (z,E_\gamma) = f_1\left[1 - e^{-\tau_\gamma(z,\,E_\gamma')}\right],
\end{equation}
where $\tau$ is the optical depth for Compton scattering, given by
\begin{equation}
\tau_\gamma(z,E_\gamma')=\frac{f_2}{H(z)}\phgcom(z,E_\gamma'),
\end{equation}
where $f_1=0.91$ and $f_2=0.6$.

\begin{center}
\begin{figure}
\includegraphics[width=0.9\linewidth,keepaspectratio,clip]{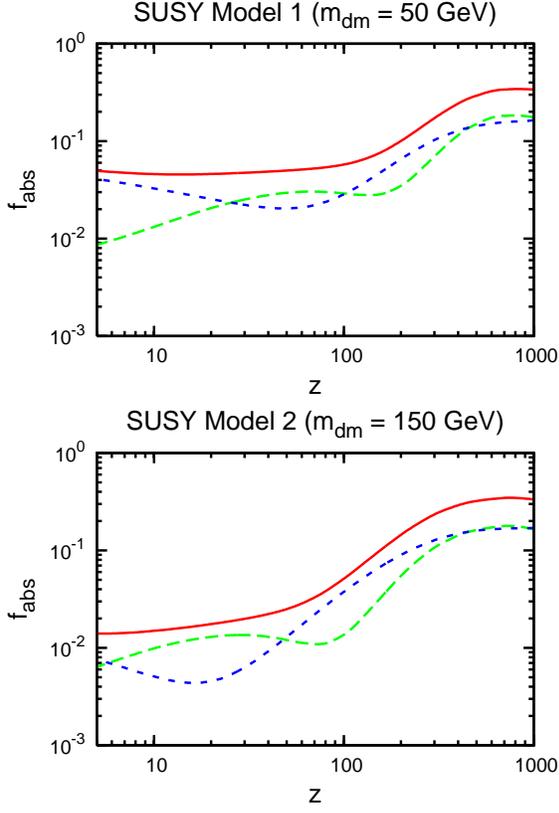}
\caption{Absorbed energy fraction for neutralino models 1 (upper panel) and 2 (lower panel). We show the total absorbed fraction (solid red line) together with the contribution from the photon (long dashed green curves) and electron/positron (short dashed blue curves) channels.}
\label{fig:fabs_susy_A}
\end{figure}
\end{center}

\begin{center}
\begin{figure}
\includegraphics[width=0.9\linewidth,keepaspectratio,clip]{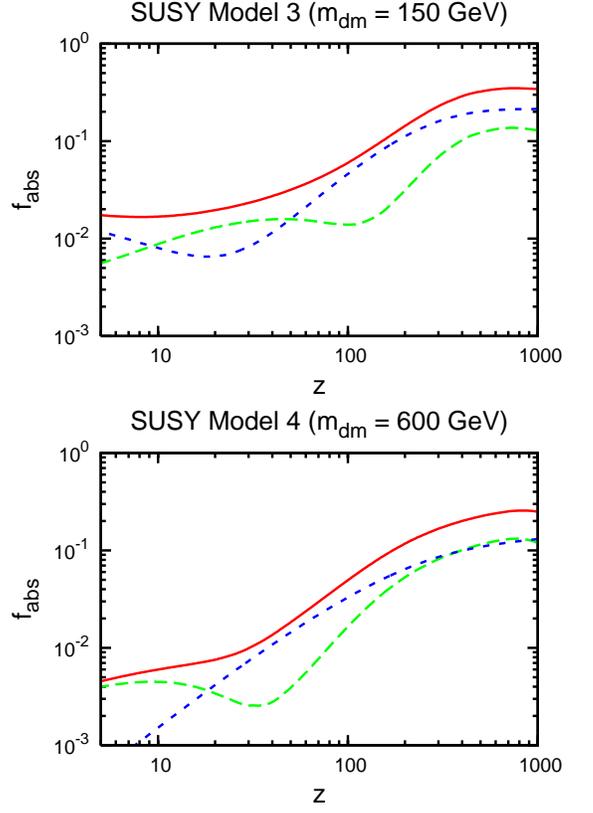}
\caption{Same as for Fig.\,\ref{fig:fabs_susy_A} except now for models 3 (upper panel) and 4 (lower panel).}
\label{fig:fabs_susy_B}
\end{figure}
\end{center}

We can now proceed with the evaluation of the photon channel contribution to $\dot\epsilon$, as given by expression (\ref{eq:epsdot}), and consequently to $\fabs$. The results are shown in Figs.\,\ref{fig:fabs_susy_A} and \ref{fig:fabs_susy_B} (long dashed green lines).

We now turn our attention to the absorption of electrons and positrons produced in neutralino annihilations. Relativistic electrons and positrons lose their energy through inverse Compton scattering off of CMB photons, while slow electrons and positrons lose their energy through collisional ionisation of neutral atoms. In addition, positrons can annihilate over thermal electrons; this process also is very effective for particles with very low kinetic energy, i.e. $E_{e^+}\simeq m_ec^2$. 
Since the relevant energy range is the same as for photons, i.e. $1-10\,\GeV$, we expect inverse Compton scattering to be the only relevant process. In any case, we included collisional ionisation and annihilation over thermal electrons (just for positrons) in our equations and explicitly checked that such terms have a negligible impact on our results.
The evolution equations for ultra-relativistic electrons and positrons then have the same form as the ones for photons:
\begin{align}
&\frac{{\rm d}N_\epm}{{\rm d}z}(z,z')=N_\epm(z,z')  \frac{\phi_{N,\epm}(z,\bar E_\epm(z,z')) }{H(z)(1+z)} \\
&\frac{{\rm d}\bar E_\epm}{{\rm d}z}(z,z')=\bar E_\epm(z,z')\left[ \frac{\phi_{E,\epm}(z,\bar E_\epm(z,z')) }{H(z)(1+z)} + \frac{1}{1+z}\right] 
\end{align}
where we take $\phi_{N,\epm} = \pheion$ and $\phi_{E,\epm} = \phecom$ for electrons, while for positrons we also include the annihilation term in $\phi_N$. As we have just mentioned, one can in practice take $\phi_N= 0$ in both cases and still obtain the same results.

The efficiency of inverse Compton scattering in transferring energy to the IGM can be estimated as follows. The average energy of the upscattered photons is given by $E_\gamma'=\gamma^2 k_{\rm B} T_\CMB^0 (1+z)$; for an electron in the 1 to 10\,GeV energy range, this gives a corresponding photon energy of $(1- 100) (1+z)\,\keV$. This means that, especially at low redshifts, the up-scattered photons can have the right energy to transfer energy to the IGM by ionising neutral atoms. We quantify this effect as follows. The energy $E_\mathrm{ion.}$ corresponding to unitary optical depth for photoionisation (i.e. $\phgion=H$) at a given redshift $z$ is given by
\begin{equation}
E_\mathrm{ion.}(z) = 0.64 \,\keV\times(1+z)^{0.45}.
\end{equation}
We consider that upscattered photons with $E_\gamma'<E_\mathrm{ion.}$ transfer their energy to the IGM, while the others are lost. We then take the efficiency of inverse Compton scattering as being equal to a fraction $F$ of the total CMB energy density carried by photons that after scattering have an energy below the ionisation threshold. The efficiency $\eta_\mathrm{Com.}$ is then given by
\begin{multline}
\eta_\mathrm{Com.}=F(E<E_\mathrm{max.}) = \left[\int_0^{E_\mathrm{max.}/c} p f(p,z) {\rm d}^3p \right] \times \\ \left[\int_0^\infty p f(p',z) {\rm d}^3p \right]^{-1},
\end{multline}
where $f(p,z)=\left[e^{pc/k_{\rm B}T(z)}-1\right]^{-1}$ is the Bose-Einstein distribution, and $E_\mathrm{max.} = E_\mathrm{ion.}(z)/\gamma^2$. Using the dimensionless variable $y=pc/k_{\rm B}T$ this can be rewritten as
\begin{equation}
\eta_\mathrm{Com.}(z) = \frac{\pi^4}{15}\left[\int_0^{y_{\rm max.}} \frac{y^3}{e^y-1}  {\rm d}y \right],
\end{equation}
where $y_{\rm max.}=E_\mathrm{max.} c/k_{\rm B}T$. Then we finally take the absorption rate in Eq.(\ref{eq:epsdot}) to be
\begin{equation}
\Phi_{e^\pm} = \eta_\mathrm{Com.} \phi_\mathrm{Com.}.
\end{equation}
The results for $\fabs$ are displayed in Figs.\,\ref{fig:fabs_susy_A} and \ref{fig:fabs_susy_B}.


In addition to supersymmetric neutralinos, we have also considered the case of MeV light dark matter annihilating directly to $e^+ e^-$ pairs. In this case the spectrum is monochromatic at $E_\epm= m_{\rm DM}c^2$.
We consider two particular cases of LDM with particle masses $m_{\rm DM}= 3$\,MeV and 20\,MeV. The treatment is basically the same as that in the previous section for electrons and positrons produced in neutralino annihilation, apart from the fact that, especially for the 3\,MeV case, the annihilation of positrons over thermal IGM electrons gives an appreciable contribution to the total absorbed energy fraction. Annihilations are only important when the positron has lost all of its kinetic energy through inverse Compton scattering; it can be seen in fact that the energy loss rate for ionisation is always larger than the loss rate for annihilations. Then the only regime where annihilations can be effective is when the kinetic energy of the positron is below the H ionisation threshold at 13.6\,eV, i.e. where $E_{e^+} \simeq m_e c^2$. 

We then use the following scheme to follow the evolution of the $N$ and $\bar E$ for positrons. Based on the argument above, we split the integral in the Eq.\,(\ref{eq:epsdot}) in two parts. The first integral, evaluated from $z' = z$ to $z' = z_1$, takes into account the contribution from positrons that still retain an appreciable fraction of their kinetic energy at redshift $z$, and are thus losing energy mainly through inverse Compton scattering and collisional ionisation. The second integral, evaluated from $z'=z_1$ to $z'=z_{\rm max.}$, takes into account the contribution of positrons that have already lost all their kinetic energy (i.e. slow positrons) and can only lose more energy by annihilating with thermal electrons. The redshift $z_1$ is simply obtained by solving the differential equation for $\bar E$ and finding the redshift where $\bar E = 13.6\,\eV$.

Finally, it should be taken into account that the photons produced in the annihilation will not necessarily transfer all of their energy to the IGM. We follow a method similar to the one described in the previous section, and assume that energy is transferred via Compton scattering, with an efficiency given by
\begin{equation}
\eta_\mathrm{ann.} (z,E_\gamma) = f_1\left[1 - e^{-\tau_\gamma(z,\,E_\gamma')}\right],
\end{equation}
where $\tau$ is the optical depth for Compton scattering defined above, the parameters $f_1$ and $f_2$ have also been defined above, and $E_\gamma = m_ec^2$ since the positron and electron annihilate basically whilst at rest.

The absorbed energy fraction for annihilating MeV DM has been computed explicitly in RMF07 for $m_{\rm DM}=1,\,3 $ and $10\,\MeV$. We have compared our results to theirs in the case of 3\,MeV and although the results are qualitatively similar (the absorbed fraction is appreciable at large redshifts, has a large decrease at intermediate redshifts, and then starts increasing again to $\fabs\sim O(0.1)$ at $z=5$) there are also some quantitative differences. As we could not track down the origin of such discrepancies, we have decided to compute the expected brightness temperature for 3\,MeV annihilating DM particles using both, the version of the function $\fabs$ that we have obtained in this paper ($\fabs^{\rm CLS}$) and the one from RMF07 ($\fabs^{\rm RMF}$), for which they provide an analytic approximation. However, for comparison we have also computed the absorbed fraction for the 10\,MeV case (not considered elsewhere in this paper) and we have found an excellent agreement between our results and those of RMF07, so the discrepancies are likely due to the different handling of positron annihilations, which only contribute to the lower masses. 

\begin{center}
\begin{figure}
\includegraphics[width=0.8\linewidth,keepaspectratio,clip]{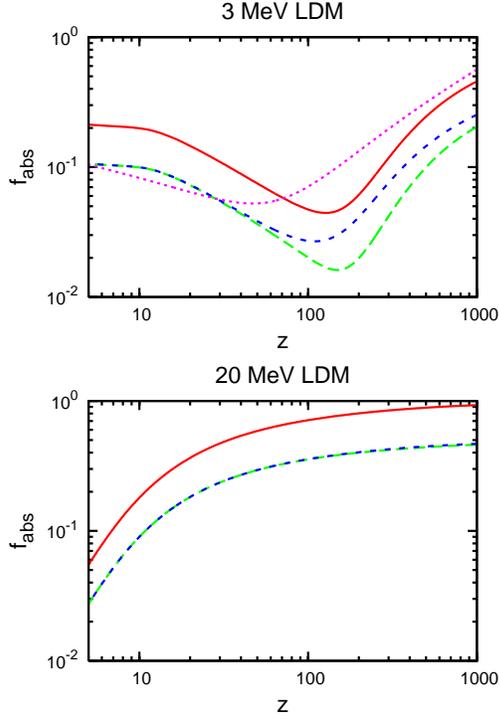}
\caption{Absorbed energy fraction for the 3\,MeV (upper panel) and 10\,MeV (lower panel) LDM. We show the total $\fabs$ (solid red line) together with the contribution from electrons and positrons (long dashed green and short dashed blue lines, respectively). For the 3\,MeV case, we also show the absorbed fraction compute in RMF07 (dotted magenta line).}
\label{fig:fabs_ldm}
\end{figure}
\end{center}

We present our results for the absorbed fraction for MeV DM in Fig.\,\ref{fig:fabs_ldm}, together with the result for $\fabs$ for 3\,MeV LDM from RMF07.


\section{Clumping factor parameters}
\label{sec:app-tables}
\noindent In this section we conveniently list the relevant parameters associated with each of the clumping factors utilised throughout this study, for clumping factors calculated using structures possessing NFW (Table\,\ref{tab:SUSY_NFW}) and Moore (Table\,\ref{tab:SUSY_Moore}) DM density profiles for our four neutralino DM models, and for Burkert profiles using our two LDM candidates (Table\,\ref{tab:LDM}).
\begin{table*}
\begin{center}
\begin{tabular}{lrrcrrrrr}
\hline
& $n_{\rm min.}$ & $n_{\rm cut}$ & $F_{\rm sub.}, F_{\rm ss}$~(\%) & $N_c$ &Model 1&Model 2 & Model 3&Model 4 \\[0.1cm]
\hline
\hline
N0\footnote{Corresponding to the absence of structures, i.e. $C(z)=1$.} 	&   -- 		& -- 		& \phantom{0.}\text{--} & -- & $-0.022$ 	& $-0.007$ 	& $-0.007$ 	& $-0.001$ \\
N1 	&$ -12$ 	& $-12$ 	&\phantom{0.}3 & 3 		& -- 			& --			& --			& --\\
N2 	& $-12$ 	& $-12$ 	&\phantom{0.}3	& 1.5 	& -- 			& $2.35$		& $3.72$ 		& $-1.12$ \\
N3 	& $-12$ 	& $-12$ 	& 0.3 		& 3 		& -- 			& $1.89$		& $3.16$ 		& $-1.07$ \\
N4 	& $-12$ 	& $-12$ 	& 0.3 		& 1.5 	&$18.0$ 		& $1.28$		& $2.41$ 		& $-0.987$ \\
N5 	& $-12$ 	& $6$ 	&\phantom{0.}3 & 3 		& -- 			& -- 			& $-0.905$ 	& $-0.671$ \\
N6 	& $-12$ 	& $6$ 	&\phantom{0.}3 & 1.5 	& $3.95$ 		& $-1.26$ 		& $-1.23$ 		& $-0.366$ \\
N7 	& $-12$ 	& $6$ 	& 0.3 		& 3 		& $4.25$ 		& $-1.28$ 		& $-1.23$ 		& $-0.379$ \\
N8 	& $-12$ 	& $6$	& 0.3 		& 1.5 	& $3.87$ 		& $-1.25$ 		& $-1.22$ 		& $-0.360$ \\
N9 	& $-4$ 	& $-4$ 	&\phantom{0.}3	& 3 		& -- 			& -- 			& -- 			& $-0.587$ \\
N10 	& $-4$ 	& $-4$	&\phantom{0.}3 & 1.5 	& $4.15$ 		& $-1.59$ 		& $-1.59$ 		& $-0.417$ \\
N11 	& $-4$ 	& $-4$	& 0.3 		& 3 		& $3.83$ 		& $-1.57$ 		& $-1.58$ 		& $-0.401$ \\
N12 	& $-4$ 	& $-4$	& 0.3 		& 1.5 	& $3.59$ 		& $-1.55$ 		& $-1.56$ 		& $-0.387$ \\
N13 	& $-4$ 	& $6$	&\phantom{0.}3 & 3 		& -- 			& $-1.10$ 		& $-1.22$ 		& $-0.199$ \\
N14 	& $-4$ 	& $6$	&\phantom{0.}3 & 1.5 	& $-0.993$	& $-0.925$ 	& $-1.04$ 		& $-0.162$ \\ 
N15 	& $-4$ 	& $6$ 	& 0.3 		& 3 		& $-1.02$ 		& $-0.927$ 	& $-1.04$ 		& $-0.162$ \\
N16 	& $-4$ 	& $6$ 	& 0.3 		& 1.5 	& $-1.04$ 		& $-0.912$ 	& $-1.03$ 		& $-0.159$ \\
N17 	& $6$ 	& $6$	&\phantom{0.}3 & 3 		& $-0.034$ 	& $-0.009$ 	& $-0.009$ 	& $-0.002$ \\
N18 	& $6$ 	& $6$	&\phantom{0.}3 & 1.5 	& $-0.034$ 	& $-0.009$ 	& $-0.009$ 	& $-0.002$ \\
N19 	& $6$ 	& $6$	& 0.3 		& 3 		& $-0.033$ 	& $-0.009$ 	& $-0.009$ 	& $-0.002$\\
N20 	& $6$ 	& $6$ 	& 0.3 		& 1.5 	& $-0.033$ 	& $-0.009$ 	& $-0.009$ 	& $-0.002$\\
\hline	
\end{tabular}
\caption{Parameters used for the calculation of the clumping factors for structures with NFW DM density profiles. \\
Column (1) - clumping factor label; \\
Column (2) - $n_{\rm min.} = \log(M_{\rm min.}/M_{\odot})$, where $M_{\rm min.}$ is the minimum halo mass considered [see Eq. (\ref{rate_z})]; \\
Column (3) - $n_{\rm cut} = \log(M_{\rm cut}/M_{\odot})$, where $M_{\rm cut}$ is the mass below which the concentration-mass relation for halos is truncated
[see Eq. (\ref{c-m})]; \\
Column (4) - percentage of the host halo (subhalo) mass contributed by each subhalo (sub-subhalo)  mass decade\\
Column (5) - ratio of concentrations for a subhalo and halo of the same mass located at the same redshift;\\
Column (6 - 9) - value of the difference in the differential brightness temperature relative to the standard ``no DM'' scenario, $\delta T_b - \delta T_{b,0}$\,(mK), evaluated at redshift $z=30$ in the four SUSY models described in the text. A dash -- indicates that the model does not satisfy the constraints on the reionization redshift and/or on the diffuse gamma background (see Sec. \ref{sec:Results} for details).}
\label{tab:SUSY_NFW}
\end{center}
\end{table*}
\begin{table*}
\begin{center}
\begin{tabular}{lrrcrrrrr}
\hline
& $n_{\rm min.}$ & $n_{\rm cut}$ & $F_{\rm sub.}, F_{\rm ss}$~(\%) & $N_c$ &Model 1&Model 2 & Model 3&Model 4 \\[0.1cm]
\hline
\hline
M0\footnote{Corresponding to the absence of structures, i.e. $C(z)=1$.} 	&   -- 		& -- 		& \phantom{0.}\text{--} & -- & $-0.022$ 	& $-0.007$ & $-0.007$ & $-0.001$ \\
M1 	&$ -12$ 	& $-12$ 	&\phantom{0.}3 & 3 		& -- 			& --			& --			&--\\
M2 	& $-12$ 	& $-12$ 	&\phantom{0.}3	& 1.5 	& -- 			& --			& --			&--\\
M3 	& $-12$ 	& $-12$ 	& 0.3 		& 3 		& --			& --			& --			&--\\
M4 	& $-12$ 	& $-12$ 	& 0.3 		& 1.5 	& --			& --			& --			& $6.69$ \\
M5 	& $-12$ 	& $6$ 	&\phantom{0.}3 & 3 		& --			& --			& --			& -- \\
M6 	& $-12$ 	& $6$ 	&\phantom{0.}3 & 1.5 	& -- 			& -- 			& -- 			& $-0.912$ \\
M7 	& $-12$ 	& $6$ 	& 0.3 		& 3 		& -- 			& -- 			& $11.3$ 		& $-0.860$ \\
M8 	& $-12$ 	& $6$	& 0.3 		& 1.5 	& -- 			& $8.61$ 		& $10.6$ 		& $-0.910$ \\
M9 	& $-4$ 	& $-4$ 	&\phantom{0.}3	& 3 		& -- 			& -- 			& -- 			& -- \\
M10 	& $-4$ 	& $-4$	&\phantom{0.}3 & 1.5 	& -- 			& -- 			& -- 			& -- \\
M11 	& $-4$ 	& $-4$	& 0.3 		& 3 		& -- 			& -- 			& -- 			& $-1.20$ \\
M12 	& $-4$ 	& $-4$	& 0.3 		& 1.5 	& -- 			& -- 			& -- 			& $-1.23$ \\
M13 	& $-4$ 	& $6$	&\phantom{0.}3 & 3 		& -- 			& -- 			& $3.62$ 		& -- \\
M14 	& $-4$ 	& $6$	&\phantom{0.}3 & 1.5 	& -- 			& -- 			& $2.09$ 		& $-1.21$ \\
M15 	& $-4$ 	& $6$ 	& 0.3 		& 3 		& -- 			& -- 			& $2.19$ 		& $-1.22$ \\
M16 	& $-4$ 	& $6$ 	& 0.3 		& 1.5 	& -- 			& $0.893$ 	& $2.06$ 		& $-1.20$ \\
M17 	& $6$ 	& $6$	&\phantom{0.}3 & 3 		& -- 			& $-0.025$	& $-0.029$ 	& $-0.004$ \\
M18 	& $6$ 	& $6$	&\phantom{0.}3 & 1.5 	& $-0.147$ 	& $-0.025$ 	& $-0.029$ 	& $-0.004$ \\
M19 	& $6$ 	& $6$	& 0.3 		& 3 		& $-0.147$ 	& $-0.025$ 	& $-0.029$ 	& $-0.004$ \\
M20 	& $6$ 	& $6$ 	& 0.3 		& 1.5 	& $-0.147$ 	& $-0.025$ 	& $-0.029$ 	& $-0.004$ \\
\hline
\end{tabular}
\caption{Same as for Table\,\ref{tab:SUSY_NFW} but for clumping factors associated with structures possessing Moore density profiles.} 
\label{tab:SUSY_Moore}
\end{center}
\end{table*}
\begin{table*}
\begin{center}
\begin{tabular}{lrrcrrrrr}
\hline
& $n_{\rm min}$ & $n_{\rm cut}$ & $F_{\rm sub, ss}(\%)$ & $N_c$&20 MeV                                &20\,MeV&$3\,$MeV                 &$3\,$MeV\\
&       &                          &                                         &             &\phantom{--}$\sv_{28}\footnote{$\sv_{28}\equiv \sv/(10^{-28}\mathrm{cm}^3\mathrm{s}^{-1})$.}=4.4$ & \phantom{--}$\sv_{28}=0.44$ & \phantom{--}$\sv_{28}=1.2$ &\phantom{--}$\sv_{28}=0.12$\\
\hline
B0\footnote{Corresponding to the absence of structures, i.e. $C(z)=1$.} 	&   -- 		& -- 		& \phantom{0.}\text{--} & -- & $-0.022$ 	& $-0.007$ 	& $-0.007$ 	& $-0.001$ \\
B1 	& $ -0.80$ & $-0.80$ &\phantom{0.}3 & 3 		&\text{ --} 			& --			& --			& --\\
B2 	& $ -0.80$ & $-0.80$ &\phantom{0.}3	& 1.5 	& -- 			& --			& --			& --\\
B3 	& $ -0.80$ & $-0.80$ & 0.3 		& 3 		& -- & $20.2$ & -- & -- \\
B4 	& $ -0.80$ & $-0.80$ & 0.3 		& 1.5 	& --  & $20.0$ & -- & -- \\
B5 	& $ -0.80$	& $6$ 	&\phantom{0.}3 & 3 		& -- & -- & -- & -- \\
B6 	& $ -0.80$ & $6$ 	&\phantom{0.}3 & 1.5 	&  -- & 13.1 & -- & -- \\
B7 	& $ -0.80$ & $6$ 	& 0.3 		& 3 		& -- & 12.8 & -- & -- \\
B8 	& $ -0.80$ & $6$	& 0.3 		& 1.5 	& -- & 12.8 & -- & -- \\
B9 	& $1.66$ 	& $1.66$ 	&\phantom{0.}3	& 3 		& -- & -- & -- & -- \\
B10	& $1.66$ 	& $1.66$	&\phantom{0.}3 & 1.5 	& -- & -- & -- & -- \\
B11 	& $1.66$ 	& $1.66$	& 0.3 		& 3 		& -- & -- & -- & $-1.81$ \\
B12 	& $1.66$ 	& $1.66$	& 0.3 		& 1.5 	&  -- & -- & -- & $-1.81$ \\
B13 	& $1.66$ 	& $6$	&\phantom{0.}3 & 3 		&  -- & -- & -- & -- \\
B14 	& $1.66$ 	& $6$	&\phantom{0.}3 & 1.5 	&  -- & -- & -- & $-2.32$ \\
B15 	& $1.66$ 	& $6$ 	& 0.3 		& 3 		& -- & -- & -- & $-2.37$ \\
B16 	& $1.66$ 	& $6$ 	& 0.3 		& 1.5 	& -- & -- & -- & $-2.37$ \\
B17 	& $6$ 	& $6$	&\phantom{0.}3 & 3 		&  -- & $-0.552$ & -- & $-0.230$ \\
B18 	& $6$ 	& $6$	&\phantom{0.}3 & 1.5 	& -- & $-0.552$ & -- & $-0.230$ \\
B19 	& $6$ 	& $6$	& 0.3 		& 3 		&  $0.284$ & $-0.542$ & $-1.01$ & $-0.225$ \\
B20 	& $6$ 	& $6$ 	& 0.3 		& 1.5 	& $0.284$ & $-0.542$ & $-1.01$ & $-0.225$\\
\hline
\end{tabular}
\caption{Same as for Table\,\ref{tab:SUSY_NFW} but for clumping factors associated with structures possessing Burkert density profiles, using LDM of mass 20\,MeV [columns (6) and (7)] and 3 MeV [columns (8) and (9)]).}
\label{tab:LDM} 
\end{center}
\end{table*}
\label{lastpage}
\end{document}